\documentclass[a4paper,11pt]{article}
\PassOptionsToPackage{unicode}{hyperref}
\usepackage{jheppub}
\usepackage[T1]{fontenc}
\usepackage{amsmath}
\usepackage{amssymb}
\usepackage{graphicx}
\usepackage{booktabs}
\usepackage{comment}
\usepackage{multirow}
\usepackage{subcaption}
\usepackage[section]{placeins}
\usepackage{float}
\usepackage{slashed}
\usepackage{xcolor}

\graphicspath{{Lambdac decay/}}
\allowdisplaybreaks[3]
\AtBeginDocument{%
  \setlength{\abovedisplayskip}{7pt plus 2pt minus 2pt}%
  \setlength{\belowdisplayskip}{7pt plus 2pt minus 2pt}%
  \setlength{\abovedisplayshortskip}{5pt plus 2pt minus 2pt}%
  \setlength{\belowdisplayshortskip}{5pt plus 2pt minus 2pt}%
}

\title{\boldmath Angular distribution of $\Lambda_c^+ \to \Lambda^*_{1405/1520}(\to\Sigma\pi)\ell^+\nu_\ell$ and implications for form factors}

\author[1]{Kaiwen Chen}
\author[1]{, Zhi-Peng Xing}
\author[2,3]{, Xudong Yu}
\author[1]{ and Ruilin Zhu}
\affiliation[1]{Department of Physics and Institute of Theoretical Physics, Nanjing Normal University, Nanjing 210023, Jiangsu, China and
Nanjing Key Laboratory of Particle Physics and Astrophysics,
Nanjing Normal University, Nanjing, Jiangsu 210023, China}
\affiliation[2]{School of Physics, Peking University, Beijing 100871, China}
\affiliation[3]{ School of Physical Sciences, University of Chinese Academy of Sciences, Beijing 100049, China}
\emailAdd{kwchen@nnu.edu.cn}
\emailAdd{zpxing@nnu.edu.cn}
\emailAdd{yuxd@stu.pku.edu.cn}
\emailAdd{rlzhu@njnu.edu.cn}
\abstract{We study the cascade semileptonic decay $\Lambda_c^+\to\Lambda^*_{1405/1520}(\to\Sigma\pi)\ell^+\nu_\ell$ using the helicity amplitude formalism, and derive the full four-body angular distribution. The contributions from the spin-$1/2$ and spin-$3/2$ intermediate resonances, their interference effects, and the corresponding angular coefficients are obtained explicitly.
Using available form-factor inputs, we provide numerical predictions for the invariant-mass distribution, helicity-angle distributions, and angular asymmetries. Based on these observables, we discuss strategies for extracting the relevant form factors and analyze the sensitivity of the angular observables to variations in the form factors. Finally, we investigate the strong phases in the $\Lambda_c^+\to\Lambda^*$
form factors and discuss how the strong-phase difference can be extracted from the angular distribution. Our results offer useful theoretical guidance for future experimental analyses.}

\begin{document}
\maketitle
\hypersetup{
  pdftitle={Angular distribution of Lambdac to Lambda*(1405/1520)(to Sigma pi) l nu},
  pdfauthor={Kaiwen Chen, Zhi-Peng Xing, Ruilin Zhu}}
\flushbottom

\section{Introduction}
\label{sec:intro}

In recent years, significant progress has been made in the study of charm hadron spectroscopy and inner structures~\cite{BESIII:2013ouc,Belle:2014nuw,LHCb:2019kea,LHCb:2020bwg,CMS:2023owd}. Several unexpected multiquark states have been discovered, which have also updated our understanding~\cite{Chen:2016qju,Lu:2016ogy,Guo:2017jvc,Cheng:2018hwl,Chen:2019asm,Zhu:2020xni,Zhu:2015bba,Wang:2025hex,Wang:2017vnc,He:2020jna}. Because the charm quark mass lies between the perturbative and non-perturbative QCD scales, the physics related to charm hadrons is remarkably rich.

As the ground state of charmed baryons, $\Lambda_c^+$ has been extensively investigated at various experimental facilities such as BESIII, Belle~II, and LHCb~\cite{Belle:2013jfq,BESIII:2015bjk,LHCb:2017xtf}.   For its decay behaviors, semi-leptonic $\Lambda_c^+$ decays have received considerable attention from both experimental and theoretical perspectives, because they are important for precision tests of the Standard Model and searches for new physics.  Since the strong and weak interactions can be cleanly separated in semi-leptonic $\Lambda_c^+$ decays, these processes offer unique opportunities to study the Cabibbo-Kobayashi-Maskawa (CKM) matrix parameters, test Lepton Flavor Universality (LFU), examine SU(3) flavor symmetry, and constrain the charmed baryon lifetime~\cite{Gutsche:2014zna,Gronau:2018vei,Cheng:2021vca,Wang:2022tcm,Geng:2023pkr,Ke:2023qzc,Xing:2024nvg,Cheng:2025oyr,Li:2025nzx,Xing:2026ohi,Cheng:2025kpp,Hu:2026drh,Duan:2026gly}.  The recent theoretical studies of $\Lambda_c^+$ can be  seen in Refs.~\cite{Lu:2016ogy,Faustov:2016yza,Geng:2017mxn,Zhao:2018zcb,Zou:2019kzq,Li:2021iwf,Li:2021qod,Jia:2024pyb}.

 The amplitudes of semi-leptonic decays can be expressed in terms of several form factors.  Since the form factors contain hadronic information, they are useful for studying the hadron structure, particularly for excited states like $\Lambda^*_{1405}$, which is controversial in spectroscopic studies~\cite{CLAS:2013rjt,Li:2021qod}. In addition to their relevance for studying hadron structure, the form factors entering the hadronic matrix element depend on $q^2$, the momentum transfer between the initial and final hadrons.  Different theoretical methods have different ranges of applicability in $q^2$, such as the high-$q^2$ region for Lattice QCD and the low-$q^2$ region for PQCD~\cite{Zhang:2021oja,Rui:2025iwa}.  Determining the form factors over the full $q^2$ region is therefore a difficult issue, and extracting them from the theoretical side remains a challenging and important problem. Conversely, experimental data can be also used to constrain or extract the form factors~\cite{CLEO:2006uty,BESIII:2018xre,BESIII:2026txt}. 
 
Primary charmed baryon semi-leptonic decays such as $\Lambda_c^+ \to \Lambda^* \ell^+ \nu_\ell$ can be studied experimentally through the cascade four-body decay. Since the angular distribution of this four-body decay provides rich observables, it is natural to use these observables to determine the form factors. Angular distribution analysis helps to understand the structure and spin-parity of hadrons and relevant references can be consulted to Refs.~\cite{Xing:2022uqu,Xing:2022phq,Zhao:2024ren,Chen:2024orv,Zhang:2025mne}. Motivated by the large branching fraction of
  $\Lambda^*(1405)\to\Sigma\pi$, we study the angular distributions of
  $\Lambda_c^+ \to \Lambda^*(1405,1520)(\to\Sigma\pi)\ell^+\nu_\ell$
  and develop a method to extract the
  $\Lambda_c^+\to \Lambda^*(1405)$ and
  $\Lambda_c^+\to \Lambda^*(1520)$ transition form factors from the
  corresponding angular observables. Since this process has not yet been measured experimentally, our work focuses on the feasibility of extracting these form factors, including phase information, and aims to guide future experimental analyses.
 
The rest of this paper is organized as follows. In Sec.~II, we construct the decay amplitude, including the kinematics and the lineshape of the excited states in this four-body decay.  The definition of the form factors, the helicity-basis conversion, and the resulting helicity amplitudes are discussed in Sec.~III. In Sec.~IV, the angular distribution analysis is presented using the helicity amplitude method, and the numerical results---including predictions for observables and the form-factor dependence---are given in Sec.~V. A brief summary is presented in the last section. Some details of the calculation are collected in the appendix.

\section{Decay amplitude}
\label{sec:amplitude}
\subsection{Helicity amplitude}
\begin{figure}[htbp]
	\centering
	\includegraphics[width=0.7\textwidth]{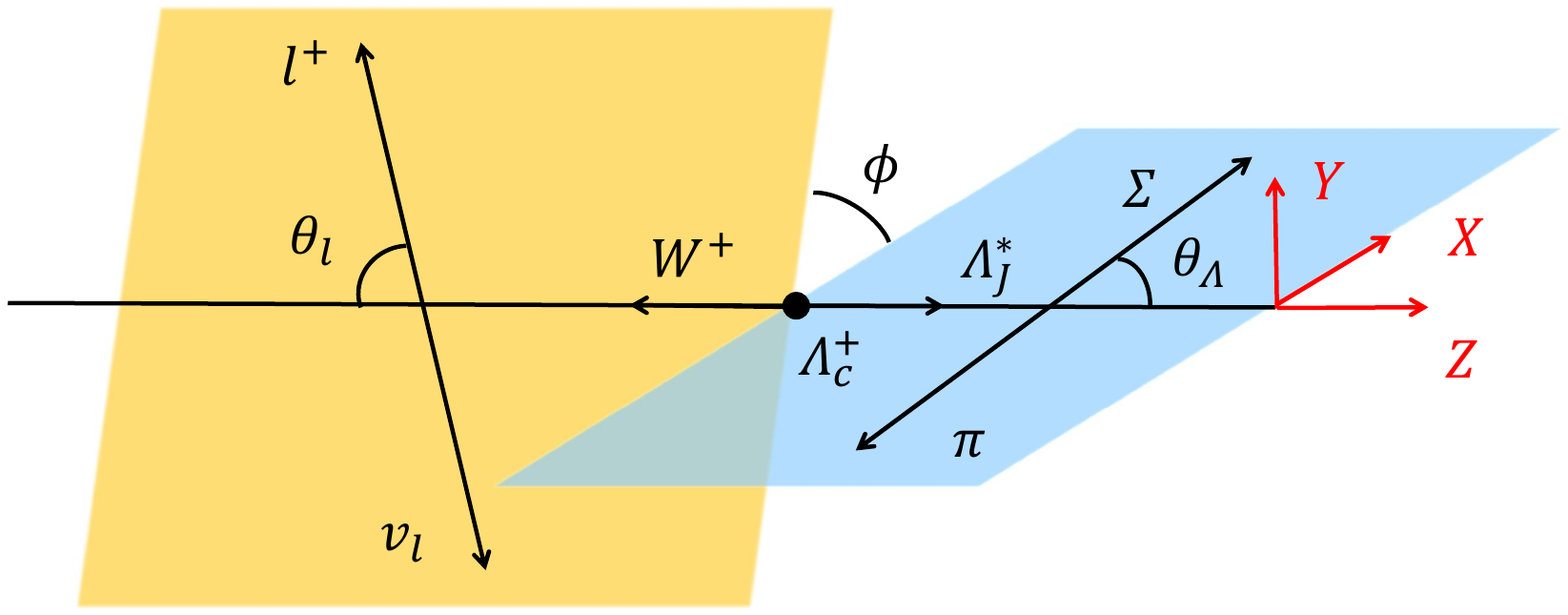}
	\caption{Kinematics and angular definitions for the decay $\Lambda_c^+ \to \Lambda^*_J(\to \Sigma\pi) \ell^+\nu_\ell$.}
	\label{Angle}
\end{figure}

For the cascade decay $\Lambda_c^+ \to \Lambda_J^*(\to \Sigma\pi)\ell^+\nu_\ell$ with the spin quantum number $J$, the kinematics can be described by five independent variables: the momentum transfer between the initial and final hadrons $q^2$, the $\Sigma\pi$ invariant mass squared $m_{\Sigma\pi}^2$, and three helicity angles $(\theta_\Lambda, \theta_\ell, \phi)$, as defined in Fig.~\ref{Angle}. In the rest frame of the $\Lambda_c^+$ baryon, $\Lambda^{*}_J$ moves along the $z$-axis. The angle $\theta_\ell$ ($\theta_\Lambda$) is defined as the angle between the negative (positive) $z$-axis and the direction of motion of $\ell^+$ ($\Sigma$) in the $W^+$ ($\Lambda^{*}_J$) rest frame. The angle $\phi$ is the angle between the $\Lambda^{*}_J$ and virtual $W^+$ cascade decay planes. 
Its decay width can be written as
\begin{eqnarray}
	d\Gamma_4 &=& \frac{1}{2m_{\Lambda_c}}d\phi_4\times \frac{1}{2}\sum_{spin}|\mathcal M|^2,\\
    &&{\rm with}\notag\\
d\phi_4&=&\frac{\sqrt{\lambda(m_{\Lambda_c},m_{\Sigma\pi},\sqrt{q^2})}\,(1-\frac{m_\ell^2}{q^2})\,\sqrt{\lambda(m_{\Sigma\pi},m_\Sigma,m_\pi)}}{16384\pi^6 m_{\Lambda_c}^2  m_{\Sigma\pi}^2}
d\cos\theta_\Lambda d\cos\theta_\ell d\phi dq^2dm_{\Sigma\pi}^2,\notag
	\label{eq:phasespace}
\end{eqnarray}
where $\lambda(m_1,m_{2},m_{3})=((m_1+m_{2})^{2}-m_{3}^{2})((m_1-m_{2})^{2}-m_{3}^{2})$.

For the decay $\Lambda_c^+ \to \Lambda^*_{1405}/\Lambda^*_{1520}(\to \Sigma\pi)\ell^+\nu_\ell$, the amplitude reads
\begin{eqnarray}
	i\mathcal{M} = -\sum_{s_{\Lambda^*}} M(\Lambda_c^+ \to \Lambda^*\ell^+\nu_\ell) L_{\Lambda^*}(p_{\Lambda^*}^2,m_{\Lambda^*}) iM(\Lambda^* \to \Sigma\pi).
	\label{eq:total_amp}
\end{eqnarray}
Here $p_{\Lambda^*}$ is the momentum of the resonance $\Lambda^*$, and it satisfies $p_{\Lambda^*}^2=m_{\Sigma\pi}^2$. The total amplitude can be divided into two parts: the semi-leptonic $\Lambda_c^+$ decay amplitude $M(\Lambda_c^+ \to \Lambda^*\ell^+\nu_\ell)$ and the $\Lambda^*$ strong decay amplitude $M(\Lambda^* \to \Sigma\pi)$. A Breit-Wigner-type lineshape $L_{\Lambda^*}(p_{\Lambda^*}^2,m_{\Lambda^*})$ is used to describe the resonance $\Lambda^*$.
The semi-leptonic decay is induced by the $c\to s \ell^+\nu_\ell$ weak decay Hamiltonian, which is 
\begin{eqnarray}
	H_{\rm eff} = \frac{G_F}{\sqrt{2}}V_{cs}^*\,[\bar{s}\gamma_{\mu}(1-\gamma_5)c]\,[\bar{\nu}_\ell\gamma^{\mu}(1-\gamma_5)\ell].
	\label{eq:weak_hamiltonian}
\end{eqnarray}
Then the matrix element can be factorized into a product of leptonic and hadronic helicity amplitudes as
\begin{eqnarray}
	M(\Lambda_c^+ \to \Lambda^*\ell^+\nu_\ell)  
	&=& \frac{G_F}{\sqrt{2}}V_{cs}^* \sum_{\lambda} \langle \ell^+\nu_\ell |[\bar{\nu}_\ell\slashed{\epsilon}(1-\gamma_5)\ell]| 0 \rangle \langle \Lambda^* |[\bar{s}\slashed{\epsilon}^*(1-\gamma_5)c]| \Lambda_c^+ \rangle \notag \\
	&=& \frac{G_F}{\sqrt{2}}V_{cs}^* \sum_{\lambda} L_{\lambda}(\theta_\ell,\phi) H_{\lambda},
	\label{eq:weak_matrix_element}
\end{eqnarray}
where $\lambda = t, \pm1, 0$ denotes the helicity of the virtual $W$ boson, and the polarization vectors satisfy $g_{\mu\nu} = \sum_{\lambda} \epsilon_\mu^*(\lambda)\epsilon_\nu(\lambda)$. The leptonic helicity amplitude is defined as
\begin{eqnarray}
	L_{\lambda}(\theta_\ell,\phi) = \langle \ell^+\nu_\ell |[\bar{\nu}_\ell\slashed{\epsilon}(1-\gamma_5)\ell]| 0 \rangle = \bar{u}_{\nu_\ell}\slashed{\epsilon}(1-\gamma_5)v_{\ell},
	\label{eq:lepton_amplitude_def}
\end{eqnarray}
and the hadronic helicity amplitude is
\begin{eqnarray}
	H_{\lambda} = \langle \Lambda^* |[\bar{s}\slashed{\epsilon}^*(1-\gamma_5)c]| \Lambda_c^+ \rangle.
	\label{eq:hadron_amplitude_def}
\end{eqnarray}

For the strong decay $\Lambda^* \to \Sigma\pi$, the matrix element can be written in terms of the Wigner $D$-function:
\begin{eqnarray}
	M(\Lambda^* \to \Sigma\pi) = \mathcal{A}_{J}\times\big(D^{J_{\Lambda^*}}_{s_{\Lambda^*},s_{\Sigma}}(\theta_\Lambda)\big)^*,
	\label{eq:strong_decay}
\end{eqnarray}
where $J_{\Lambda^*}$ is the spin of the resonance, and $\mathcal{A}_J$ is the amplitude without angular dependence. Using the experimentally measured decay width of the two-body decay $\Lambda^* \to \Sigma\pi$, the amplitude $\mathcal{A}_J$ can be extracted as
\begin{eqnarray}
	\mathcal{A}_J &=& \sqrt{\Gamma(\Lambda_J^* \to \Sigma\pi)\,(2 J_{\Lambda^*}+1)4\pi m_{\Lambda_J^*}^2/|\vec{p}_\Sigma|}, 
	\label{eq:coupling}
\end{eqnarray}
where $J_{\Lambda^*}$ is the spin of the resonance $\Lambda^*$.
Combining Eqs.~(\ref{eq:total_amp})--(\ref{eq:strong_decay}), the total decay amplitude for the four-body cascade decay reads
\begin{eqnarray}
	i\mathcal{M} 
    &=& -i\frac{G_FV_{cs}^*}{\sqrt{2}}\sum_{s_{\Lambda^*},\lambda} L_{\lambda}(\theta_\ell,\phi) H_{\lambda}  \mathcal{A}_{J} \big(D^{J_{\Lambda^*}}_{s_{\Lambda^*},s_{\Sigma}}(\theta_\Lambda)\big)^*L_{\Lambda^*}(p_{\Lambda^*}^2,m_{\Lambda^*}) \notag \\
    &=&-i\frac{G_FV_{cs}^*}{\sqrt{2}}A^{s_{\Lambda_c}}_{s_{\Sigma},\lambda}(\theta_\Lambda)L_{\lambda}(\theta_\ell,\phi),
	\label{eq:full_amplitude}
\end{eqnarray}
where $A^{s_{\Lambda_c}}_{s_{\Sigma},\lambda}(\theta_\Lambda)$ is the combined hadronic helicity amplitude defined for convenience. For $\Lambda^*_{1405}$ and $\Lambda^*_{1520}$, it can be expressed as 
\begin{eqnarray}
	A^{s_{\Lambda_c}}_{s_{\Sigma},\lambda}(\theta_\Lambda) &=& \sum_{J=\frac{1}{2},\frac{3}{2}} \mathcal{H}^{J}_{s_{\Lambda_c},\lambda}(D^{J_{\Lambda^*}}_{s_{\Lambda^*},s_{\Sigma}}(\theta_\Lambda))^*,
	\notag \\
	\mathcal{H}^{\frac{3}{2}}_{s_{\Lambda_c},\lambda} &=&\mathcal{A}_{J} L_{\Lambda^*_{1520}}(m_{\Sigma\pi}^2,m_{\Lambda^*})\,H_{\lambda}(\Lambda_c^+ \to \Lambda^*_{1520}),
	\notag \\
	\mathcal{H}^{\frac{1}{2}}_{s_{\Lambda_c},\lambda} &=&\mathcal{A}_{J} L_{\Lambda^*_{1405}}(m_{\Sigma\pi}^2,m_{\Lambda^*})\,H_{\lambda}(\Lambda_c^+ \to \Lambda^*_{1405}),
	\label{eq:A_def}
\end{eqnarray}
where angular momentum conservation requires $s_{\Lambda_c} = s_{\Lambda^*} - \lambda$.
The coherent sum over the spin-$\frac{1}{2}$ and spin-$\frac{3}{2}$ intermediate
states is essential for studying the angular distribution of the four-body channel. After squaring the amplitude, it
generates not only the individual $\Lambda^*_{1405}$ and $\Lambda^*_{1520}$
contributions, but also interference terms proportional to the product of their
lineshapes and helicity amplitudes, which are sensitive to angular observables such as the forward-backward asymmetry~\cite{Xing:2022uqu}. 

\subsection{The lineshapes for resonances $\Lambda^*_{1405}$ and $\Lambda^*_{1520}$}
\label{sec:flatte}

For the lineshape of the $\Lambda^*_{1405}$ resonance, the simple Breit--Wigner form is known to be insufficient due to the proximity of the $\bar{K}N$ threshold~\cite{CLAS:2013rjt}. Therefore, we adopt the Flatt\'{e} parametrization of Ref.~\cite{CLAS:2013rjt}, in which the propagator is modified to
\begin{eqnarray}
	L_{\Lambda^*_{1405}}(m_{\Sigma\pi}^2,m_{\Lambda^*})=\frac{1}{m_{1405}^2-m_{\Sigma\pi}^2-im_{1405}(\Gamma_1(q_1)+\Gamma_2(q_2))},
	\label{eq:flatte}
\end{eqnarray}
where 
\begin{eqnarray}
	\Gamma_i(q) = \Gamma^0_i\,\frac{m_{1405}|\vec{q}_i|}{m_{\Sigma\pi}|\vec{q}_0|}.
	\label{eq:flatte_gamma}
\end{eqnarray}
In the above equation, $|\vec{q}_0|$ is the magnitude of the momentum of the final-state particle when the resonance decays at its central mass $m_{1405}$, while $|\vec{q}_i|$ is the magnitude of the momentum of the final-state particle when the resonance decays at a given invariant mass. The index 
$i$ can take the value 1 or 2, representing the momenta of the final-state particles in the two decay channels $\Sigma\pi$ and $\bar{K}N$, respectively. The quantities $\Gamma_1$ and $\Gamma_2$ correspond to the decay widths of the $\Sigma\pi$ and $\bar{K}N$ channels, respectively. Since only the $\Sigma\pi$ final state is studied in our work, when $\sqrt{q_2^2}$ lies below the $\bar{K}N$ threshold, the momentum $|\vec{q}_2|$ corresponding to the invariant mass $m_{\bar K N}$ is not physical and becomes purely imaginary, following a treatment similar to that used in the PDG~\cite{ParticleDataGroup:2020ssz}, which provides the characteristic threshold cusp behavior. The parameters $\Gamma^0_1$ and $\Gamma^0_2$ are taken from the experimental fit~\cite{CLAS:2013rjt}:
\begin{eqnarray}
	\Gamma^0_1 = 0.085\,\text{GeV}, \qquad \gamma = \Gamma^0_2/\Gamma^0_1 = 0.91.
	\label{eq:flatte_params}
\end{eqnarray}
For consistency, we adopt the optimal-fit mass $m_{1405} = 1.338\,\text{GeV}$ from the same analysis. For the lineshape of $\Lambda^*_{1520}$, the standard Breit-Wigner parametrization is sufficient:
\begin{eqnarray}
L_{\Lambda^*_{1520}}(m_{\Sigma\pi}^2,m_{\Lambda^*})=\frac{1}{m_{1520}^2-m_{\Sigma\pi}^2-im_{1520}\Gamma_{1520}},
\end{eqnarray}
where $m_{1520}$ is the central mass of $\Lambda^*_{1520}$ and $\Gamma_{1520}$ is the decay width of $\Lambda^*_{1520}$.
\section{Semi-leptonic decay form factors }
\label{sec:ff}
For the semi-leptonic decay $\Lambda_c^+ \to \Lambda^*\ell^+\nu_\ell$, the hadronic matrix element can be expressed in terms of several form factors corresponding to the spin and parity quantum numbers.  

\subsection{Semi-leptonic decay form factors of spin-$\frac{1}{2}^+$ to spin-$\frac{3}{2}^-$ processes}

For the spin-$\frac{3}{2}$ channel, the helicity-basis parametrization is particularly useful because the form factors are directly tied to definite virtual-$W$ helicity components. This makes the connection between the form-factor inputs and the angular coefficients transparent.
In our work, we adopt the helicity-basis form factor decomposition of Ref.~\cite{Meinel:2020owd}:
\begin{eqnarray}
	\langle \Lambda^*(p',s')|\bar{s}\gamma^\mu c|\Lambda_c^+(p,s) \rangle &=&
	\bar{u}_{\lambda}(p',s')\Bigg(
	f_0^{(\frac{3}{2}^-)} \frac{ m_{\Lambda^*}}{s_+}\,\frac{(m_{\Lambda_c}-m_{\Lambda^*})\,p^\lambda q^\mu}{q^2}   \notag \\
    &&+ f_+^{(\frac{3}{2}^-)} \frac{m_{\Lambda^*}}{s_-}
    \,\frac{(m_{\Lambda_c}+m_{\Lambda^*})\, p^\lambda}{q^2\, s_+}
    \Big(q^2(p^\mu+p'^{\mu}) - (m_{\Lambda_c}^2-m_{\Lambda^*}^2) q^\mu\Big)   \notag \\
	&&+ f_\perp^{(\frac{3}{2}^-)} \frac{m_{\Lambda^*}}{s_-} \left(p^\lambda \gamma^\mu - \frac{2\, p^\lambda(m_{\Lambda_c}p'^{\mu} + m_{\Lambda^*} p^\mu)}{s_+}    \right)  \notag \\
	&&+ f_{\perp'}^{(\frac{3}{2}^-)} \frac{m_{\Lambda^*}}{s_-}
	\left( p^\lambda \gamma^\mu - \frac{2\, p^\lambda p'^{\mu}}{m_{\Lambda^*}}\right.
	\notag \\
	&&\quad\left.
	+\frac{2\, p^\lambda(m_{\Lambda_c}p'^{\mu} + m_{\Lambda^*} p^\mu  )}{s_+}
	+ \frac{s_-\,  g^{\lambda\mu}}{m_{\Lambda^*}}\right)
	\Bigg)u(p,s),\notag\\
\langle \Lambda^*(p',s')|\bar{s}\gamma^\mu\gamma_5 c|\Lambda_c^+(p,s) \rangle &=&
	\bar{u}_{\lambda}(p',s')\Bigg(
	-g_0^{(\frac{3}{2}^-)}\gamma_5\,\frac{m_{\Lambda^*}}{s_-}\frac{(m_{\Lambda_c}+m_{\Lambda^*})\,p^\lambda q^\mu}{q^2}  \notag \\
    &&-g_+^{(\frac{3}{2}^-)}\gamma_5\,\frac{m_{\Lambda^*}}{s_+}
    \frac{(m_{\Lambda_c}-m_{\Lambda^*})\, p^\lambda}{q^2\, s_-}
    \Big(q^2(p^\mu+p'^{\mu}) \notag\\
    &&- (m_{\Lambda_c}^2-m_{\Lambda^*}^2) q^\mu\Big)  \notag \\
	&&-g_\perp^{(\frac{3}{2}^-)}\gamma_5 \frac{m_{\Lambda^*}}{s_+}\left(p^\lambda \gamma^\mu - \frac{2\, p^\lambda(m_{\Lambda_c}p'^{\mu} - m_{\Lambda^*} p^\mu  )}{s_-}    \right)  \notag \\
	&&-g_{\perp'}^{(\frac{3}{2}^-)}\gamma_5 \frac{m_{\Lambda^*}}{s_+}
	\left(p^\lambda \gamma^\mu + \frac{2\, p^\lambda p'^{\mu}}{m_{\Lambda^*}}\right.
	\notag \\
	&&\quad\left.
	+ \frac{2\, p^\lambda(m_{\Lambda_c}p'^{\mu}  - m_{\Lambda^*} p^\mu )}{s_-}
	- \frac{s_+\,  g^{\lambda\mu}}{m_{\Lambda^*}}\right)
	\Bigg)u(p,s),
	\label{eq:ff_32_axial}
\end{eqnarray}
with
\begin{eqnarray}
	s_\pm = (m_{\Lambda_c}\pm m_{\Lambda^*})^2 - q^2.\notag
	\label{eq:splusminus}
\end{eqnarray}
Each form factor can be parametrized as
\begin{eqnarray}
	f(q^2) = \frac{1}{1-q^2/(m_{\rm pole}^f)^2} \sum_{n=0}^{2} a_n^f (w-1)^n,
	\label{eq:ff_param}
\end{eqnarray}
where
\begin{eqnarray}
	 w(q^2) = (m_{\Lambda_c}^2+m_{1520}^2-q^2)/(2m_{\Lambda_c}m_{1520}).\notag
	\label{eq:w_def}
\end{eqnarray}
The pole masses are listed in Table~\ref{tab:polemassesLcLstar}. The fit parameters $a_n^f$ from the Lattice QCD calculation of Ref.~\cite{Meinel:2021mdj} are given in Table~\ref{tab:LcLStarFFparams}. The unlisted $a_0^f$ parameters can be determined using
\begin{eqnarray}
	a_0^{f_{\perp}^{(\frac{3}{2}^-)}} &=& -a_0^{f_{\perp'}^{(\frac{3}{2}^-)}},\quad
	a_0^{f_{+}^{(\frac{3}{2}^-)}} = 2\frac{m_{\Lambda_c}-m_{1520}}{m_{\Lambda_c}+m_{1520}} a_0^{f_{\perp'}^{(\frac{3}{2}^-)}}, \notag\\
	a_0^{g_0^{(\frac{3}{2}^-)}} &=& 0,\quad a_0^{g_{+}^{(\frac{3}{2}^-)}} = a_0^{g_{\perp}^{(\frac{3}{2}^-)}}-a_0^{g_{\perp'}^{(\frac{3}{2}^-)}},\label{eq:a0fstart}
\end{eqnarray}
and the $a_2^f$ parameters can be determined using
\begin{eqnarray}
	 a_2^{f_0^{(\frac{3}{2}^-)}} &=& -\frac{1}{(w_0-1)^2}\Bigg[ a_0^{f_0^{(\frac{3}{2}^-)}}+ a_1^{f_0^{(\frac{3}{2}^-)}} (w_0-1) -\frac{(m_{\Lambda_c}+m_{1520})^2 }{(m_{\Lambda_c}-m_{1520})^2}
	\notag\\
    &&\times\left(2 a_0^{f_{\perp'}^{(\frac{3}{2}^-)}} \frac{m_{\Lambda_c}-m_{1520}}{m_{\Lambda_c}+m_{1520}}
	+(w_0-1)
	\Big(a_1^{f_{+}^{(\frac{3}{2}^-)}}
	+a_2^{f_{+}^{(\frac{3}{2}^-)}} (w_0-1)\Big)\right)\Bigg], \label{eq:a2fstart}\notag \\
	 a_2^{g_0^{(\frac{3}{2}^-)}} &=&
	 -\frac{1}{(w_0-1)^2}\Bigg[
	 a_1^{g_0^{(\frac{3}{2}^-)}} (w_0-1)
	-\frac{(m_{\Lambda_c}-m_{1520})^2}
	{(m_{\Lambda_c}+m_{1520})^2}\notag\\
    &&\times
	\left(a_0^{g_{\perp}^{(\frac{3}{2}^-)}}
	-a_0^{g_{\perp'}^{(\frac{3}{2}^-)}}
	+(w_0-1)\Big(
	a_1^{g_{+}^{(\frac{3}{2}^-)}}
	+a_2^{g_{+}^{(\frac{3}{2}^-)}}
	(w_0-1)\Big)\right)\Bigg].
	\label{eq:a0fend}
\end{eqnarray}

\begin{table}[ht]
	\centering
	\caption{\label{tab:polemassesLcLstar}Pole masses $m_{\rm pole}^f$ used in the $\Lambda_c^+\to\Lambda^*_{1520}$ form-factor parametrization\cite{Zyla:2020zbs}. They correspond to the nearest $D_s$ meson states with the quantum numbers of current and $J^P$ shown.}
	\begin{tabular}{p{0.67\textwidth}cc}
		\hline\hline
		$f$ & $J^P$ & $m_{\rm pole}^f$ [GeV]  \\
		\hline
		$f_+^{(\frac32^-)}$, $f_\perp^{(\frac32^-)}$, $f_{\perp'}^{(\frac32^-)}$, $h_+^{(\frac32^-)}$, $h_\perp^{(\frac32^-)}$, $h_{\perp'}^{(\frac32^-)}$ & $1^-$ & $2.112$  \\
		$f_0^{(\frac32^-)}$ & $0^+$ & $2.318$  \\
		$g_+^{(\frac32^-)}$, $g_\perp^{(\frac32^-)}$, $g_{\perp'}^{(\frac32^-)}$, $\widetilde{h}_+^{(\frac32^-)}$, $\widetilde{h}_\perp^{(\frac32^-)}$, $\widetilde{h}_{\perp'}^{(\frac32^-)}$ & $1^+$ & $2.460$  \\
		$g_0^{(\frac32^-)}$ & $0^-$ & $1.968$  \\
		\hline\hline
	\end{tabular}
\end{table}

\begin{table}[ht]
	\caption{\label{tab:LcLStarFFparams} $\Lambda_c^+ \to \Lambda^*_{1520}$ form-factor parameters \cite{Meinel:2021mdj}. The unlisted parameters should be determined using Eq.~\ref{eq:a0fstart} and Eq.~\ref{eq:a0fend}.}
	\centering
	\renewcommand\arraystretch{1.12}
	\setlength{\tabcolsep}{2pt}
	\begin{tabular}{@{}lccc@{\qquad}lccc@{}}
		\hline\hline
		$f^{(\frac32^-)}$ & $a_0^f$ & $a_1^f$ & $a_2^f$
		& $g^{(\frac32^-)}$ & $a_0^g$ & $a_1^g$ & $a_2^g$ \\
		\hline
		$f_0$ & $\phantom{-}4.71(40)$ & $-26.2(9.5)$ &
		& $g_0$ & & $\phantom{-}1.16(14)$ & \\
		$f_+$ & & $\phantom{-}1.28(24)$ & $-3.1(2.4)$
		& $g_+$ & & $-10.4(4.9)$ & $\phantom{-}26(44)$ \\
		$f_{\perp}$ & & $\phantom{-}4.02(76)$ & $-22.1(8.7)$
		& $g_{\perp}$ & $\phantom{-}2.26(25)$ & $\phantom{-}0.4(5.6)$ & $-60(58)$ \\
		$f_{\perp'}$ & $\phantom{-}0.1444(94)$ & $-0.42(32)$ & $\phantom{-}3.2(3.8)$
		& $g_{\perp'}$ & $-0.156(67)$ & $\phantom{-}4.9(2.3)$ & $-44(24)$ \\
		\hline\hline
	\end{tabular}
\end{table}

\subsection{Semi-leptonic decay form factors of spin-$\frac{1}{2}^+$ to spin-$\frac{1}{2}^-$ processes}

For the transition to a spin-$\frac{1}{2}$ negative-parity baryon, the helicity-basis form factors for $\Lambda_c^+(\frac{1}{2}^+) \to \Lambda^*(\frac{1}{2}^-)$ are defined as~\cite{Meinel:2021rbm}
\begin{eqnarray}
	\langle \Lambda^*(p',s')|\bar{s}\gamma^\mu c|\Lambda_c^+(p,s) \rangle
	&=&\,\bar{u}(p',s')\gamma_5\Bigg[(m_{\Lambda_c}+m_{\Lambda^*})\frac{q^\mu}{q^2}g_0^{-1/2} \notag \\
	&&+\frac{m_{\Lambda_c}-m_{\Lambda^*}}{s_-}\left(p^\mu+p'^\mu-\left(m_{\Lambda_c}^2-m_{\Lambda^*}^2\right)
	\frac{q^\mu}{q^2}\right)g_{+}^{-1/2} \notag \\
	&&+\left(\gamma^\mu+\frac{2m_{\Lambda^*}}{s_-}p^\mu-\frac{2m_{\Lambda_c}}{s_-}p'^\mu\right)g_\perp^{-1/2}\Bigg]
	u\left(p,s\right), \\
	\langle \Lambda^*(p',s')|\bar{s}\gamma^\mu\gamma_5 c|\Lambda_c^+(p,s) \rangle
	&=&-\bar{u}(p',s')\Bigg[(m_{\Lambda_c}-m_{\Lambda^*})\frac{q^\mu}{q^2}f_0^{-1/2} \notag \\
	&&+\frac{m_{\Lambda_c}+m_{\Lambda^*}}{s_+}\left(p^\mu+p'^\mu-\left(m_{\Lambda_c}^2-m_{\Lambda^*}^2\right)
	\frac{q^\mu}{q^2}\right)f_{+}^{-1/2} \notag \\
	&&+\left(\gamma^\mu-\frac{2m_{\Lambda^*}}{s_+}p^\mu-\frac{2m_{\Lambda_c}}{s_+}p'^\mu\right)f_\perp^{-1/2}\Bigg]
	u\left(p,s\right).
	\label{eq:helicity_ff_12}
\end{eqnarray}

However, for the numerical prediction, we use the form factors in another basis, parametrized as~\cite{Li:2021qod}
\begin{eqnarray}
	\langle \Lambda^*(p',s')|\bar{s}\gamma^\mu c|\Lambda_c^+(p,s) \rangle &=& \bar{u}(p',s')
	\Big(
	g_1\gamma^{\mu}+i\frac{g_2}{m_{\Lambda_c}}\sigma^{\mu\nu}q_{\nu}+\frac{g_3}{m_{\Lambda_c}}q^{\mu}
	\Big)\gamma_5 u(p,s), \\
	\langle \Lambda^*(p',s')|\bar{s}\gamma^\mu\gamma_5 c|\Lambda_c^+(p,s) \rangle &=& \bar{u}(p',s')
	\Big(
	f_1\gamma^{\mu}+i\frac{f_2}{m_{\Lambda_c}}\sigma^{\mu\nu}q_{\nu}+\frac{f_3}{m_{\Lambda_c}}q^{\mu}
	\Big) u(p,s).
	\label{eq:ff_12}
\end{eqnarray}
Here each form factor is parametrized as
\begin{eqnarray}
	F(q^2) = \frac{F(0)}{(1-q^2/m_{\Lambda_c}^2)(1-b_{1}q^2/m_{\Lambda_c}^2+b_{2}(q^2/m_{\Lambda_c}^2)^2)}.
	\label{eq:ff_12_param}
\end{eqnarray}
The numerical parameters taken from Ref.~\cite{Li:2021qod} are listed in Table~\ref{c2sformfactor}.

\begin{table}[ht]
	\centering \caption{The form factors for $\Lambda_{c}\to \Lambda^*_{1405}$ \cite{Li:2021qod}.}
	\label{c2sformfactor}
	\renewcommand\arraystretch{1.5}
	\setlength{\tabcolsep}{8pt}
	\begin{tabular}{ccccc}
         \hline\hline
		&$F(0)$   &$F(q^2_{\rm max})$   &$b_1$   &$b_2$   \\
        \hline
		$g_1$ &$0.46\pm0.05$  &$0.49\pm0.05$  &$-0.56\pm0.03$ &$1.13\pm0.35$\\
		$g_2$ &$-0.41\pm0.04$ &$-0.53\pm0.05$ &$0.81\pm0.03$  &$1.54\pm0.38$\\
		$g_3$ &$-1.45\pm0.15$ &$-1.92\pm0.19$ &$1.04\pm0.04$  &$1.94\pm0.39$\\
		$f_1$ &$0.63\pm0.06$  &$0.73\pm0.07$  &$0.05\pm0.03$  &$0.64\pm0.37$\\
		$f_2$ &$-0.56\pm0.06$ &$-0.71\pm0.07$ &$0.61\pm0.03$  &$0.75\pm0.38$\\
		$f_3$ &$-0.75\pm0.08$ &$-0.98\pm0.10$ &$0.86\pm0.03$  &$1.38\pm0.38$\\
        \hline\hline
	\end{tabular}
\end{table}

For the full angular analysis, it is convenient to convert the form factors to the helicity basis, since each form factor directly corresponds to a definite helicity. 
Therefore, we derive the conversion relations between the two bases and transform the form factors given in Ref.~\cite{Li:2021qod} to the helicity basis. The conversion between the two bases is
\begin{eqnarray}
	\begin{pmatrix}
		g_0 \\ g_+ \\ g_\perp
	\end{pmatrix}
	&=&
	\begin{pmatrix}
		-1 & 0 & \frac{q^2}{m_{\Lambda_c}(m_{\Lambda_c}+m_{\Lambda^*})} \\
		-1 & -\frac{q^2}{m_{\Lambda_c}(m_{\Lambda_c}-m_{\Lambda^*})} & 0 \\
		-1 & -\frac{m_{\Lambda_c}-m_{\Lambda^*}}{m_{\Lambda_c}} & 0
	\end{pmatrix}
	\begin{pmatrix}
		g_1 \\ g_2 \\ g_3
	\end{pmatrix},
	\\
	\begin{pmatrix}
		f_0 \\ f_+ \\ f_\perp
	\end{pmatrix}
	&=&
	\begin{pmatrix}
		-1 & 0 & -\frac{q^2}{m_{\Lambda_c}(m_{\Lambda_c}-m_{\Lambda^*})} \\
		-1 & \frac{q^2}{m_{\Lambda_c}(m_{\Lambda_c}+m_{\Lambda^*})} & 0 \\
		-1 & \frac{m_{\Lambda_c}+m_{\Lambda^*}}{m_{\Lambda_c}} & 0
	\end{pmatrix}
	\begin{pmatrix}
		f_1 \\ f_2 \\ f_3
	\end{pmatrix}.
	\label{eq:conversion_matrix}
\end{eqnarray}

With these form factors defined above, we evaluate the hadronic helicity amplitudes
$H_{\lambda}=H_V(\lambda)-H_A(\lambda)$ in the $\Lambda_c^+$ rest frame. The
explicit expressions for the $\Lambda_c^+\to\Lambda^*(1/2^-)$ and
$\Lambda_c^+\to\Lambda^*(3/2^-)$ transitions are collected in
Appendix~\ref{sec:app_helicity_amplitudes}.

\section{General form of the four-body angular distribution}
\label{sec:angular}
As shown in Eq.~\ref{eq:full_amplitude}, the amplitude of the $\Lambda_c^+ \to \Lambda^*_{1405/1520}(\to\Sigma\pi)\ell^+\nu_\ell$ decay can be divided into hadronic and leptonic parts. Since each part is Lorentz invariant, they can be evaluated in convenient reference frames. 
In the rest frame of the $W$ boson, the leptonic helicity amplitude $L_\lambda(\theta_\ell,\phi)$ defined in Eq.~(\ref{eq:lepton_amplitude_def}) can be calculated straightforwardly. Since the neutrino mass is taken to be zero in our work, its helicity can only be $-\frac{1}{2}$. The leptonic helicity amplitudes then take the form listed in Appendix~\ref{sec:app_helicity_amplitudes}, with $\beta = \sqrt{1-\frac{m_\ell^2}{q^2}}$.

Substituting the lepton and hadron amplitudes into the differential decay width formula and performing the phase-space decomposition of Eq.~(\ref{eq:phasespace}), the full four-body angular distribution can be written as
\begin{eqnarray}
	\frac{d^5\Gamma}{d\cos\theta_\Lambda d\cos\theta_\ell d\phi dq^2 dm_{\Sigma\pi}^2} =
	&&\frac{G_F^2|V_{cs}|^2(q^2-m_\ell^2)\sqrt{\lambda(m_{\Lambda_c},m_{\Sigma\pi},\sqrt{q^2})}\sqrt{\lambda(m_{\Sigma\pi},m_\Sigma,m_\pi)}}{65536 \pi^6 m_{\Lambda_c}^3 q^2 m_{\Sigma\pi}^2} \notag \\
	&&\times \frac{1}{2}\sum_{s_{\Sigma},s_{\Lambda_c},s_\ell}\Bigl|\sum_{\lambda}A_{s_{\Sigma},\lambda}^{s_{\Lambda_c}}(\theta_\Lambda)L_{\lambda}(\theta_\ell,\phi)\Bigr|^2,
	\label{eq:master_formula}
\end{eqnarray}
with $A_{s_{\Sigma},\lambda}^{s_{\Lambda_c}}(\theta_\Lambda)$ defined in Eq.~(\ref{eq:A_def}). Defining the phase-space prefactor
\begin{eqnarray}
	P = \frac{G_F^2|V_{cs}|^2(q^2-m_\ell^2)\sqrt{\lambda(m_{\Lambda_c},m_{\Sigma\pi},\sqrt{q^2})}\sqrt{\lambda(m_{\Sigma\pi},m_\Sigma,m_\pi)}}{131072 \pi^6 m_{\Lambda_c}^3 q^2 m_{\Sigma\pi}^2},
	\label{eq:P_factor}
\end{eqnarray}
the differential decay width can be expressed in terms of several angular coefficients $L_i$:
\begin{eqnarray}
	&&\frac{d^5\Gamma}{d\cos\theta_\Lambda d\cos\theta_\ell d\phi dq^2 dm_{\Sigma\pi}^2}
	\notag \\
	&&\quad\quad\quad =
P\Big(L_1+L_2\cos\theta_\ell+L_3\cos2\phi+L_4\cos2\theta_\ell
	+L_5\cos2\theta_\ell\cos2\phi
	\notag \\
	&&\quad\quad\quad+L_6\cos2\theta_\ell\sin2\phi +L_7\sin2\phi+L_8\sin\theta_\ell\cos\phi+L_9\sin\theta_\ell\sin\phi \notag \\
&&\quad\quad\quad+L_{10}\sin2\theta_\ell\cos\phi+L_{11}\sin2\theta_\ell\sin\phi\Big).
	\label{eq:angular_distribution}
\end{eqnarray}
Here we first expand only the leptonic helicity amplitude $L_\lambda(\theta_\ell,\phi)$; the $\theta_\Lambda$ dependence can then be studied by expanding the hadronic helicity amplitude $A_{s_{\Sigma},\lambda}^{s_{\Lambda_c}}(\theta_\Lambda)$ as follows.

The angular coefficients $L_i$ are expressed in terms of the combined helicity amplitudes $A_{s_{\Sigma},\lambda}^{s_{\Lambda_c}}$ as follows:
\begin{eqnarray}
	L_1 &=& \beta^2\sum_{s_{\Lambda_c},s_{\Sigma}}
	\Big(
	(2q^2+2m_\ell^2)|A_{s_{\Sigma},0}^{s_{\Lambda_c}}|^2
	+(3q^2+m_\ell^2)|A_{s_{\Sigma},1}^{s_{\Lambda_c}}|^2
	\notag \\
	&&
	+(3q^2+m_\ell^2)|A_{s_{\Sigma},-1}^{s_{\Lambda_c}}|^2
	+4m_\ell^2|A_{s_{\Sigma},t}^{s_{\Lambda_c}}|^2
	\Big), \notag \\
	L_2 &=& 4\beta^2\sum_{s_{\Lambda_c},s_{\Sigma}}
	\Big(
	q^2|A_{s_{\Sigma},1}^{s_{\Lambda_c}}|^2
	-q^2|A_{s_{\Sigma},-1}^{s_{\Lambda_c}}|^2-2m_\ell^2\,\mathrm{Re}
	\big(A_{s_{\Sigma},t}^{s_{\Lambda_c}}
	A_{s_{\Sigma},0}^{s_{\Lambda_c}*}\big)
	\Big), \notag \\
	L_3 &=& 2\beta^2\sum_{s_{\Lambda_c},s_{\Sigma}}
	\Big(
	(q^2-m_\ell^2)\,\mathrm{Re}
	\big(A_{s_{\Sigma},-1}^{s_{\Lambda_c}}
	A_{s_{\Sigma},1}^{s_{\Lambda_c}*}\big)
	\Big), \notag \\
	L_4 &=& \beta^2\sum_{s_{\Lambda_c},s_{\Sigma}}
	\Big(
	(q^2-m_\ell^2)|A_{s_{\Sigma},1}^{s_{\Lambda_c}}|^2
	-2(q^2-m_\ell^2)|A_{s_{\Sigma},0}^{s_{\Lambda_c}}|^2
	+(q^2-m_\ell^2)|A_{s_{\Sigma},-1}^{s_{\Lambda_c}}|^2
	\Big), \notag \\
	L_5 &=& -L_3, L_6 = -L_7 = -L_5(\mathrm{Re}\rightarrow\mathrm{Im}), \notag\\
	L_8 &=& -4\sqrt{2}\beta^2\sum_{s_{\Lambda_c},s_{\Sigma}}
	\Big(
	q^2\,\mathrm{Re}
	\big(A_{s_{\Sigma},0}^{s_{\Lambda_c}}
	A_{s_{\Sigma},1}^{s_{\Lambda_c}*}
	+A_{s_{\Sigma},-1}^{s_{\Lambda_c}}
	A_{s_{\Sigma},0}^{s_{\Lambda_c}*}\big)
	\notag \\
	&&
	-m_\ell^2\,\mathrm{Re}
	\big(A_{s_{\Sigma},-1}^{s_{\Lambda_c}}
	A_{s_{\Sigma},t}^{s_{\Lambda_c}*}
	-A_{s_{\Sigma},t}^{s_{\Lambda_c}}
	A_{s_{\Sigma},1}^{s_{\Lambda_c}*}\big)
	\Big),
	\notag \\
    L_9 &=& -L_8(\mathrm{Re}\rightarrow\mathrm{Im}),  L_{11} = -L_{10}(\mathrm{Re}\rightarrow\mathrm{Im}). \notag \\
	L_{10} &=& -2\sqrt{2}\beta^2\sum_{s_{\Lambda_c},s_{\Sigma}}
	\Big(
	q^2\,\mathrm{Re}
	\big(A_{s_{\Sigma},0}^{s_{\Lambda_c}}
	A_{s_{\Sigma},1}^{s_{\Lambda_c}*}
	-A_{s_{\Sigma},-1}^{s_{\Lambda_c}}
	A_{s_{\Sigma},0}^{s_{\Lambda_c}*}\big)
	\notag\\
    &&+m_\ell^2\,\mathrm{Re}
	\big(A_{s_{\Sigma},-1}^{s_{\Lambda_c}}
	A_{s_{\Sigma},0}^{s_{\Lambda_c}*}
	-A_{s_{\Sigma},0}^{s_{\Lambda_c}}
	A_{s_{\Sigma},1}^{s_{\Lambda_c}*}\big)
	\Big), 
\end{eqnarray}

The angular coefficients $L_i$ still contain an implicit dependence on $\cos\theta_\Lambda$ through the Wigner $D$-functions in $A_{s_{\Sigma},\lambda}^{s_{\Lambda_c}}$. This dependence can be made explicit by expanding in $\theta_\Lambda$:
\begin{eqnarray}
	\frac{d^5\Gamma}{d\cos\theta_\Lambda d\cos\theta_\ell d\phi dq^2 dm_{\Sigma\pi}^2} &=&
	P\beta^2\Big[
	(L_{1(1)}+L_{1(2)}\cos\theta_{\Lambda}
	+L_{1(3)}\cos2\theta_{\Lambda})
	\notag \\
	&&
	+(L_{2(1)}+L_{2(2)}\cos\theta_{\Lambda}
	+L_{2(3)}\cos2\theta_{\Lambda})\cos\theta_\ell
	\notag \\
	&&
	+(L_{3(1)}+L_{3(2)}\cos2\theta_{\Lambda})\cos2\phi
	\notag \\
	&&
	+(L_{4(1)}+L_{4(2)}\cos\theta_{\Lambda}
	+L_{4(3)}\cos2\theta_{\Lambda})\cos2\theta_\ell
	\notag \\
	&&
	+(L_{5(1)}+L_{5(2)}\cos2\theta_{\Lambda})
	\cos2\theta_\ell\cos2\phi
	\notag \\
	&&
	+(L_{6(1)}+L_{6(2)}\cos2\theta_{\Lambda})
	\cos2\theta_\ell\sin2\phi 
	\notag \\
	&&
	+(L_{7(1)}+L_{7(2)}\cos2\theta_{\Lambda})\sin2\phi
	\notag \\
	&&
	+(L_{8(1)}\sin\theta_{\Lambda}
	+L_{8(2)}\sin2\theta_{\Lambda})\sin\theta_\ell\cos\phi
	\notag \\
	&&
	+(L_{9(1)}\sin\theta_{\Lambda}
	+L_{9(2)}\sin2\theta_{\Lambda})\sin\theta_\ell\sin\phi
	\notag \\
	&&
	+(L_{10(1)}\sin\theta_{\Lambda}
	+L_{10(2)}\sin2\theta_{\Lambda})\sin2\theta_\ell\cos\phi
	\notag \\
	&&
	+(L_{11(1)}\sin\theta_{\Lambda}
	+L_{11(2)}\sin2\theta_{\Lambda})\sin2\theta_\ell\sin\phi
	\Big].
	\label{eq:angular_expansion}
\end{eqnarray}
The full expressions for the coefficients $L_{i(j)}$ in terms of the helicity amplitudes $\mathcal{H}^{J}_{s_{\Lambda_c},\lambda}$ are collected in Appendix~\ref{sec:app_Lij}.

One can see that, for the integrated decay width, the angular coefficients multiplying $\sin\theta_{\Lambda(\ell)}$ or any trigonometric functions of $\phi$ do not contribute. These coefficients can be measured only through the angular distributions. Although their contributions are not numerically dominant, they remain important in studies of multi-body decays, because they encode interference effects between different resonances, different helicity amplitudes, and amplitudes associated with different weak currents.

For hadronic decays, strong interactions can generate strong phases, making the amplitude $\cal H$ defined in Eq.\eqref{eq:A_def} complex and leading to complex form factors. In our work, however, we omit these strong phases because they disappear from the squared helicity amplitudes and give smaller contributions to the interference terms than the imaginary part of the lineshape. By assuming that the form factors and ${\cal A}_J$ are real, the angular distribution coefficients can be simplified. We give the simplest form of these coefficients in Appendix~\ref{sec:app_simple}, where the helicity amplitudes are taken to be real and the lepton mass is neglected ($m_\ell \to 0$).
These simplified expressions are useful for qualitative understanding, while the full results are used in the numerical analysis.

\section{Phenomenological results}
\label{sec:results}

In this section we present the numerical predictions obtained from the angular analysis, using the form-factor inputs and helicity amplitudes defined in the previous sections. We begin with the three-body decay widths as a consistency check, then proceed to the full four-body cascade decay: invariant-mass spectrum, angular distributions in $\theta_\Lambda$, $\theta_\ell$, and $\phi$, their forward-backward asymmetries, and finally the sensitivity of the angular observables to variations of the form-factor parameters.
The numerical values of the parameters used in our calculation are as follows~\cite{ParticleDataGroup:2020ssz}:
\begin{eqnarray}
	G_F &=& 1.166 \times 10^{-5}\,\text{GeV}^{-2}, \qquad V_{cs} = 0.975, \qquad m_e = 5.1 \times 10^{-4}\,\text{GeV}, \notag \\
	m_{1520} &=& 1.52\,\text{GeV}, \qquad \Gamma_{1520} = 1.573 \times 10^{-2}\,\text{GeV}, \notag \\
	m_{\Lambda_c} &=& 2.286\,\text{GeV}, \qquad m_{\Sigma} = 1.192\,\text{GeV}, \qquad m_{\pi} = 0.135\,\text{GeV}.
	\label{eq:parameters}
\end{eqnarray}
The form factors can be evaluated using the parameters given in Tables~\ref{tab:LcLStarFFparams} and \ref{c2sformfactor} for the $\Lambda^*_{1520}$ and $\Lambda^*_{1405}$ transitions, respectively.

\subsection{Three-body decay widths and angular distribution}

After integrating over the kinematic variables associated with $\Lambda^*\to \Sigma \pi$, such as $\theta_\Lambda$, $m_{\Sigma\pi}$, and $\phi$, one obtains the formula for the semi-leptonic three-body decay of $\Lambda_c^+$. The angular distribution can be obtained by integrating over $\phi$ in Eq.~(\ref{eq:angular_distribution}). The differential decay width with respect to $\cos\theta_\ell$ reads
\begin{eqnarray}
	\frac{d^2\Gamma}{d\cos\theta_\ell dq^2}=
	P_{\rm 3b}\Big( L^3_1+L^3_2\cos\theta_\ell+L^3_4\cos2\theta_\ell\Big),
	\label{eq:threebody_angular}
\end{eqnarray}
where the three-body phase-space factor is
\begin{eqnarray}
	P_{\rm 3b} = \frac{G_F^2|V_{cs}|^2}{2} \times \frac{1}{2} \times \frac{\sqrt{\lambda(m_{\Lambda_c},\sqrt{q^2},m_{\Lambda^*})}(q^2-m_\ell^2)}{512\pi^3 m_{\Lambda_c}^3 q^2}.
	\label{eq:threebody_P}
\end{eqnarray}
The  angular coefficients $L_i^3$ are
\begin{eqnarray}
	L^3_1 &=& \beta^2
	\Big(
	(2q^2+2m_\ell^2)(|H_{\frac{1}{2}0}|^2+|H_{-\frac{1}{2}0}|^2)
	+(3q^2+m_\ell^2)(|H_{-\frac{1}{2}1}|^2+|H_{\frac{1}{2}1}|^2)
	\notag \\
	&&\qquad +(3q^2+m_\ell^2)(|H_{\frac{1}{2}-1}|^2+|H_{-\frac{1}{2}-1}|^2)
	+4m_\ell^2(|H_{\frac{1}{2}t}|^2+|H_{-\frac{1}{2}t}|^2)
	\Big), \notag \\
	L^3_2 &=& 4\beta^2
	\Big(
	q^2(|H_{-\frac{1}{2}1}|^2+|H_{\frac{1}{2}1}|^2)-q^2(|H_{\frac{1}{2}-1}|^2+|H_{-\frac{1}{2}-1}|^2) \notag \\
	&&\qquad - 2m_\ell^2(H_{\frac{1}{2}t}H_{\frac{1}{2}0}+H_{-\frac{1}{2}t}H_{-\frac{1}{2}0})
	\Big), \notag \\
	L^3_4 &=& \beta^2
	\Big(
	(q^2-m_\ell^2)(|H_{-\frac{1}{2}1}|^2+|H_{\frac{1}{2}1}|^2)
	-2(q^2-m_\ell^2)(|H_{\frac{1}{2}0}|^2+|H_{-\frac{1}{2}0}|^2) \notag \\
	&&\qquad +(q^2-m_\ell^2)(|H_{\frac{1}{2}-1}|^2+|H_{-\frac{1}{2}-1}|^2)
	\Big).
	\label{eq:threebody_Li}
\end{eqnarray}
Here $H_{s_{\Lambda_c},\lambda}$ is a shorthand for the hadronic helicity amplitude $H_\lambda(\Lambda_c^+\to \Lambda^*\ell^+\nu_\ell)$, and the terms $|H_{\frac{1}{2}1}|^2$ and $|H_{-\frac{1}{2}-1}|^2$ appear only for the $\Lambda^*_{1520}$ transition due to its spin-$\frac{3}{2}$ nature.

The total three-body decay width and branching ratio are obtained by integrating over $q^2$ and $\cos\theta_\ell$ as
\begin{eqnarray}
	\Gamma(\Lambda_c^+\to \Lambda^*\ell^+\nu_\ell) &=& \int_{m_\ell^2}^{(m_{\Lambda_c}-m_{\Lambda^*})^2}dq^2\int^{1}_{-1}d\cos\theta_\ell\,P_{\rm 3b}\Big( L_1+L_2\cos\theta_\ell+L_4\cos2\theta_\ell\Big).\notag\\
    \mathcal{B}(\Lambda_c^+\to \Lambda^*\ell^+\nu_\ell)&=&\Gamma(\Lambda_c^+\to \Lambda^*\ell^+\nu_\ell)/\Gamma_{\rm total}.
\end{eqnarray}
 Using the numerical values of the form factors in Refs.~\cite{Meinel:2021mdj,Li:2021qod}, which are also given in Sec.~3, the numerical predictions are summarized in Table~\ref{result}.

The experimental data in Table~\ref{result} are collected from the PDG~\cite{Xing:2026ohi}
. 
There is a sizable discrepancy between our predictions and the experimental results, especially for $\Lambda_c^+\to\Lambda^*_{1405}\ell^+\nu_\ell$. This comparison indicates that the currently available form-factor inputs may still have sizable uncertainties for these excited-state transitions. Moreover, the experimental result for $\Lambda_c^+\to\Lambda^*_{1405}\ell^+\nu_\ell$ is obtained through the final state $pK$, and therefore cannot be directly identified with the $\Sigma\pi$ channel studied in this work. In the following analysis, we use the form factors from Refs.~\cite{Li:2021qod,Meinel:2021mdj} without further rescaling.

\begin{table}[ht]
	\caption{\label{result}Numerical results for the three-body and four-body branching ratios, in units of $(10^{-4})$.  The experimental results are taken from Ref.~\cite{Zyla:2020zbs,BESIII:2025kna}. It should be noted that the experimental result for the decay channel $\Lambda_c^+ \to  \Lambda^*_{1405}({\rm pK})e^+\nu_e$ refers to the case where $\Lambda^*_{1405}$ decays into ${\rm pK}$. And the experimental result of $\mathcal{B}(\Lambda_c^+ \to \Sigma\pi(\Lambda^*_{1520}) e^+\nu_e)$ are estimated from $\mathcal{B}(\Lambda_c^+ \to \Lambda^*_{1520} e^+\nu_e)\times\mathcal{B}(\Lambda^*_{1520} \to \Sigma\pi)$. The inputting form factors are taken from Refs.~\cite{Li:2021qod,Meinel:2021mdj}.}
	\begin{tabular}{ccccc}
		\hline\hline
        \multicolumn{4}{c}{Three-body decay processes}\\
		\hline
		Channel
		& Theo.   
		& Channel 
		& Exp. \\
		\hline
		$\Lambda_c^+ \to \Lambda^*_{1405}e^+\nu_e$ 
		& $75.2  \pm 8$      
		&  $\Lambda_c^+ \to  \Lambda^*_{1405}({ pK})e^+\nu_e$  
		&  $4.2 \pm 1.9 \pm 0.4$ \\
		\hline
		$\Lambda_c^+ \to  \Lambda^*_{1520}e^+\nu_e$ 
		& $5.12  \pm 0.83 $
		& $\Lambda_c^+ \to  \Lambda^*_{1520}e^+\nu_e$    
		& $10.2 \pm 5.2 \pm 1.1$ \\
		\hline
		\hline\multicolumn{4}{c}{Four-body decay processes}\\
		\hline
		Channel 
		& Theo.   
		& Channel 
		& Exp. \\
		\hline
		$ \Lambda_c^+ \to  \Sigma^{\pm}\pi^{\mp}(\Lambda^*_{1405}+\Lambda^*_{1520})e^+\nu_e$ 
		& $6.8 \pm 0.83$
		&  $\Lambda_c^+\to\Sigma^{\pm}\pi^{\mp} e^+\nu_e$ 
		&  $7.6^{+2.5}_{-2.3} \pm 1.3$   \\
		\hline
		$\Lambda_c^+ \to \Sigma\pi(\Lambda^*_{1405})e^+\nu_e$ 
		& $8.11  \pm1.21$
		&  $\Lambda_c^+ \to  \Sigma\pi(\Lambda^*_{1405}) e^+\nu_e$ 
		&  ------\\
		\hline
		$\Lambda_c^+ \to \Sigma\pi(\Lambda^*_{1520}) e^+\nu_e$ 
		& $2.08 \pm 0.35 $
		& $\Lambda_c^+ \to  \Sigma\pi(\Lambda^*_{1520}) e^+\nu_e$ 
		& $4.4 \pm 2.2 \pm 0.5$ \\
		\hline\hline
	\end{tabular}
\end{table}
One can see from Table~\ref{result} that the theoretical prediction for
  $\Lambda_c^+ \to \Lambda^*_{1405}e^+\nu_e$ is much larger than the experimental
  measurement quoted for the $pK$ final state. However, the current experimental
  result is measured through the channel
  $\Lambda_c^+ \to \Lambda^*_{1405}(\to pK)e^+\nu_e$, rather than the inclusive
  three-body mode $\Lambda_c^+ \to \Lambda^*_{1405}e^+\nu_e$. For a rough comparison,
  one may write
  \[\mathcal{B}(\Lambda_c^+ \to \Lambda^*_{1405}(\to pK)e^+\nu_e)
  \simeq
  \mathcal{B}(\Lambda_c^+ \to \Lambda^*_{1405}e^+\nu_e)
  \mathcal{B}(\Lambda^*_{1405}\to pK).  \]
  Although the decay $\Lambda^*_{1405}\to pK$ has been observed experimentally,
  its branching fraction has not yet been measured. As a conservative estimate,
  if one assumes
  $\mathcal{B}(\Lambda^*_{1405}\to pK)\lesssim 10\%$, the predicted branching
  fraction for the $pK$ final state is smaller than $7.52\times 10^{-4}$, which is
  compatible with the experimental value
  $\mathcal{B}(\Lambda_c^+ \to \Lambda^*_{1405}(\to pK)e^+\nu_e)
  =(4.19\pm1.9)\times 10^{-4}$ within the present uncertainties. 
  The other theoretical predictions are generally compatible with the corresponding experimental measurements within the
  present $1\sigma$ uncertainties.

\subsection{Four-body decay widths and angular distribution}
\begin{figure}[htbp]
	\centering
    \begin{subfigure}{0.45\textwidth}
		\centering
		\includegraphics[width=0.9\linewidth]{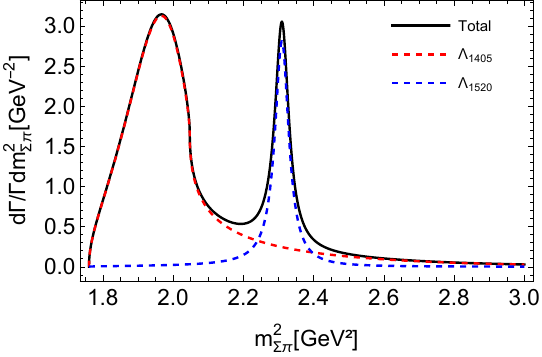}
		\caption{$m_{\Sigma\pi}^2$}
        \label{2a}
	\end{subfigure}
	\begin{subfigure}{0.45\textwidth}
		\centering
		\includegraphics[width=0.9\linewidth]{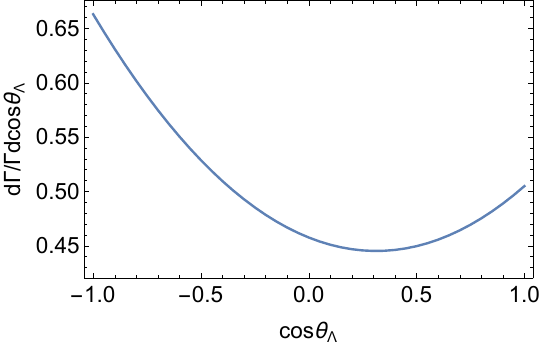}
		\caption{$\cos\theta_\Lambda$}
	\end{subfigure}
	\begin{subfigure}{0.45\textwidth}
		\centering
		\includegraphics[width=0.9\linewidth]{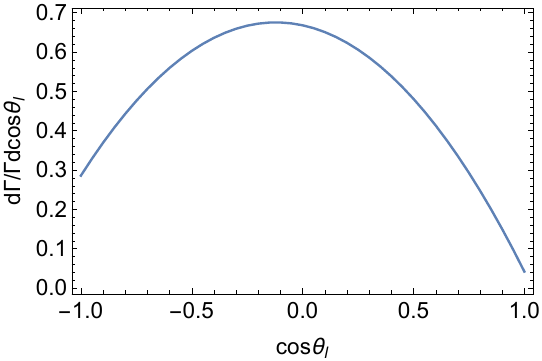}
		\caption{$\cos\theta_\ell$}
	\end{subfigure}
	\begin{subfigure}{0.45\textwidth}
		\centering
		\includegraphics[width=0.9\linewidth]{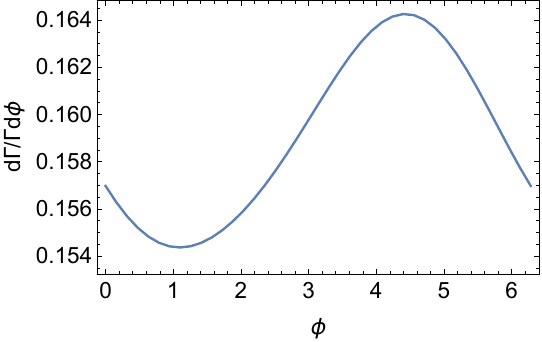}
		\caption{$\phi$}
	\end{subfigure}
	\caption{The differential decay width as a function of the invariant mass $m_{\Sigma\pi}$ and one-dimensional angular distributions in $\cos\theta_\Lambda$, $\cos\theta_\ell$, and $\phi$.}
	\label{fig:angular_distributions}
\end{figure}

For the four-body decay $\Lambda_c^+ \to \Lambda^*(\Sigma\pi)\ell^+\nu_\ell$, the total decay width and branching ratio can be obtained by integrating over all kinematic variables in Eq.~(\ref{eq:angular_expansion}) as
\begin{eqnarray}
\Gamma(\Lambda_c^+\to \Lambda^*(\to \Sigma\pi)\ell^+\nu_\ell) &=& \int_{(m_{\Sigma}+m_\pi)^2}^{(m_{\Lambda_c}-m_\ell)^2}dm_{\Sigma\pi}^2\,\int_{m_\ell^2}^{(m_{\Lambda_c}-m_{\Lambda^*})^2}dq^2\int^{1}_{-1}d\cos\theta_\ell \notag \\
&&
\times\int^{1}_{-1}d\cos\theta_\Lambda\,\int^{2\pi}_{0}d\phi\,\frac{d^5\Gamma}{d\cos\theta_\Lambda d\cos\theta_\ell d\phi dq^2 dm_{\Sigma\pi}^2}.\notag\\
    \mathcal{B}(\Lambda_c^+\to \Lambda^*(\to \Sigma\pi)\ell^+\nu_\ell)&=&\Gamma(\Lambda_c^+\to \Lambda^*(\to \Sigma\pi)\ell^+\nu_\ell)/\Gamma_{\rm total}.
\end{eqnarray}
The numerical results are summarized in Table~\ref{result}. 
The differential decay width can be derived by integrating out some variables in Eq.~(\ref{eq:angular_expansion}).
The differential decay width with respect to $m_{\Sigma\pi}^2$ is
\begin{eqnarray}
	\frac{d\Gamma}{dm_{\Sigma\pi}^2} = \int_{m_\ell^2}^{(m_{\Lambda_c}-m_{\Sigma\pi})^2}dq^2\,\frac{8\pi P \beta^2}{9}(9L_{1(1)}-3L_{1(3)}-3L_{4(1)}+L_{4(3)}).
	\label{eq:invariant_mass}
\end{eqnarray}
The resulting distribution is shown in Fig.~\ref{fig:angular_distributions}(a). 

Integrating Eq.~(\ref{eq:angular_expansion}) over $\theta_\ell$ and $\phi$, the differential decay width as a function of $\cos\theta_\Lambda$ can be expressed as
\begin{eqnarray}
	\frac{d^2\Gamma}{dm_{\Sigma\pi}^2 d\cos\theta_\Lambda} = \int dq^2\big(L_{\Lambda}+L_{\Lambda c}\cos\theta_{\Lambda}+L_{\Lambda 2c}\cos2\theta_{\Lambda}\big),
	\label{eq:thetaLambda_dist}
\end{eqnarray}
where
\begin{eqnarray}
	L_{\Lambda} &=& \frac{4\pi P \beta^2}{3}(3L_{1(1)}-L_{4(1)}), \;L_{\Lambda c} = \frac{4\pi P \beta^2}{3}(3L_{1(2)}-L_{4(2)}), \notag \\
	L_{\Lambda 2c} &=& \frac{4\pi P \beta^2}{3}(3L_{1(3)}-L_{4(3)}).
	\label{eq:L_Lambda}
\end{eqnarray}
Since the angle $\theta_\Lambda$ reflects the orientation of the $\Lambda^*$ decay plane relative to the $\Lambda_c^+$ decay, the behavior of the corresponding coefficients reflects the resonance structure of $\Lambda^*$. As shown in the appendix, the coefficient $L_\Lambda$ receives contributions from the squared helicity amplitudes of both resonances, whereas $L_{\Lambda 2c}$ receives contributions only from the $\Lambda^*_{1520}$ resonance. The interference between the two resonances is reflected by the coefficient $L_{\Lambda c}$.

Similarly, after integrating over $\theta_\Lambda$ and $\phi$ ($\theta_\ell$), the differential decay width with respect to $\cos\theta_\ell$ ($\phi$) is
\begin{eqnarray}
	&&\frac{d^2\Gamma}{dm_{\Sigma\pi}^2 d\cos\theta_\ell} = \int dq^2\big(L_{\ell}+L_{\ell c}\cos\theta_{\ell}+L_{\ell 2c}\cos2\theta_{\ell}\big),\notag\\
   && 	\frac{d^2\Gamma}{dm_{\Sigma\pi}^2 d\phi} = \int dq^2\big(L_{\phi}+L_{\phi c}\cos\phi+L_{\phi 2c}\cos2\phi+L_{\phi s}\sin\phi+L_{\phi 2s}\sin2\phi\big),
	\label{eq:thetal_dist}
\end{eqnarray}
with
\begin{eqnarray}
	L_{\ell} &=& \frac{4\pi P \beta^2}{3}(3L_{1(1)}-L_{1(3)}), \; L_{\ell c} = \frac{4\pi P \beta^2}{3}(3L_{2(1)}-L_{2(3)}), \notag \\
	L_{\ell 2c} &=& \frac{4\pi P \beta^2}{3}(3L_{4(1)}-L_{4(3)}),\notag\\
    	L_\phi &=& \frac{4P\beta^2}{9}(9L_{1(1)}-3L_{1(3)}-3L_{4(1)}+L_{4(3)}), \notag \\
	L_{\phi c} &=& \frac{\pi^2P\beta^2}{4}L_{8(1)}, \qquad L_{\phi s} = \frac{\pi^2P\beta^2}{4}L_{9(1)}, \notag \\
	L_{\phi 2c} &=& \frac{4P\beta^2}{9}L_{3(1)}, \qquad
	L_{\phi 2s} = \frac{4P\beta^2}{9}L_{7(1)}.
	\label{eq:L_ell}
\end{eqnarray}
Since $\theta_\ell$ is the angle between the leptonic decay and the $\Lambda_c^+$ decay plane, the corresponding angular distribution contains interference terms between the vector helicity amplitude $H_V$ and the axial-vector helicity amplitude $H_A$. If we set $m_\ell\to 0$, only the interference contribution remains, and it should be proportional to $f_{\perp^{(\prime)}}\times g_{\perp^{(\prime)}}$.
The angular distribution of $\phi$ provides a more complete picture of the resonance. For the coefficients $L_{\phi c}$, $L_{\phi s}$, $L_{\phi 2c}$, and $L_{\phi 2s}$, only resonance interference
effects contribute, and both the real and imaginary parts can be probed. $L_{\phi c}$ and $L_{\phi 2c}$ correspond to the real part of the interference term, whereas $L_{\phi s}$ and $L_{\phi 2s}$ correspond to its imaginary part. Since the imaginary part of the decay amplitude can arise simultaneously from the resonance lineshape of $\Lambda^*$ and from the hadronic decay part of $\Lambda_c^+\to\Lambda^*$, it is challenging to identify its source through purely theoretical means. However, the angular distribution provides a way to study it. The coefficient $L_{\phi 2s}$ contains only the interference term between different helicity amplitudes in the $\Lambda^*_{1520}$ decay. Therefore, this term receives no contribution from the imaginary part of the lineshape, and the corresponding coefficient is sensitive to the strong phase in the $\Lambda_c^+\to\Lambda^*_{1520}\ell^+\nu_\ell$ process, which will be discussed in detail later. 

The $\cos\theta_{\Lambda}$, $\cos\theta_{\ell}$, and $\phi$ distributions are shown in Fig.~\ref{fig:angular_distributions}(b), (c), and (d), respectively. One can see that both the $\cos{\theta_\ell}$ and $\cos{\theta_\Lambda}$ distributions have a slight asymmetry about the zero axis. Since the constant and $\cos{2\theta_{\ell(\Lambda)}}$ terms are symmetric, only the $\cos\theta_{\ell(\Lambda)}$ term can contribute to this asymmetry. In our work, their contributions to the differential decay width are small.  However, for the $\phi$ distribution, the asymmetry contributed by $L_{\phi s}$ and $L_{\phi 2s}$ is large. Since the form factors used in our work are real, $L_{\phi 2s}$, which corresponds to the imaginary part of the helicity amplitude $H_\lambda$, vanishes. The large asymmetry shows that $L_{\phi s}$ gives a sizable contribution to this differential decay width. These angular coefficients will be discussed later in detail.

\begin{figure}[htbp]
	\centering
	\begin{subfigure}{0.24\textwidth}
		\centering
		\includegraphics[width=\linewidth]{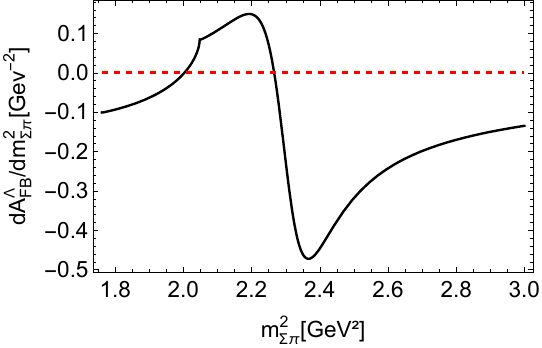}
		\caption{$A^\Lambda_{FB}$}
	\end{subfigure}
	\hfill
	\begin{subfigure}{0.24\textwidth}
		\centering
		\includegraphics[width=\linewidth]{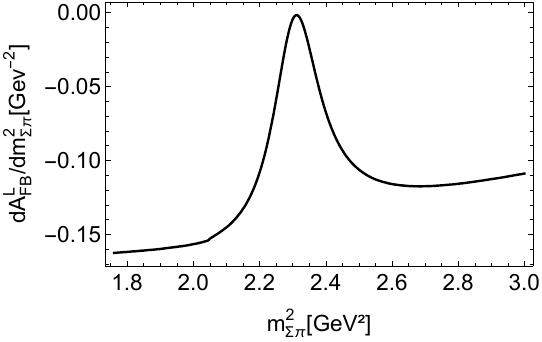}
		\caption{$A^\ell_{FB}$}
	\end{subfigure}
    \hfill
    \begin{subfigure}{0.24\textwidth}
    	\centering
    	\includegraphics[width=\linewidth]{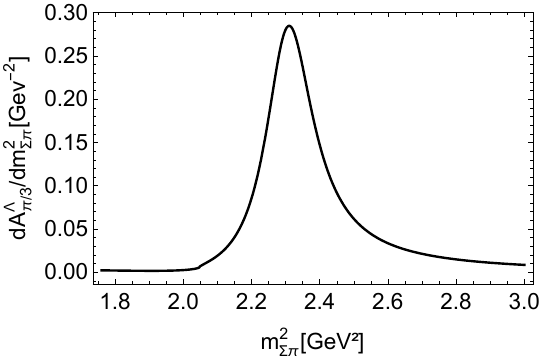}
    	\caption{$A^\Lambda_{\frac{\pi}{3}}$}
    \end{subfigure}
	\hfill
    \begin{subfigure}{0.24\textwidth}
    	\centering
    	\includegraphics[width=\linewidth]{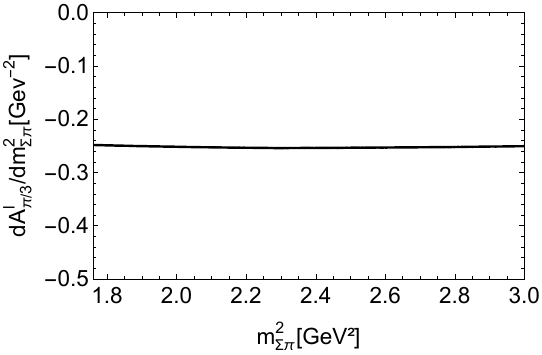}
    	\caption{$A^\ell_{\frac{\pi}{3}}$}
    \end{subfigure}
	\begin{subfigure}{0.24\textwidth}
		\centering
		\includegraphics[width=\linewidth]{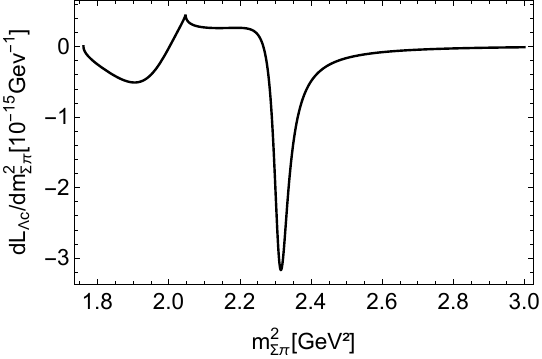}
		\caption{$L_{\Lambda c}$}
	\end{subfigure}
    \hfill 
	\begin{subfigure}{0.24\textwidth}
		\centering
		\includegraphics[width=\linewidth]{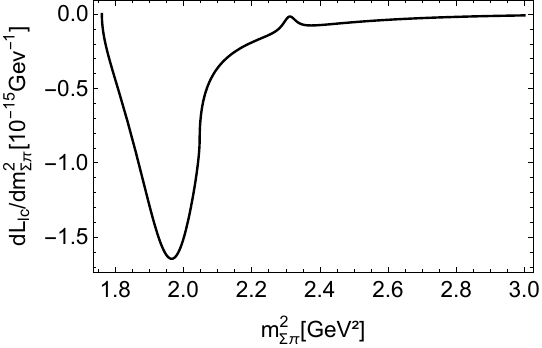}
		\caption{$L_{\ell c}$}
	\end{subfigure}
    \hfill
    \begin{subfigure}{0.24\textwidth}
    	\centering
    	\includegraphics[width=\linewidth]{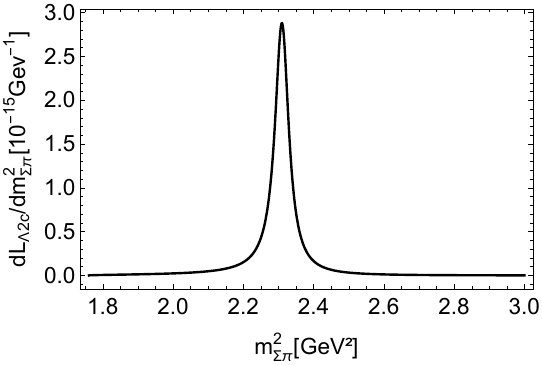}
    	\caption{$L_{\Lambda 2c}$}
    \end{subfigure}
    \hfill
    \begin{subfigure}{0.24\textwidth}
    	\centering
    	\includegraphics[width=\linewidth]{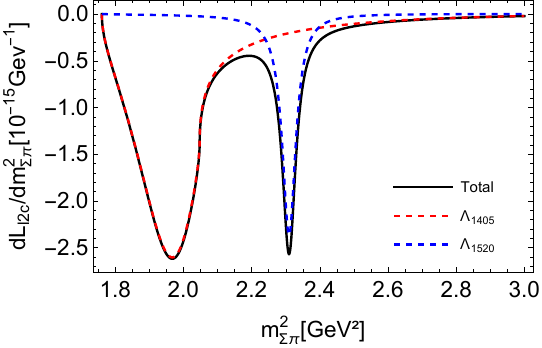}
    	\caption{$L_{\ell 2c}$}
    \end{subfigure}
	\caption{$A_{FB}$ and $A_{\frac{\pi}{3}}$ asymmetries and their corresponding angular coefficients as functions of $m_{\Sigma\pi}$.}
	\label{fig:afb_distributions}
\end{figure}

To make these effects, especially the interference effects, easier to measure experimentally, special observables can be defined. First, we define the forward-backward asymmetry $A_{FB}$. After normalization, the $A_{FB}$ observables corresponding to $\theta_\Lambda$ and $\theta_\ell$ are defined as
\begin{eqnarray}
	A^\Lambda_{FB} &=& \frac{[\int_{0}^{1}-\int_{-1}^{0}]d\cos\theta_{\Lambda} \frac{d^2\Gamma}{dm_{\Sigma\pi}^2 d\cos\theta_{\Lambda}}}{[\int_{0}^{1}+\int_{-1}^{0}]d\cos\theta_{\Lambda} \frac{d^2\Gamma}{dm_{\Sigma\pi}^2 d\cos\theta_{\Lambda}}}
	= \frac{\int dq^23(3L_{1(2)}-L_{4(2)})}{\int dq^22(9L_{1(1)}-3L_{1(3)}-3L_{4(1)}+L_{4(3)})}.\notag\\
    	A^\ell_{FB} &=& \frac{[\int_{0}^{1}-\int_{-1}^{0}]d\cos\theta_{\ell}  \frac{d^2\Gamma}{dm_{\Sigma\pi}^2 d\cos\theta_{\ell}}}{[\int_{0}^{1}+\int_{-1}^{0}]d\cos\theta_{\ell}\frac{d^2\Gamma}{dm_{\Sigma\pi}^2 d\cos\theta_{\ell}}}
	= \frac{\int dq^23(3L_{2(1)}-L_{2(3)})}{\int dq^22(9L_{1(1)}-3L_{1(3)}-3L_{4(1)}+L_{4(3)})}.
	\label{eq:AFB_Lambda}
\end{eqnarray}
One can see that $A_{FB}$ is proportional to the coefficient of the $\cos\theta$ term, with $A^\Lambda_{FB}\propto L_{\Lambda c}$, and can therefore directly reflect the interference behavior.  The numerical results are
\begin{eqnarray}
\langle A^\Lambda_{FB}\rangle = -0.079,\;\;\; \langle A^\ell_{FB}\rangle = -0.122.
\end{eqnarray}
The $m_{\Sigma\pi}$ dependence of $A_{FB}$ and the corresponding angular coefficients are shown in Fig.~\ref{fig:afb_distributions}.

For the forward-backward asymmetry, the zero point in its distribution is an important issue. As discussed in our previous work~\cite{Xing:2022uqu}, since we consider real form factors, the zero point of $A^\Lambda_{FB}$ corresponds to each resonance central mass, with $A^\Lambda_{FB}\propto {\rm Re}(L_{\Lambda^*_{1405}}L^*_{\Lambda^*_{1520}})\propto (m_{\Sigma\pi}^2-m_{1405}^2)(m_{\Sigma\pi}^2-m_{1520}^2)$, which is consistent with our numerical results in Fig.~\ref{fig:afb_distributions}.  If we consider a small strong phase in the helicity amplitude $H_\lambda$, we find that
\begin{eqnarray}
&&A^\Lambda_{FB}\propto Re({\cal H}^\frac{1}{2}{\cal H}^{\frac{3}{2}*})\sim |H_\lambda(1405)||H_{\lambda}(1520)|\notag\\
&&\quad\times\bigg({\rm Re}(L_{\Lambda^*_{1405}}L^*_{\Lambda^*_{1520}})-(\delta^{\frac{1}{2}}-\delta^{\frac{3}{2}}){\rm Im}(L_{\Lambda^*_{1405}}L^*_{\Lambda^*_{1520}})\bigg),
\end{eqnarray}
where $\delta^\frac{1}{2}$ and $\delta^{\frac{3}{2}}$ are the helicity-amplitude strong phases of $\Lambda^*_{1405}$ and $\Lambda^*_{1520}$, respectively.
Therefore, the strong phase generated in the $\Lambda_c^+\to \Lambda^* \ell^+\nu_\ell$
  decay can shift the zero point of $A^\Lambda_{FB}$ away from the central resonance mass. The size of this shift may serve as a criterion for measuring these strong-phase differences.
For the $A^\ell_{FB}$, the numerator can be expanded by form factors as
\begin{eqnarray}
L_{\ell c}=24q^2\sqrt{s^+s^-}
\Big( 
|L_{\Lambda^*_{1520}}|^2\mathcal{A}_{\frac{3}{2}}^2(f^{\frac{3}{2}}_{\perp^\prime}g^{\frac{3}{2}}_{\perp^\prime}-\frac{1}{3}f^{\frac{3}{2}}_{\perp}g^{\frac{3}{2}}_{\perp})-4|L_{\Lambda^*_{1405}}|^2\mathcal{A}_{\frac{1}{2}}^2f^{\frac{1}{2}}_{\perp}g^{\frac{1}{2}}_{\perp}
\Big).
\end{eqnarray}
Consistent with the above discussion, this observable arises from the interference between the vector helicity
  amplitude $H_V$ and the axial-vector helicity amplitude $H_A$. The zero crossing of $A^\ell_{FB}$ as a function of
  $m_{\Sigma\pi}^2$ is affected by the contributions from different resonances. In our analysis, the $\Lambda^*_{1405}$
  gives the dominant contribution, leading to a mostly negative $A^\ell_{FB}$, although its maximum value is reached
  near the $\Lambda^*_{1520}$ mass region.

In addition, asymmetries can provide information on other coefficients. In our work, we define a normalized asymmetry observable $A_{\frac{\pi}{3}}$ as
\begin{eqnarray}
	A^\Lambda_{\frac{\pi}{3}} &=& \frac{[\int_{1/2}^{1}+\int_{-1}^{-1/2}-\int_{-1/2}^{1/2}]d\cos\theta_{\Lambda} \frac{d^2\Gamma}{dm_{\Sigma\pi}^2 d\cos\theta_{\Lambda}}}{[\int_{1/2}^{1}+\int_{-1}^{-1/2}+\int_{-1/2}^{1/2}]d\cos\theta_{\Lambda}\frac{d^2\Gamma}{dm_{\Sigma\pi}^2 d\cos\theta_{\Lambda}}}
	= \frac{\int dq^23(3L_{1(3)}-L_{4(3)})}{\int dq^22(9L_{1(1)}-3L_{1(3)}-3L_{4(1)}+L_{4(3)})}.\notag\\
    	A^\ell_{\frac{\pi}{3}} &=& \frac{[\int_{1/2}^{1}+\int_{-1}^{-1/2}-\int_{-1/2}^{1/2}]d\cos\theta_{\ell} \frac{d^2\Gamma}{dm_{\Sigma\pi}^2 d\cos\theta_{\ell}}}{[\int_{1/2}^{1}+\int_{-1}^{-1/2}+\int_{-1/2}^{1/2}]d\cos\theta_{\ell}\frac{d^2\Gamma}{dm_{\Sigma\pi}^2 d\cos\theta_{\ell}}}
	= \frac{\int dq^23(3L_{4(1)}-L_{4(3)})}{\int dq^22(9L_{1(1)}-3L_{1(3)}-3L_{4(1)}+L_{4(3)})}.\notag\\
\end{eqnarray}
This asymmetry $A_{\frac{\pi}{3}}$ is proportional to the coefficient of the $\cos2\theta$ term. The numerical results are given as 
\begin{eqnarray}
\langle A^\Lambda_{\frac{\pi}{3}}\rangle=0.0632 , \qquad  \langle A^\ell_{\frac{\pi}{3}}\rangle=-0.252.
\end{eqnarray}
The $m_{\Sigma\pi}$ dependence of $A^\Lambda_{\frac{\pi}{3}}$, together with the corresponding angular coefficient, is shown in Fig.~\ref{fig:afb_distributions}. 
For $A^\Lambda_{\frac{\pi}{3}}$, the behavior is similar to that of $L_{\Lambda c}$, especially the position of the resonance pole. As discussed above, since $L_{\Lambda c}$ receives a contribution from $\Lambda^*_{1520}$, the pole position corresponds to $m_{1520}^2$.
For $A^\ell_{\frac{\pi}{3}}$, one can see that the pole structure disappears.  Since the ratio $R=\frac{-3L_{4(1)}+L_{4(3)}}{9L_{1(1)}-3L_{1(3)}}$ can be estimated as $R=-0.176<1$, the formula for $A^\ell_{\frac{\pi}{3}}$ can be expanded as
$A^\ell_{\frac{\pi}{3}}=\frac{3}{2}(1+R+R^2+\cdot\cdot\cdot),$
where both the numerator and denominator of the ratio contain the same pole structure. Therefore, $A^\ell_{\frac{\pi}{3}}$ does not reflect the pole structure.

For the angle $\phi$, the angular distribution is more complex. However, we can still define observables that are helpful for determining the angular coefficients experimentally. Similar to the forward-backward asymmetry, the behavior of the $\phi$-dependent angular coefficients can be reflected by collecting data from selected regions of the $\phi$ angle. These observables are defined as
\begin{eqnarray}
	A^\phi_{\pi} &=& \frac{[\int_{0}^{\pi}-\int_{\pi}^{2\pi}]d\phi \frac{d^2\Gamma}{dm_{\Sigma\pi}^2 d\phi}}{[\int_{0}^{\pi}+\int_{\pi}^{2\pi}]d\phi \frac{d^2\Gamma}{dm_{\Sigma\pi}^2 d\phi}}
	= \frac{\int dq^29\pi L_{9(1)}}{\int dq^28(9L_{1(1)}-3L_{1(3)}-3L_{4(1)}+L_{4(3)})}.\notag\\
    A^\phi_{\frac{\pi}{2}} &=& \frac{[\int_{0}^{\pi/2}-\int_{\pi/2}^{\pi}-\int_{\pi}^{3\pi/2}+\int_{3\pi/2}^{2\pi}]d\phi \frac{d^2\Gamma}{dm_{\Sigma\pi}^2 d\phi}}{[\int_{0}^{\pi/2}+\int_{\pi/2}^{\pi}+\int_{\pi}^{3\pi/2}+\int_{3\pi/2}^{2\pi}]d\phi \frac{d^2\Gamma}{dm_{\Sigma\pi}^2 d\phi}}
	= \frac{\int dq^2 9\pi L_{8(1)}}{\int dq^2 8(9L_{1(1)}-3L_{1(3)}-3L_{4(1)}+L_{4(3)})}.\notag\\
        A^\phi_{-\frac{\pi}{2}} &=& \frac{[\int_{0}^{\pi/2}-\int_{\pi/2}^{\pi}+\int_{\pi}^{3\pi/2}-\int_{3\pi/2}^{2\pi}]d\phi \frac{d^2\Gamma}{dm_{\Sigma\pi}^2 d\phi}}{[\int_{0}^{\pi/2}+\int_{\pi/2}^{\pi}+\int_{\pi}^{3\pi/2}+\int_{3\pi/2}^{2\pi}]d\phi \frac{d^2\Gamma}{dm_{\Sigma\pi}^2 d\phi}}
	= \frac{\int dq^216L_{7(1)}}{\int dq^2\pi(9L_{1(1)}-3L_{1(3)}-3L_{4(1)}+L_{4(3)})}.\notag\\
            A^\phi_{ \frac{\pi}{3}} &=& \frac{[(\int_{0}^{\pi/3}+\int_{2\pi/3}^{\pi}+\int_{\pi}^{4\pi/3}+\int_{5\pi/3}^{2\pi})-2(\int_{\pi/3}^{2\pi/3}+\int_{4\pi/3}^{5\pi/3})]d\phi \frac{d^2\Gamma}{dm_{\Sigma\pi}^2 d\phi}}{[(\int_{0}^{\pi/3}+\int_{2\pi/3}^{\pi}+\int_{\pi}^{4\pi/3}+\int_{5\pi/3}^{2\pi})+2(\int_{\pi/3}^{2\pi/3}+\int_{4\pi/3}^{5\pi/3})]d\phi \frac{d^2\Gamma}{dm_{\Sigma\pi}^2 d\phi}} \notag\\
        &=&\frac{\int dq^218\sqrt{3}L_{3(1)}}{\int dq^2(-9\pi L_{1(1)}+3\pi L_{1(3)}+3\pi L_{4(1)}-\pi L_{4(3)}+6\sqrt{3}L_{3(1)})},
\end{eqnarray}
These observables can directly determine the coefficients $L_{\phi c}$, $L_{\phi 2c}$, $L_{\phi s}$, and $L_{\phi 2s}$. The numerical values are
\begin{eqnarray}
\langle A^{\phi}_{\pi} \rangle=-0.018, \quad \langle A^{\phi}_{\frac{\pi}{2}}\rangle=-0.0069, \quad \langle A^{\phi}_{-\frac{\pi}{2}} \rangle =0, \quad \langle A^{\phi}_{\frac{\pi}{3}} \rangle =-0.0014.
\end{eqnarray}
Because the form factors used in our work are real, the observable $A^{\phi}_{-\frac{\pi}{2}}$, which depends only on the strong phase, vanishes.  A non-trivial $A^{\phi}_{-\frac{\pi}{2}}$ will be discussed in the next section.

\begin{figure}[htbp]
    \begin{subfigure}{0.3\textwidth}
	\centering
	\includegraphics[width=\linewidth]{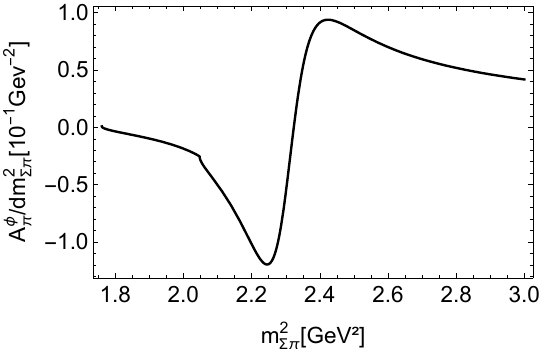}
    \caption{$A^{\phi}_{\pi}$}
    \end{subfigure}
    \hfill
    \begin{subfigure}{0.3\textwidth}
	\centering
	\includegraphics[width=\linewidth]{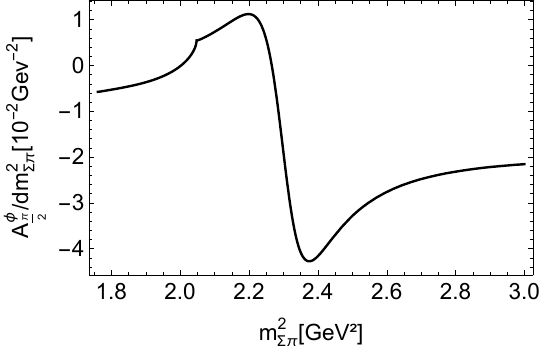}
    \caption{$A^{\phi}_{\frac{\pi}{2}}$}
    \end{subfigure}
    \hfill
    \begin{subfigure}{0.3\textwidth}
	\centering
	\includegraphics[width=\linewidth]{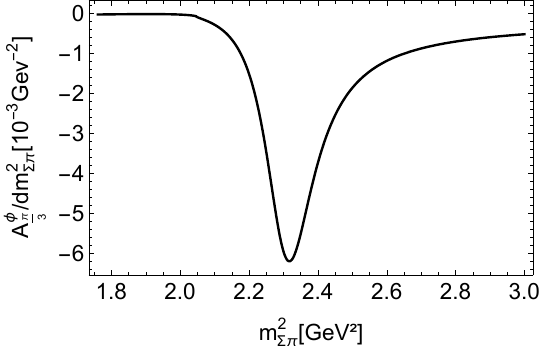}
    \caption{$A^{\phi}_{\frac{\pi}{3}}$}
    \end{subfigure}
    \begin{subfigure}{0.3\textwidth}
	\centering
	\includegraphics[width=\linewidth]{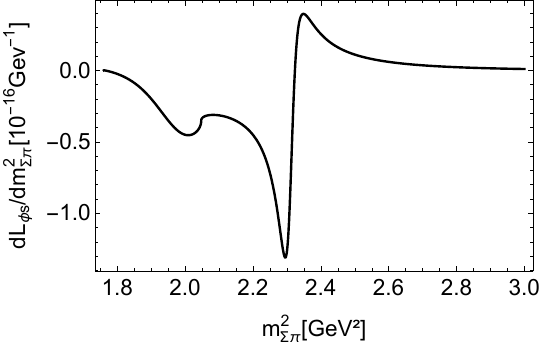}
    \caption{$L_{\phi s}$}
    \end{subfigure}
    \hfill
    \begin{subfigure}{0.3\textwidth}
	\centering
	\includegraphics[width=\linewidth]{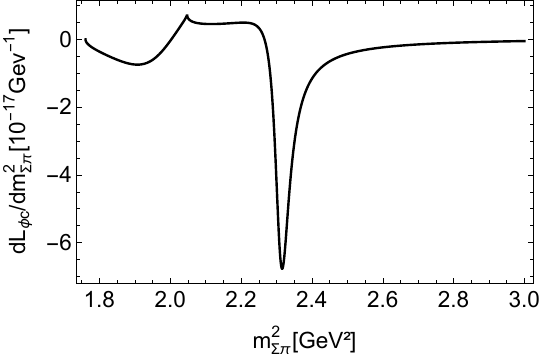}
    \caption{$L_{\phi c}$}
    \end{subfigure}
    \hfill
    \begin{subfigure}{0.3\textwidth}
	\centering
	\includegraphics[width=\linewidth]{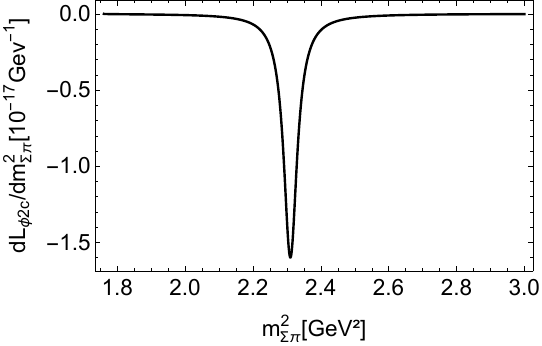}
    \caption{$L_{\phi 2c}$}
    \end{subfigure}
	\caption{$A^{\phi}_{\pi}$, $A^{\phi}_{\frac{\pi}{2}}$, and $A^{\phi}_{\frac{\pi}{3}}$ asymmetries and their corresponding angular coefficients as functions of $m_{\Sigma\pi}$.}
	\label{phico}
\end{figure}

The $m_{\Sigma\pi}$ dependence of these observables and their corresponding angular coefficients is shown in Fig.~\ref{phico}. One can see that the trends of the normalized asymmetries and the corresponding angular coefficients are similar, which is consistent with our previous analysis. However, there are non-smooth regions in the curves around $m_{\Sigma\pi}^2=2.05{\rm GeV}^2$. This is because when $m_{\Sigma\pi}^2>2.05{\rm GeV}^2$, the on-shell final state $\bar K N$ can be produced, leading to a non-smooth point in the lineshape function of $\Lambda^*_{1405}$ in Eq.~\eqref{eq:flatte}.  

\begin{figure}[htbp]
	\centering
	\begin{subfigure}{0.45\textwidth}
		\centering
		\includegraphics[width=\linewidth]{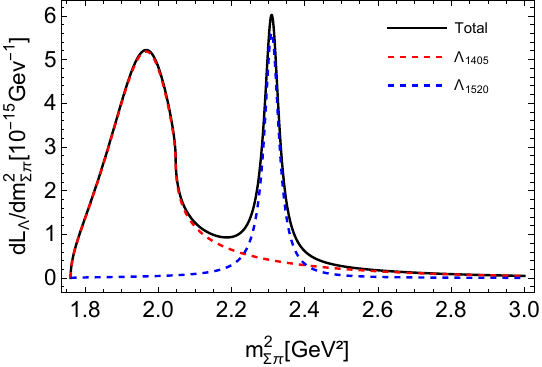}
		\caption{$L_\Lambda$}
	\end{subfigure}
	\begin{subfigure}{0.45\textwidth}
		\centering
		\includegraphics[width=\linewidth]{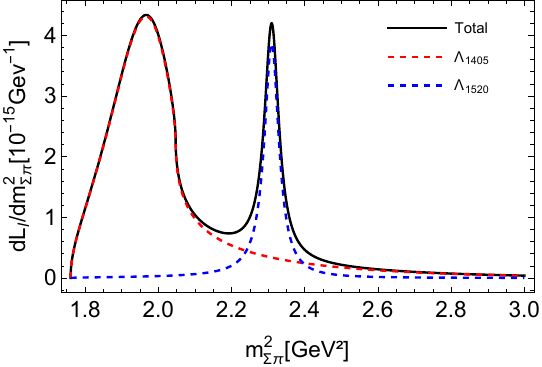}
		\caption{$L_\ell$}
	\end{subfigure}
	\caption{Angular coefficients for the $L_\Lambda$ and $L_\ell$ distributions as functions of $m_{\Sigma\pi}$.}
	\label{fig:angular_coefficients}
\end{figure}

Apart from being determined through these asymmetry observables, the angular coefficients can also be obtained by fitting the experimental data over the full angular region. Therefore, we also show the $m_{\Sigma\pi}$ dependence of the remaining angular coefficients $L_\Lambda$ and $L_\ell$ in Fig.~\ref{fig:angular_coefficients}. Their integrated values are
\begin{eqnarray}
\langle L_\Lambda \rangle=1.72 \times 10^{-15}  \rm GeV, \quad
\langle L_\ell \rangle=1.38 \times 10^{-15} \rm GeV .
\end{eqnarray}

\section{Implications for form factors}
\label{sec:ff_sensitivity}
As mentioned above, the angular coefficients in Eq.~\eqref{eq:angular_expansion} can be determined by fitting experimental data over the full angular region when sufficient data are available. After determining these coefficients, the next important task is to determine the form factors from them. Form factors contain complex non-perturbative effects in the hadronic matrix element and are important for understanding hadron structure and decay properties.

In general, the hadronic amplitude can contain an imaginary part generated by the strong interaction. Therefore, the form factors defined here can have strong phases, which can be written as
\begin{eqnarray}
f=|f|e^{i\delta^f},\;\;\;g=|g|e^{i\delta^g}.
\end{eqnarray}
Since the strong phase provides important information about the strong interaction, determining it experimentally is of great interest.
However, the strong phase increases the difficulty of fitting form factors experimentally. 
Therefore, in this section, we first omit these complex phases and discuss the strategy for extracting form factors using the angular coefficients. Then we discuss how to extract the strong phase itself.
\subsection{Form-factor sensitivity of angular observables}
First, using these real input form factors, we investigate the sensitivity of the angular observables to variations in the form-factor parameters.  We define the derivatives of the angular coefficients as $\frac{1}{L_{i}}\frac{\partial L_{i}}{\partial f_i}$ and $\frac{1}{L_{i}}\frac{\partial L_{i}}{\partial g_i}$ to quantify this sensitivity. Since the form factors, which characterize semi-leptonic decays, depend on $q^2$, the $q^2$ distributions are also given in our work. To simplify the calculation, the lepton mass $m_\ell$ is omitted, and the simplified formulae for the coefficients are given in Appendix~\ref{sec:app_simple}. The integrated derivatives for $\Lambda^*_{1405}$ and $\Lambda^*_{1520}$ are shown in Table~\ref{sensitivity1} and Table~\ref{sensitivity2}, respectively.  

\begin{table}[htbp]
\centering
	\caption{Numerical results of the sensitivity test of the angular coefficients to the form factors of $\Lambda^*_{1405}$.}
	\label{sensitivity1}
	\begin{tabular}{cccccccc}
		\hline\hline
		&\hspace{1ex}
		&$g_0$\hspace{1ex} 
		&$g_+$\hspace{1ex} 
		&$g_\perp$\hspace{1ex} 
		&$f_0$\hspace{1ex} 
		&$f_+$\hspace{1ex} 
		&$f_\perp$\hspace{1ex} \\
		\hline
		&$L_{1(1)}$\hspace{1ex} 
		&0.000\hspace{1ex} 
		&0.412\hspace{1ex} 
		&0.352\hspace{1ex} 
		&0.000\hspace{1ex} 
		&0.673\hspace{1ex} 
		&0.240\hspace{1ex}  \\
		\hline
		&$L_{1(2)}$\hspace{1ex} 
		&0.000\hspace{1ex} 
		&0.228\hspace{1ex} 
		&0.238\hspace{1ex} 
		&0.000\hspace{1ex} 
		&0.553\hspace{1ex} 
		&-0.019\hspace{1ex}\\
		\hline
		&$L_{2(1)}$\hspace{1ex} 
		&0.000\hspace{1ex} 
		&0.000\hspace{1ex} 
		&1.029\hspace{1ex} 
		&0.000\hspace{1ex} 
		&0.000\hspace{1ex} 
		&1.029\hspace{1ex}\\
		\hline
		&$L_{2(2)}$\hspace{1ex} 
		&0.000\hspace{1ex} 
		&0.000\hspace{1ex} 
		&-0.416\hspace{1ex} 
		&0.000\hspace{1ex} 
		&0.000\hspace{1ex} 
		&1.416\hspace{1ex} \\
		\hline
		&$L_{4(1)}$\hspace{1ex} 
		&-----\hspace{1ex} 
		&0.763\hspace{1ex} 
		&-0.217\hspace{1ex} 
		&-----\hspace{1ex} 
		&1.247\hspace{1ex} 
		&-0.149\hspace{1ex}\\
		\hline
		&$L_{4(2)}$\hspace{1ex} 
		&-----\hspace{1ex} 
		&0.322\hspace{1ex} 
		&-0.112\hspace{1ex} 
		&-----\hspace{1ex} 
		&0.781\hspace{1ex} 
		&0.009\hspace{1ex}\\
		\hline
		&$L_{8(1)}$\hspace{1ex} 
		&0.000\hspace{1ex} 
		&1.157\hspace{1ex} 
		&1.251\hspace{1ex} 
		&0.000\hspace{1ex} 
		&-0.902\hspace{1ex} 
		&-0.506\hspace{1ex} \\
		\hline
		&$L_{9(1)}$\hspace{1ex} 
		&0.000\hspace{1ex} 
		&0.481\hspace{1ex} 
		&0.330\hspace{1ex} 
		&0.000\hspace{1ex} 
		&0.730\hspace{1ex} 
		&-0.541\hspace{1ex} \\
		\hline
		&$L_{10(1)}$\hspace{1ex} 
		&-----\hspace{1ex} 
		&0.333\hspace{1ex} 
		&-0.233\hspace{1ex} 
		&-----\hspace{1ex} 
		&0.545\hspace{1ex} 
		&0.355\hspace{1ex} \\
		\hline
		&$L_{11(1)}$\hspace{1ex} 
		&-----\hspace{1ex} 
		&-2.368\hspace{1ex} 
		&2.515\hspace{1ex} 
		&-----\hspace{1ex} 
		&2.003\hspace{1ex} 
		&-1.150\hspace{1ex} \\
		\hline\hline
	\end{tabular}	
\end{table}
 Owing to the suppression by the lepton mass, the scalar form factors $f_0$ and $g_0$ give negligible contributions for
  both $\Lambda^*_{1405}$ and $\Lambda^*_{1520}$. For the $\Lambda_c^+\to\Lambda^*_{1405}\ell^+\nu_\ell$ transition, the form
  factors $g_+$ and $f_+$ are mainly constrained by the coefficients $L_{8(1)}$ and $L_{11(1)}$. The form factor
  $g_\perp$ is most sensitive to $L_{2(1)}$, $L_{8(1)}$, and $L_{11(1)}$, while $f_\perp$ is primarily constrained by
  $L_{2(1)}$ and $L_{2(2)}$.
\begin{table}[htbp]
	\caption{Numerical results of the sensitivity test of the angular coefficients to the form factors of $\Lambda^*_{1520}$.}
	\label{sensitivity2}
	\begin{tabular}{cccccccccc}
		\hline\hline
		&\hspace{1ex} 
		&$f_0$\hspace{1ex} 
		&$f_+$\hspace{1ex} 
		&$f_\perp$\hspace{1ex} 
		&$f_{\perp^\prime}$\hspace{1ex} 
		&$g_0$\hspace{1ex} 
		&$g_+$\hspace{1ex} 
		&$g_\perp$\hspace{1ex}
		&$g_{\perp^\prime}$\hspace{1ex} \\
		\hline
		&$L_{1(1)}$\hspace{1ex} 
		&0.000\hspace{1ex} 
		&0.095\hspace{1ex} 
		&0.009\hspace{1ex} 
		&0.036\hspace{1ex} 
		&0.000\hspace{1ex} 
		&0.058\hspace{1ex} 
		&0.063\hspace{1ex}
		&0.000\hspace{1ex} \\
		\hline
		&$L_{1(2)}$\hspace{1ex} 
		&0.000\hspace{1ex} 
		&0.283\hspace{1ex} 
		&-0.095\hspace{1ex} 
		&-----\hspace{1ex} 
		&0.000\hspace{1ex} 
		&0.294\hspace{1ex} 
		&0.255\hspace{1ex}
		&-----\hspace{1ex} \\
		\hline
		&$L_{1(3)}$\hspace{1ex} 
		&0.000\hspace{1ex} 
		&0.809\hspace{1ex} 
		&0.074\hspace{1ex} 
		&-0.505\hspace{1ex} 
		&0.000\hspace{1ex} 
		&0.495\hspace{1ex} 
		&0.535\hspace{1ex}
		&-0.002\hspace{1ex} \\
		\hline
		&$L_{2(1)}$\hspace{1ex} 
		&0.000\hspace{1ex} 
		&0.000\hspace{1ex} 
		&-0.124\hspace{1ex} 
		&0.006\hspace{1ex} 
		&0.000\hspace{1ex} 
		&0.000\hspace{1ex} 
		&-0.034\hspace{1ex}
		&0.015\hspace{1ex} \\
		\hline
		&$L_{2(2)}$\hspace{1ex} 
		&0.000\hspace{1ex} 
		&0.000\hspace{1ex} 
		&-1.14\hspace{1ex} 
		&-----\hspace{1ex} 
		&0.000\hspace{1ex} 
		&0.000\hspace{1ex} 
		&1.559\hspace{1ex}
		&-----\hspace{1ex} \\
		\hline
		&$L_{2(3)}$\hspace{1ex} 
		&0.000\hspace{1ex} 
		&0.000\hspace{1ex} 
		&2.886\hspace{1ex} 
		&0.221\hspace{1ex} 
		&0.000\hspace{1ex} 
		&0.000\hspace{1ex} 
		&0.786\hspace{1ex}
		&0.581\hspace{1ex} \\
		\hline
		&$L_{3(1)}$\hspace{1ex} 
		&-----\hspace{1ex} 
		&-----\hspace{1ex} 
		&2.046\hspace{1ex} 
		&0.894\hspace{1ex} 
		&-----\hspace{1ex} 
		&-----\hspace{1ex} 
		&0.137\hspace{1ex}
		&0.449\hspace{1ex} \\
		\hline
		&$L_{4(1)}$\hspace{1ex} 
		&-----\hspace{1ex} 
		&0.176\hspace{1ex} 
		&-0.005\hspace{1ex} 
		&-0.022\hspace{1ex} 
		&-----\hspace{1ex} 
		&0.108\hspace{1ex} 
		&-0.039\hspace{1ex}
		&0.000\hspace{1ex} \\
		\hline
		&$L_{4(2)}$\hspace{1ex} 
		&-----\hspace{1ex} 
		&0.399\hspace{1ex} 
		&0.044\hspace{1ex} 
		&-----\hspace{1ex} 
		&-----\hspace{1ex} 
		&0.415\hspace{1ex} 
		&-0.120\hspace{1ex}
		&-----\hspace{1ex} \\
		\hline
		&$L_{4(3)}$\hspace{1ex} 
		&-----\hspace{1ex} 
		&0.856\hspace{1ex} 
		&-0.026\hspace{1ex} 
		&0.178\hspace{1ex} 
		&-----\hspace{1ex} 
		&0.524\hspace{1ex} 
		&-0.189\hspace{1ex}
		&0.001\hspace{1ex} \\
		\hline
		&$L_{8(1)}$\hspace{1ex} 
		&0.000\hspace{1ex} 
		&0.747\hspace{1ex} 
		&-0.450\hspace{1ex} 
		&1.350\hspace{1ex} 
		&0.000\hspace{1ex} 
		&-0.642\hspace{1ex} 
		&-1.119\hspace{1ex}
		&0.232\hspace{1ex} \\
		\hline
		&$L_{8(2)}$\hspace{1ex} 
		&0.000\hspace{1ex} 
		&0.048\hspace{1ex} 
		&-----\hspace{1ex} 
		&0.992\hspace{1ex} 
		&0.000\hspace{1ex} 
		&1.091\hspace{1ex} 
		&-----\hspace{1ex}
		&0.272\hspace{1ex} \\
		\hline
		&$L_{9(1)}$\hspace{1ex} 
		&0.000\hspace{1ex} 
		&-0.290\hspace{1ex} 
		&0.258\hspace{1ex} 
		&0.774\hspace{1ex} 
		&0.000\hspace{1ex} 
		&0.413\hspace{1ex} 
		&0.496\hspace{1ex}
		&0.103\hspace{1ex} \\
		\hline
		&$L_{10(1)}$\hspace{1ex} 
		&-----\hspace{1ex} 
		&0.194\hspace{1ex} 
		&0.185\hspace{1ex} 
		&0.554\hspace{1ex} 
		&-----\hspace{1ex} 
		&-0.285\hspace{1ex} 
		&0.337\hspace{1ex}
		&0.070\hspace{1ex} \\
		\hline
		&$L_{10(2)}$\hspace{1ex} 
		&-----\hspace{1ex} 
		&0.604\hspace{1ex} 
		&-----\hspace{1ex} 
		&1.050\hspace{1ex} 
		&-----\hspace{1ex} 
		&0.024\hspace{1ex} 
		&-----\hspace{1ex}
		&0.074\hspace{1ex} \\
		\hline
		&$L_{11(1)}$\hspace{1ex} 
		&-----\hspace{1ex} 
		&1.452\hspace{1ex} 
		&0.893\hspace{1ex} 
		&-2.680\hspace{1ex} 
		&-----\hspace{1ex} 
		&-1.477\hspace{1ex} 
		&2.463\hspace{1ex}
		&-0.510\hspace{1ex} \\
		\hline\hline
	\end{tabular}	
\end{table}

  For the $\Lambda_c^+\to\Lambda^*_{1520}\ell^+\nu_\ell$ transition, the dominant constraints on $f_+$ and $g_+$ come from the
  coefficient $L_{8(1)}$. The form factor $f_\perp$ is mainly constrained by $L_{2(3)}$ and $L_{3(1)}$, whereas
  $g_\perp$ is most sensitive to $L_{2(2)}$, $L_{2(3)}$, $L_{8(1)}$, and $L_{11(1)}$. In addition, $f_{\perp\prime}$ and
  $g_{\perp\prime}$ are primarily constrained by $L_{2(3)}$.
  
\begin{figure}[htbp]
	\centering
	\begin{subfigure}{0.19\textwidth}
		\centering
		\includegraphics[width=1.05\linewidth]{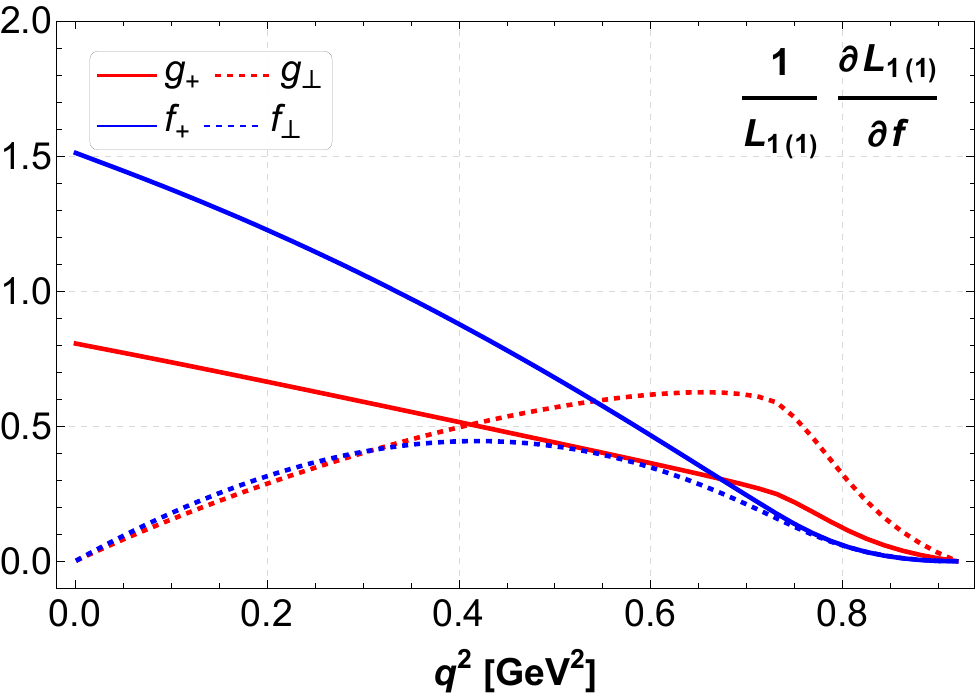}
	\end{subfigure}
	\hfill
	\begin{subfigure}{0.19\textwidth}
		\centering
		\includegraphics[width=1.05\linewidth]{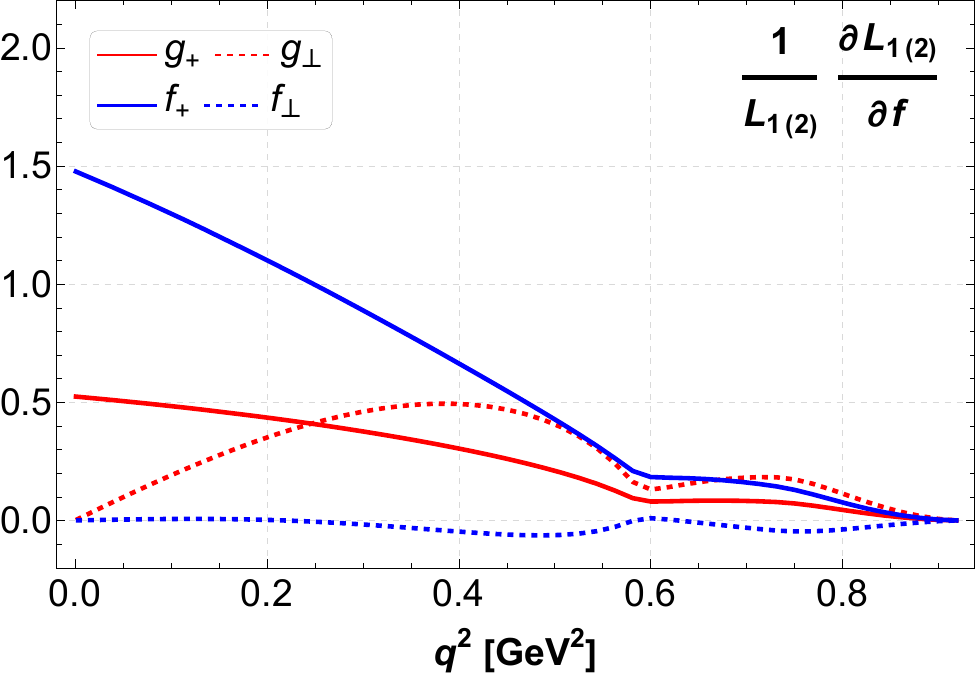}
	\end{subfigure}
	\hfill
	\begin{subfigure}{0.19\textwidth}
		\centering
		\includegraphics[width=1.05\linewidth]{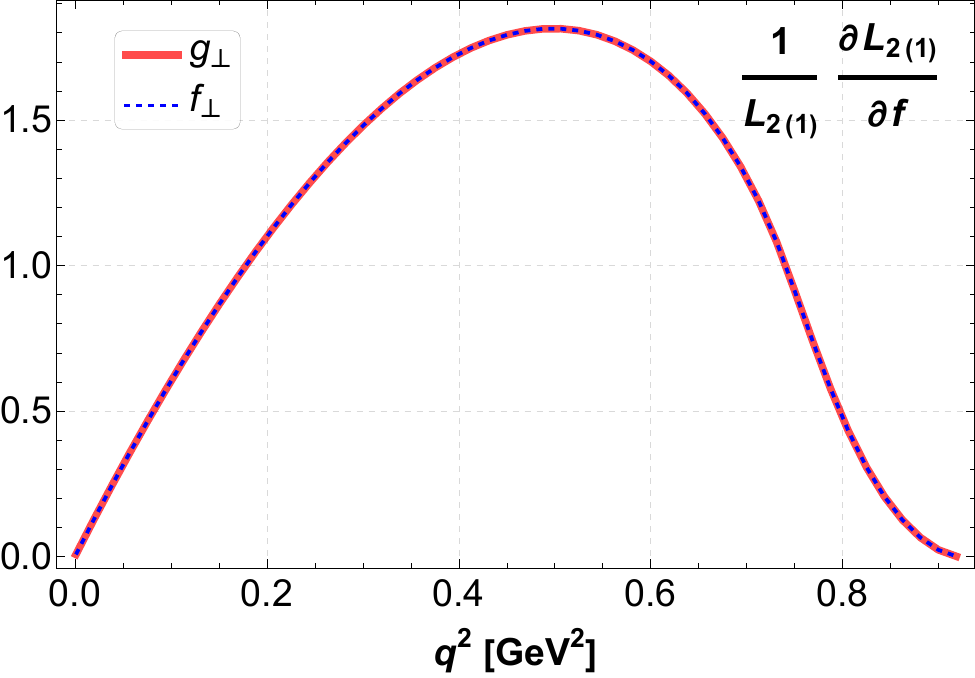}
	\end{subfigure}
	\hfill
	\begin{subfigure}{0.19\textwidth}
		\centering
		\includegraphics[width=1.05\linewidth]{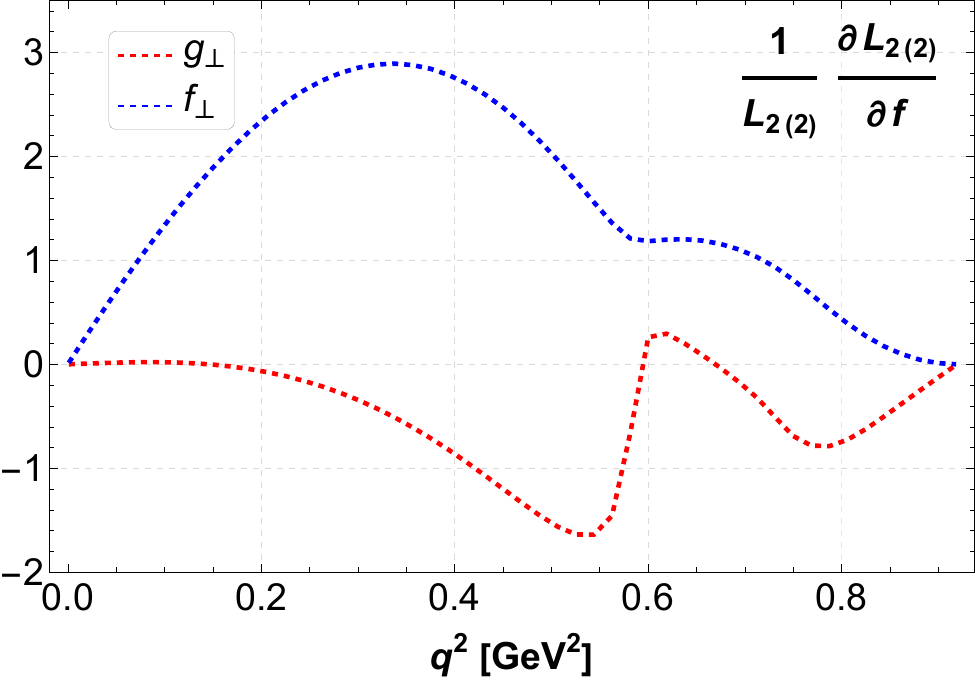}
	\end{subfigure}
    \hfill
    \begin{subfigure}{0.19\textwidth}
    	\centering
    	\includegraphics[width=1.05\linewidth]{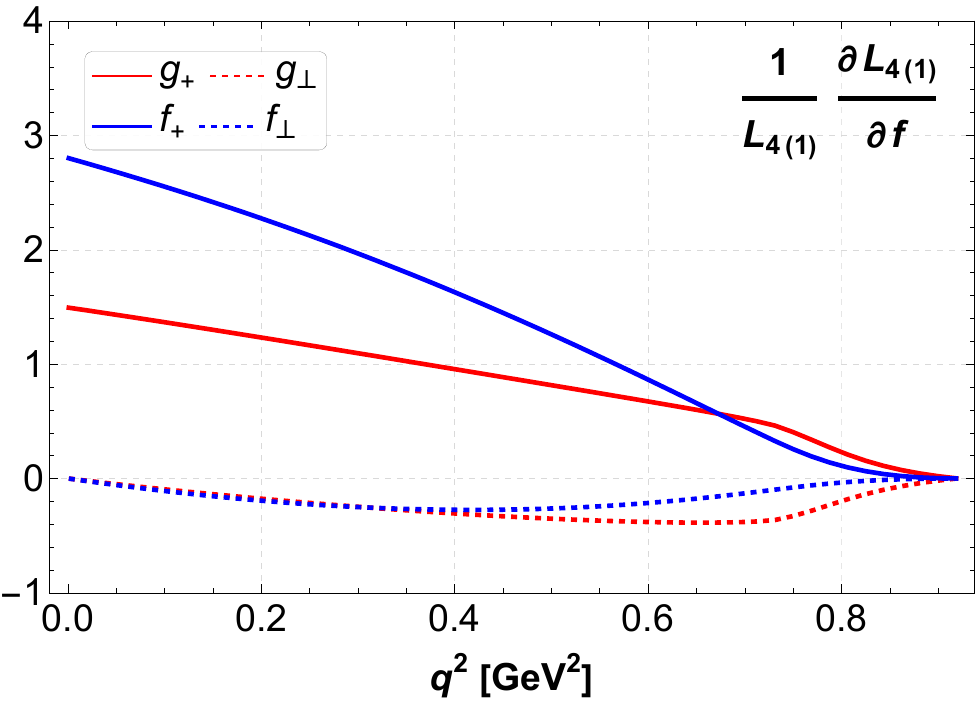}
    \end{subfigure}
    \hfill
    \begin{subfigure}{0.19\textwidth}
    	\centering
    	\includegraphics[width=1.05\linewidth]{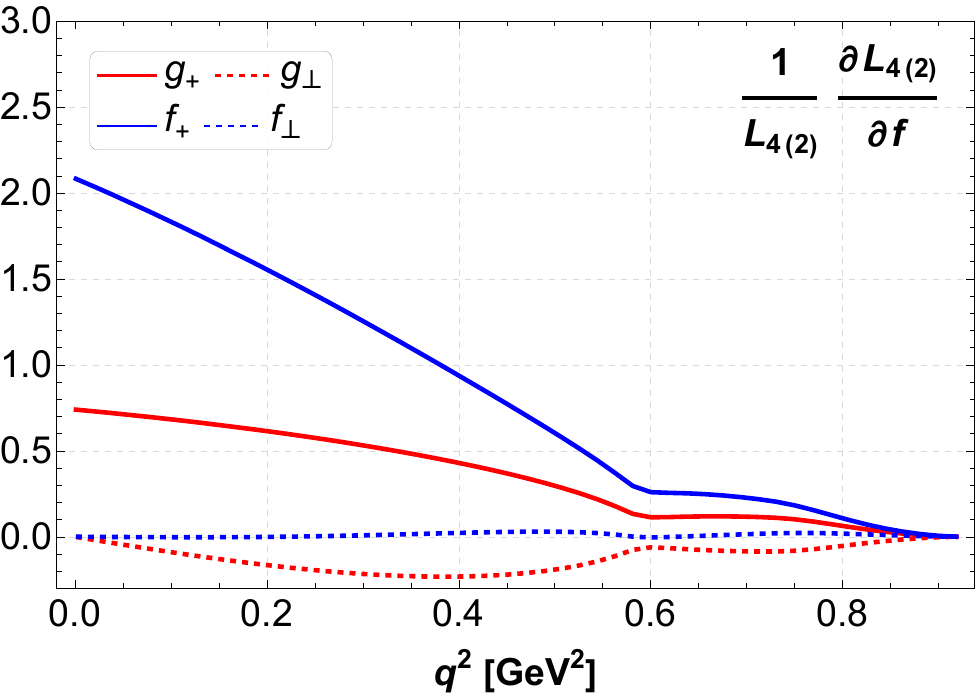}
    \end{subfigure}
    \hfill
    \begin{subfigure}{0.19\textwidth}
    	\centering
    	\includegraphics[width=1.05\linewidth]{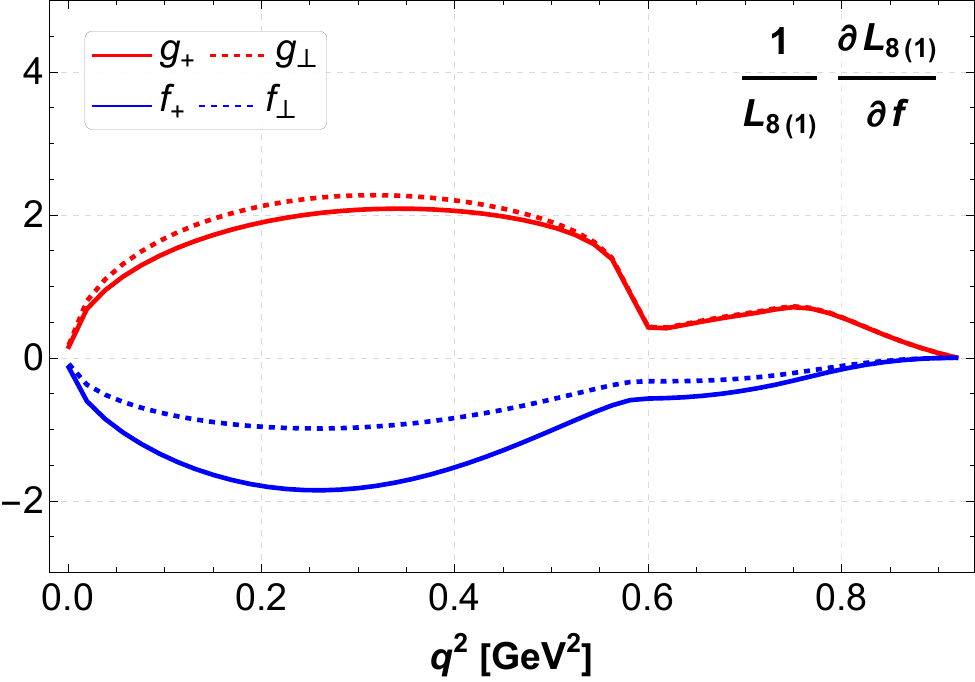}
    \end{subfigure}
    \hfill
    \begin{subfigure}{0.19\textwidth}
    	\centering
    	\includegraphics[width=1.05\linewidth]{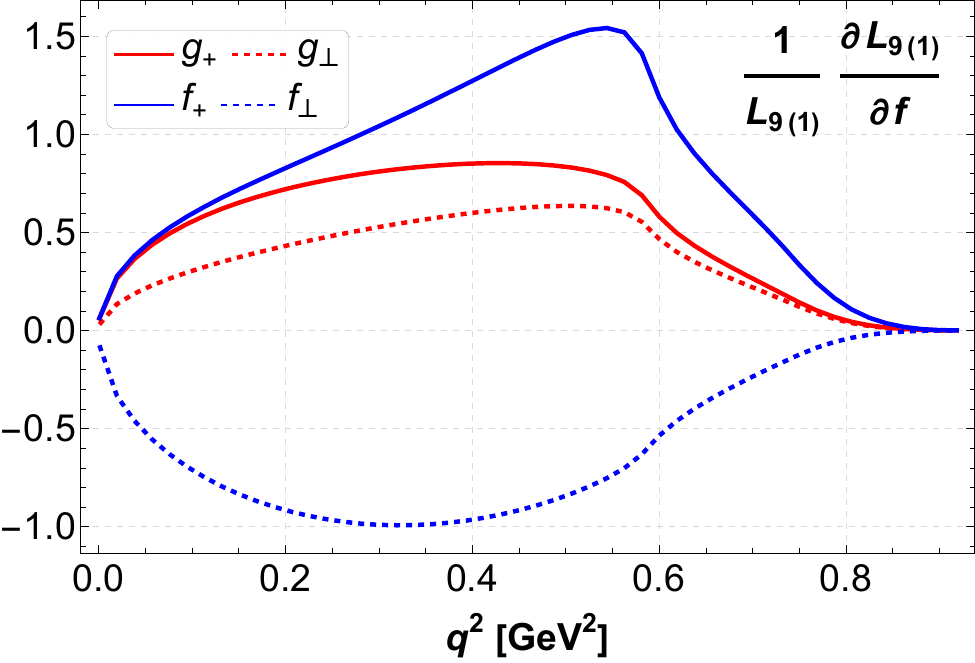}
    \end{subfigure}
     \hfill
     \begin{subfigure}{0.19\textwidth}
     	\centering
     	\includegraphics[width=1.05\linewidth]{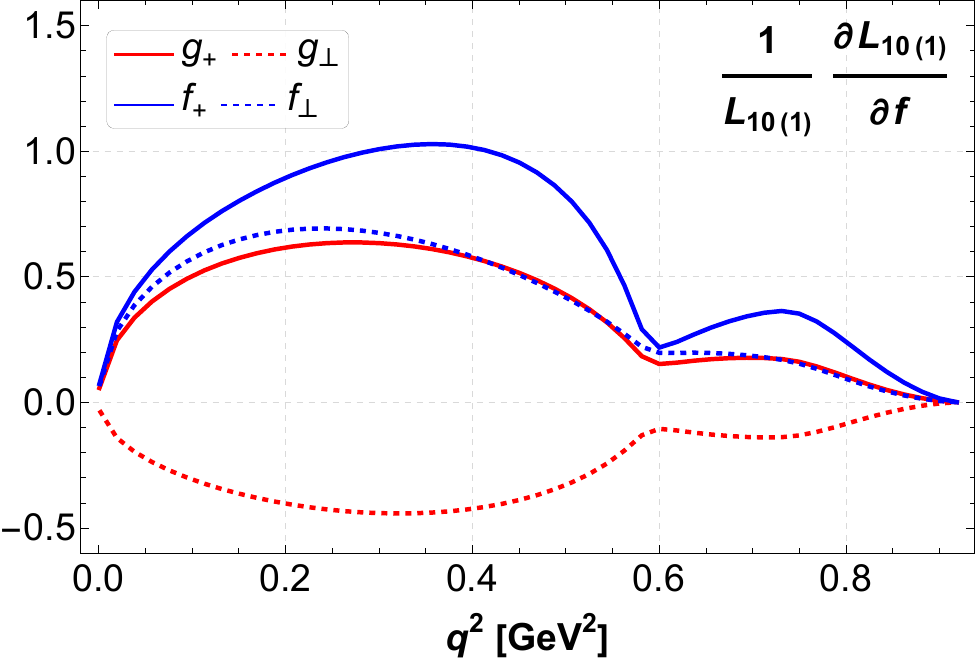}
     \end{subfigure}
     \hfill
     \begin{subfigure}{0.19\textwidth}
     	\centering
     	\includegraphics[width=1.05\linewidth]{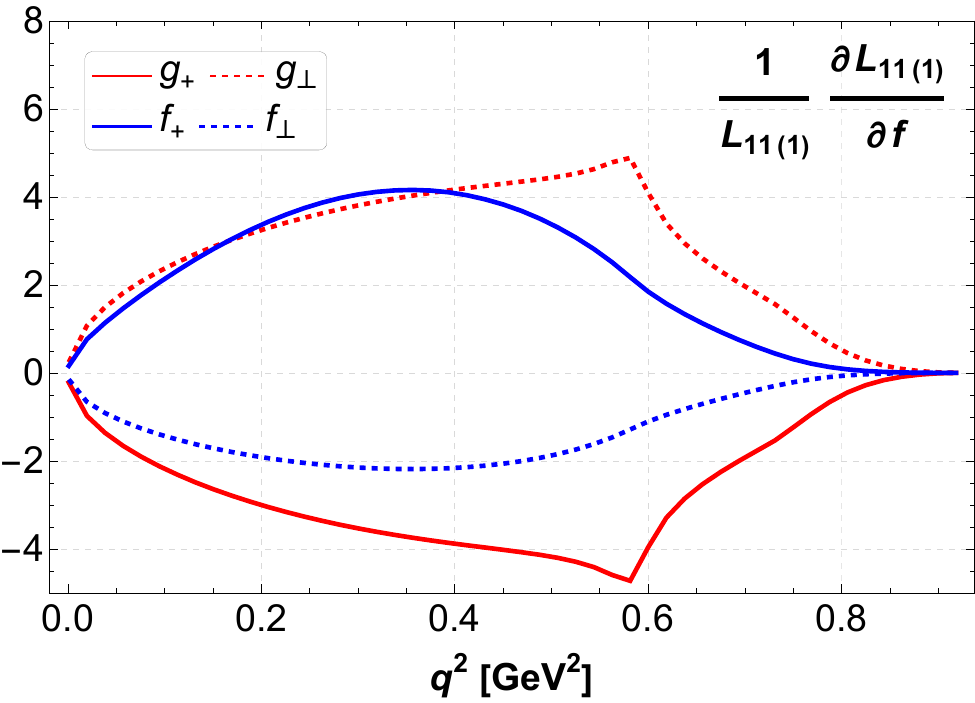}
     \end{subfigure}
	\caption{Form-factor sensitivity of the angular coefficients for the $\Lambda^*_{1405}$ transition in different $q^2$ regions. The normalization coefficient in the denominator is the value obtained after integrating over both $q^2$ and $m_{\Lambda^*}$.}
	\label{fig:ff_sens_1405}
\end{figure}

\begin{figure}[htbp]
	\centering
	\begin{subfigure}{0.24\textwidth}
		\centering
		\includegraphics[width=1.05\linewidth]{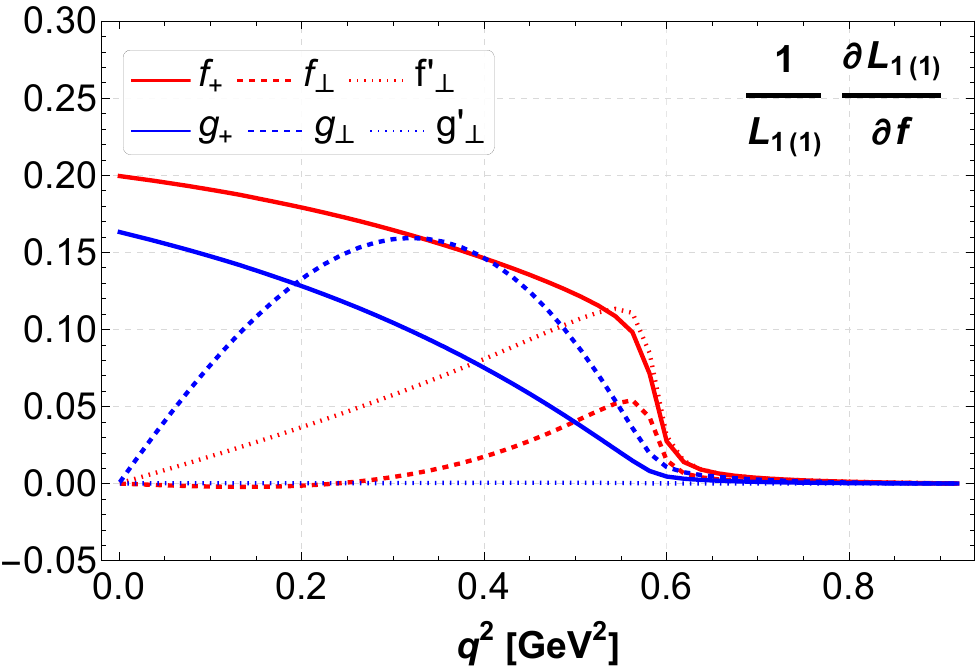}
	\end{subfigure}
	\hfill
	\begin{subfigure}{0.24\textwidth}
		\centering
		\includegraphics[width=1.05\linewidth]{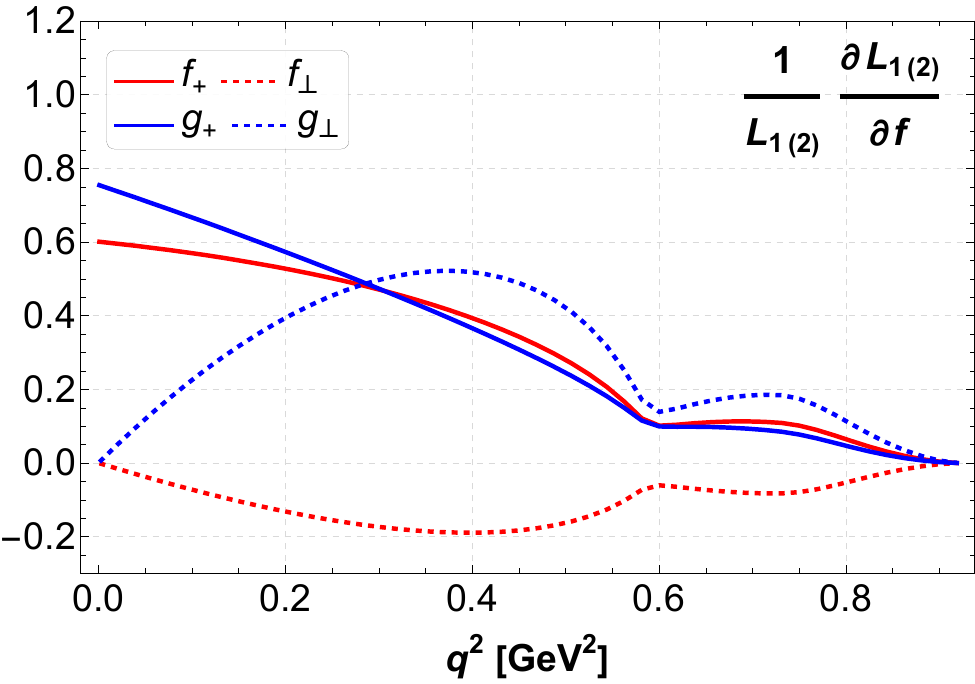}
	\end{subfigure}
	\hfill
	\begin{subfigure}{0.24\textwidth}
		\centering
		\includegraphics[width=1.05\linewidth]{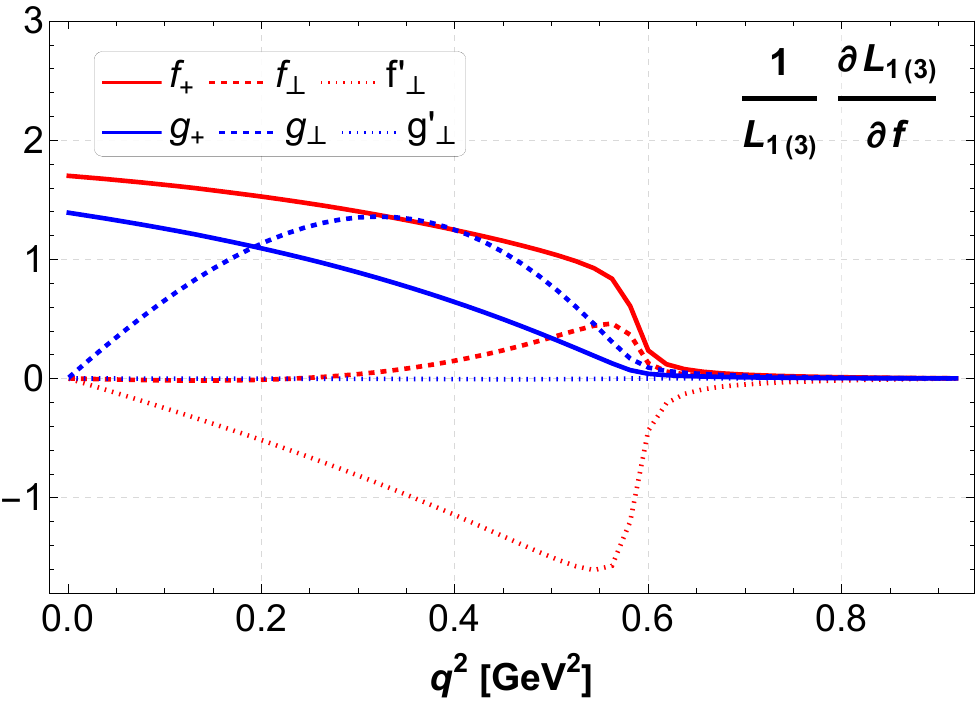}
	\end{subfigure}
	\hfill
	\begin{subfigure}{0.24\textwidth}
		\centering
		\includegraphics[width=1.05\linewidth]{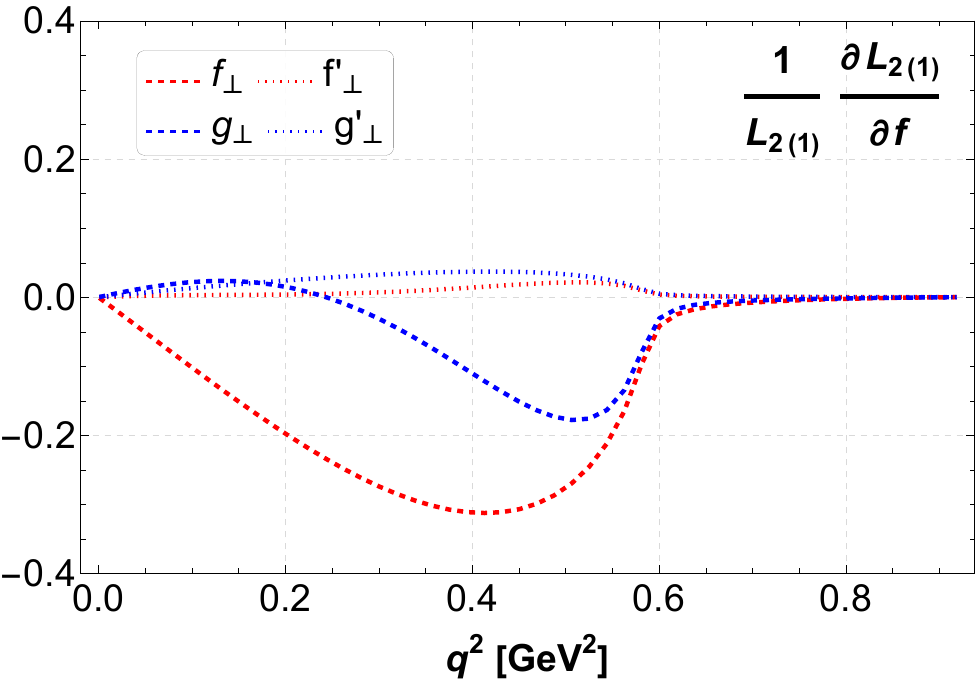}
	\end{subfigure}
	\hfill
	\begin{subfigure}{0.24\textwidth}
		\centering
		\includegraphics[width=1.05\linewidth]{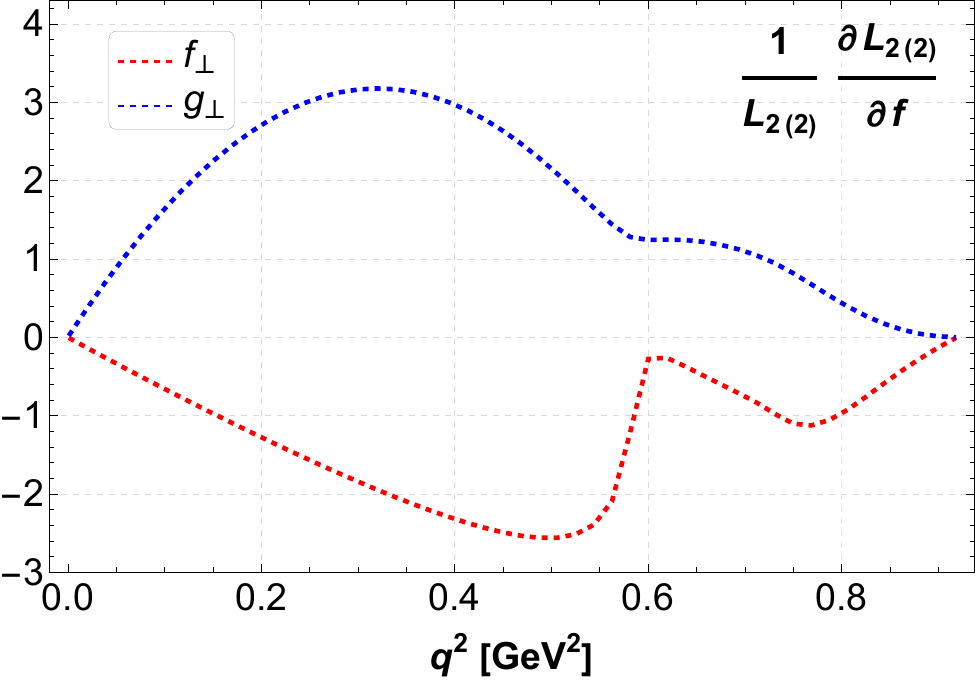}
	\end{subfigure}
	\hfill
	\begin{subfigure}{0.24\textwidth}
		\centering
		\includegraphics[width=1.05\linewidth]{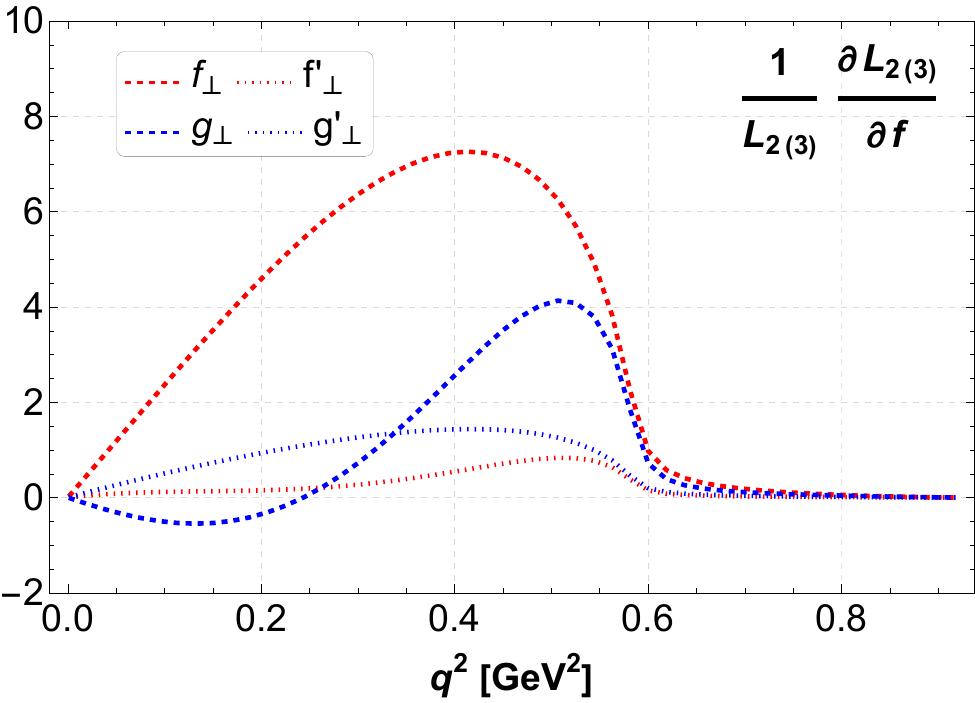}
	\end{subfigure}
	\hfill
	\begin{subfigure}{0.24\textwidth}
		\centering
		\includegraphics[width=1.05\linewidth]{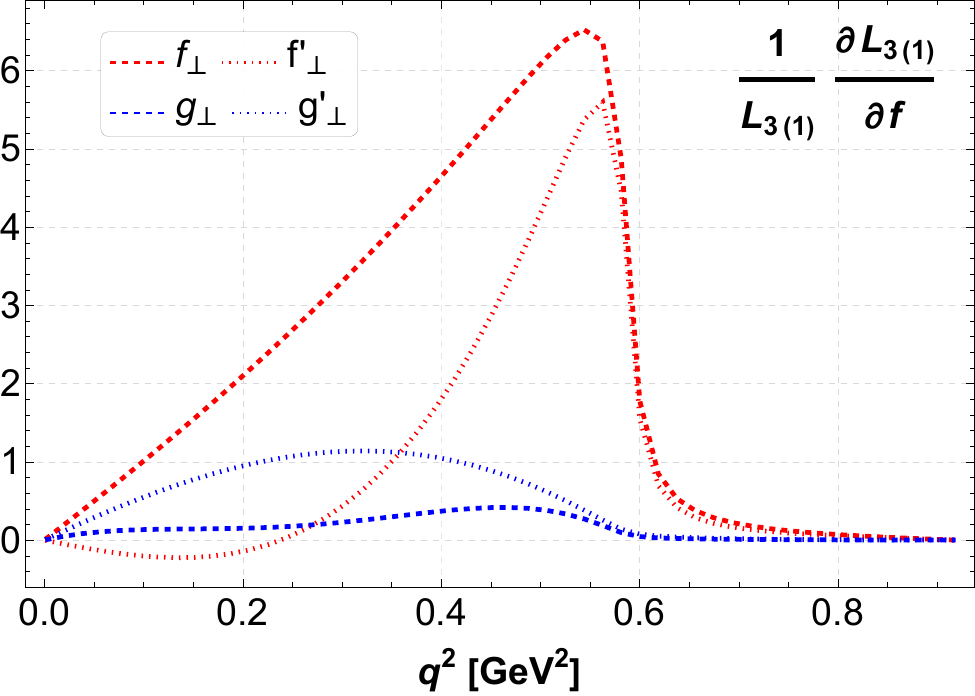}
	\end{subfigure}
	\hfill
	\begin{subfigure}{0.24\textwidth}
		\centering
		\includegraphics[width=1.05\linewidth]{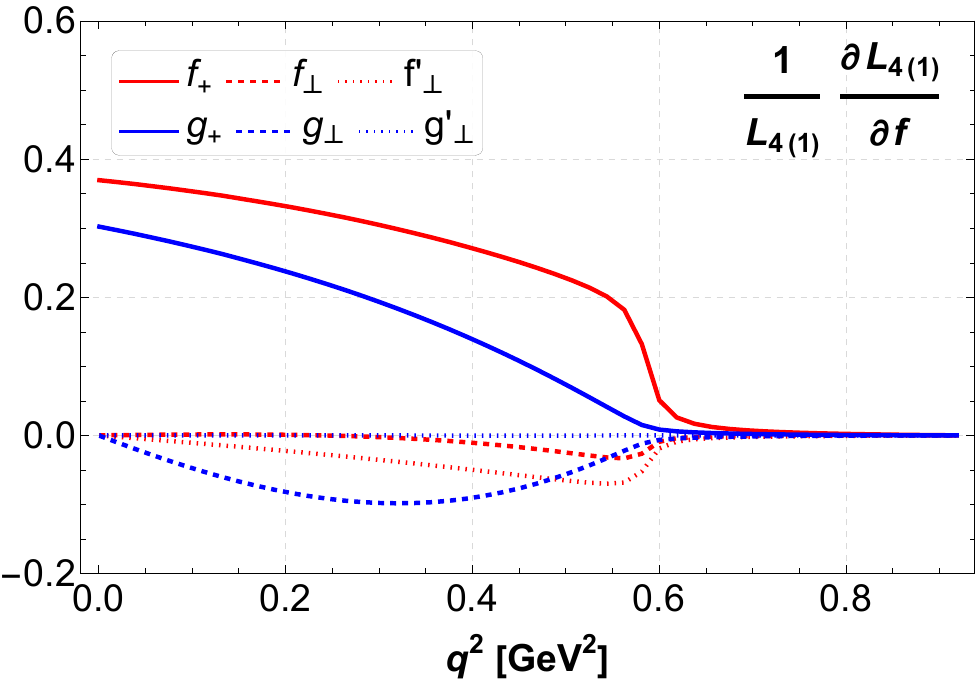}
	\end{subfigure}
	\hfill
	\begin{subfigure}{0.24\textwidth}
		\centering
		\includegraphics[width=1.05\linewidth]{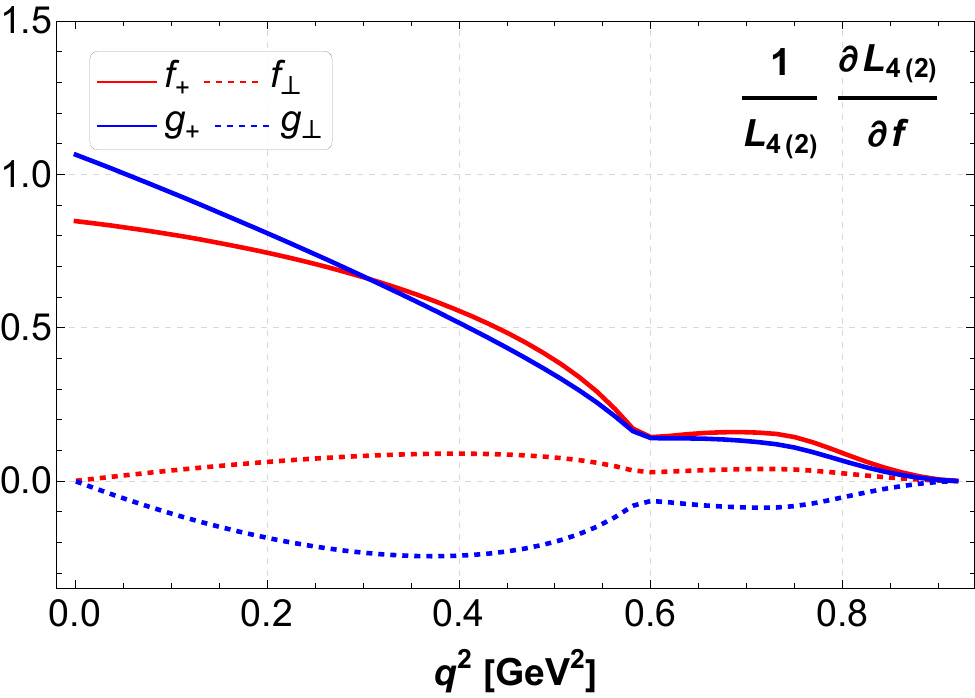}
	\end{subfigure}
	\hfill
	\begin{subfigure}{0.24\textwidth}
		\centering
		\includegraphics[width=1.05\linewidth]{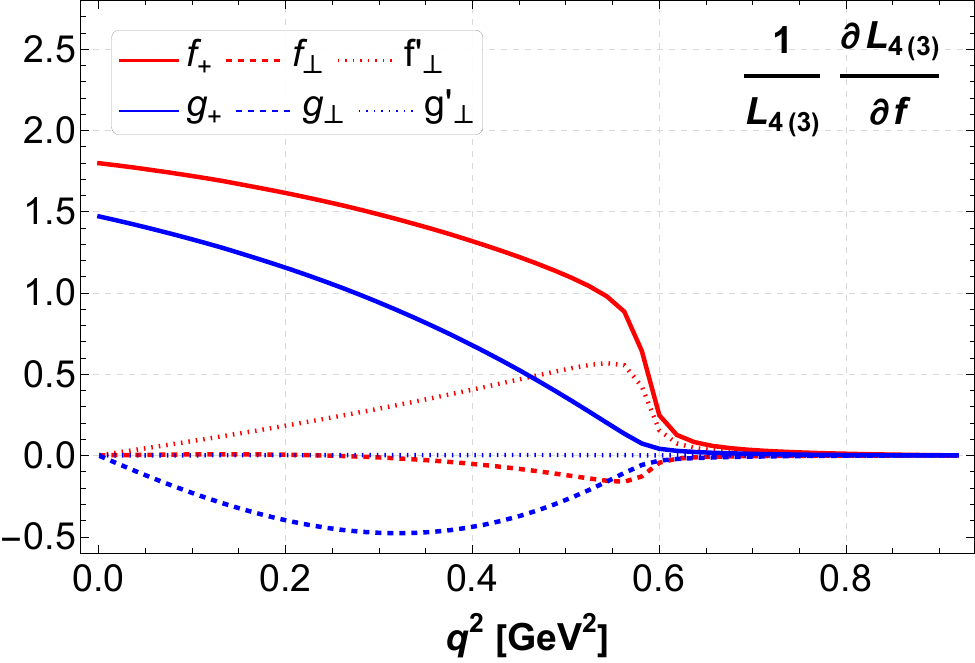}
	\end{subfigure}
    \hfill
     \begin{subfigure}{0.24\textwidth}
    	\centering
    	\includegraphics[width=1.05\linewidth]{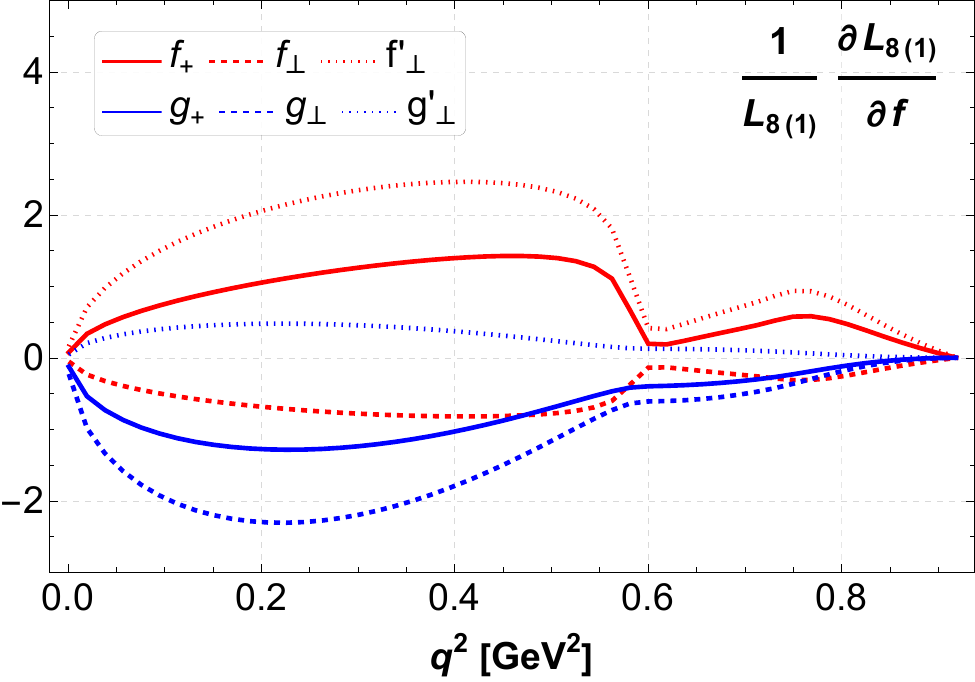}
    \end{subfigure}
    \hfill
    \begin{subfigure}{0.24\textwidth}
    	\centering
    	\includegraphics[width=1.05\linewidth]{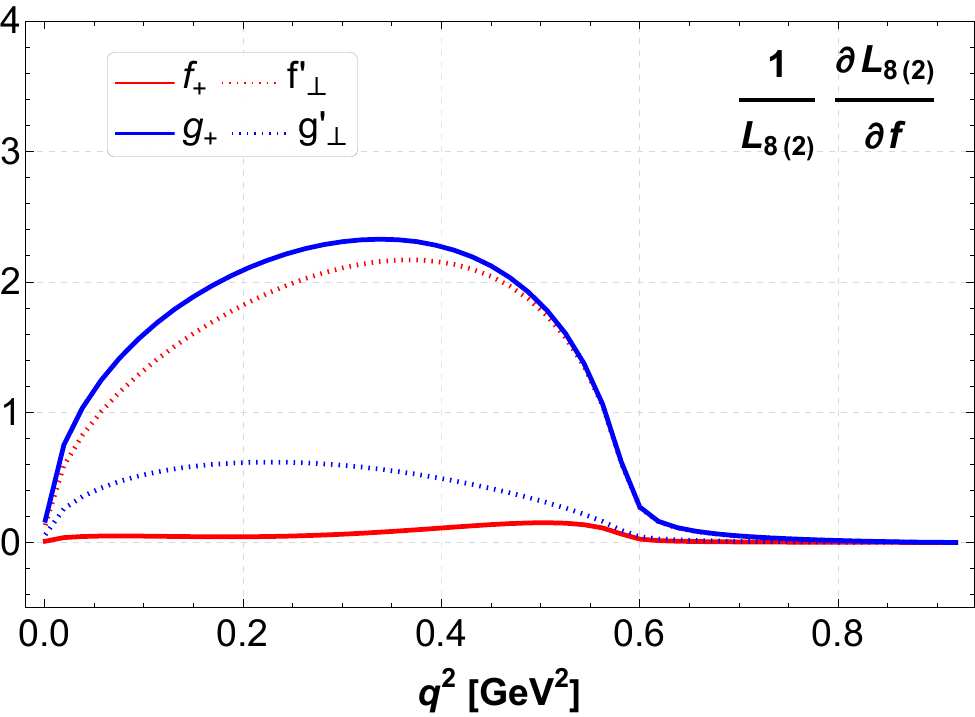}
    \end{subfigure}
    \begin{subfigure}{0.24\textwidth}
    	\centering
    	\includegraphics[width=1.05\linewidth]{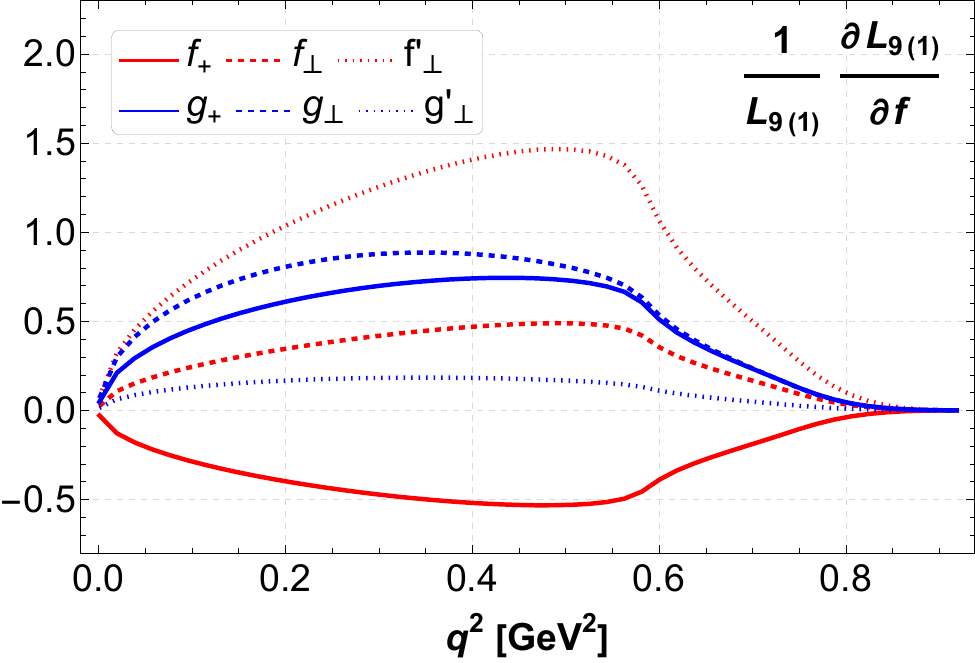}
    \end{subfigure}
    \hfill
    \begin{subfigure}{0.24\textwidth}
    	\centering
    	\includegraphics[width=1.05\linewidth]{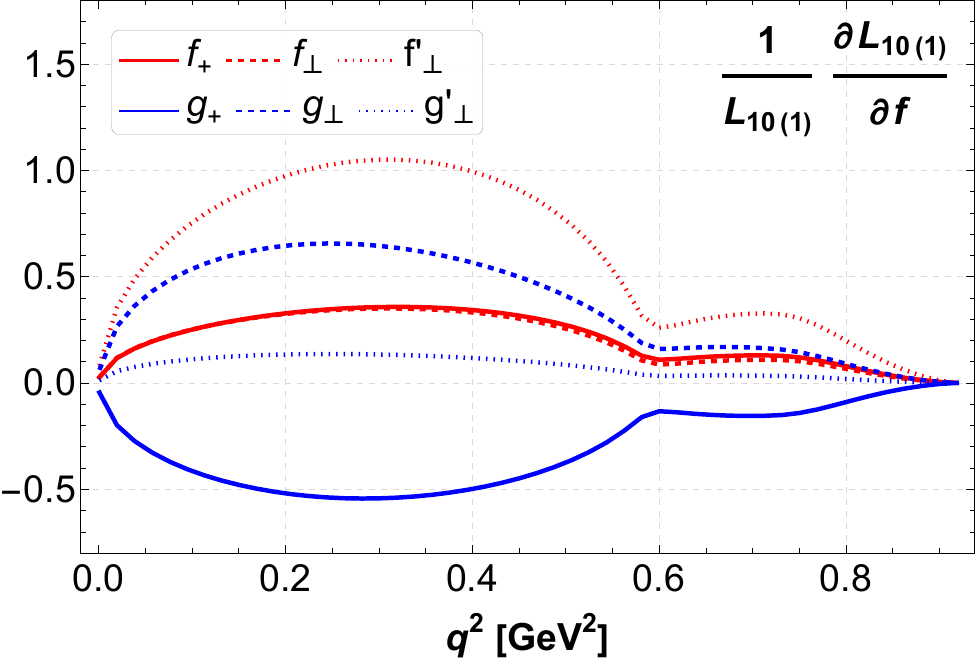}
    \end{subfigure}
    \hfill
    \begin{subfigure}{0.24\textwidth}
    	\centering
    	\includegraphics[width=1.05\linewidth]{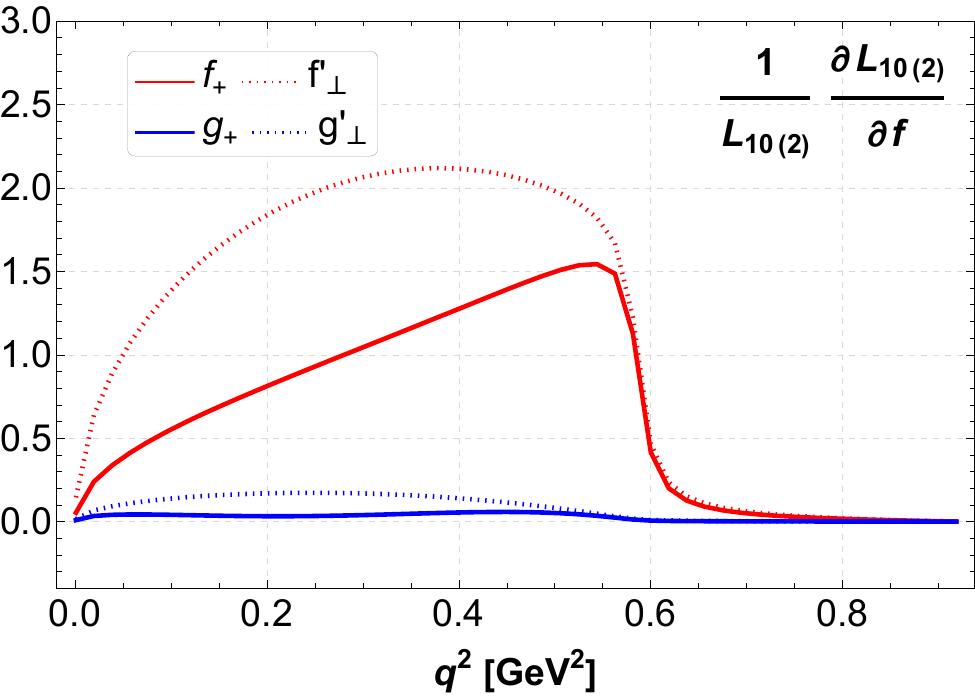}
    \end{subfigure}
    \hfill
    \begin{subfigure}{0.24\textwidth}
    	\centering
    	\includegraphics[width=1.05\linewidth]{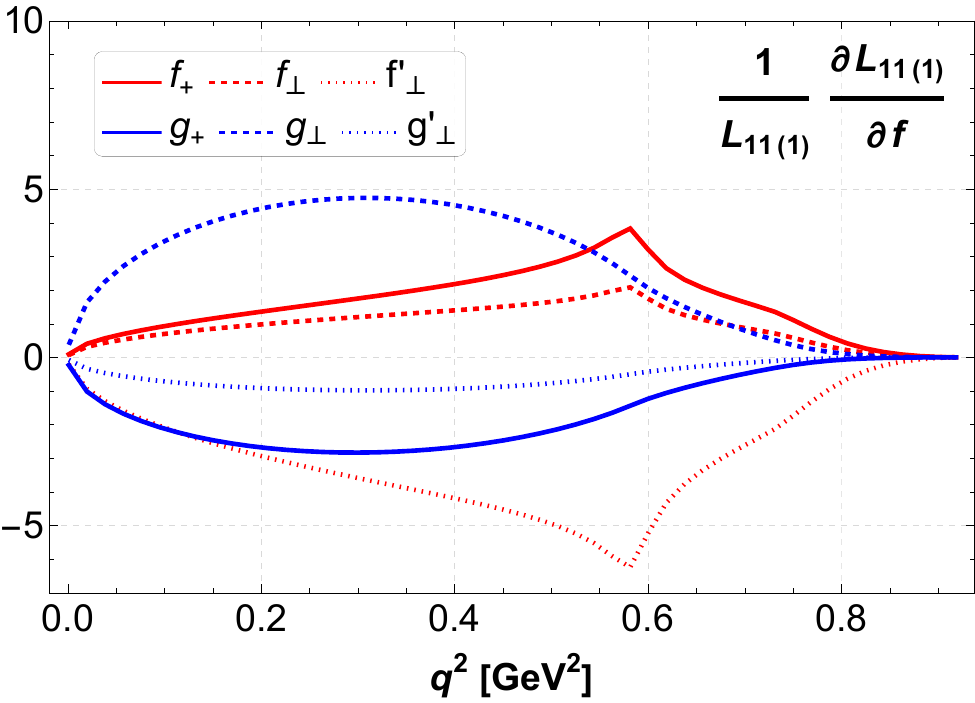}
    \end{subfigure}
	\caption{Form-factor sensitivity of the angular coefficients for the $\Lambda^*_{1520}$ transition in different $q^2$ regions. The normalization coefficient in the denominator is the value obtained after integrating over both $q^2$ and $m_{\Lambda^*}$.}
	\label{fig:ff_sens_1520}
\end{figure}

 Since experimental data can be collected event by event in specific $q^2$ regions, the $q^2$ dependence of the form-factor sensitivity may provide more information for determining form factors experimentally. We present the $q^2$ dependence of the form-factor sensitivity in Fig.~\ref{fig:ff_sens_1405} for $\Lambda_c^+\to\Lambda^*_{1405}$ and in Fig.~\ref{fig:ff_sens_1520} for $\Lambda_c^+\to\Lambda^*_{1520}$, respectively.

As shown in Fig.~\ref{fig:ff_sens_1405}, the low-$q^2$ region is particularly important for determining the $\Lambda_c^+\to\Lambda^*_{1405}$ form factors associated with $L_{1(1)}$, $L_{1(2)}$, $L_{4(1)}$, and $L_{4(2)}$. For the other coefficients, the dominant contributions mainly come from the intermediate-$q^2$ region, approximately $q^2\in[0.1,0.6]~{\rm GeV}^2$.

For the $\Lambda_c^+\to\Lambda^*_{1520}$ form factors, the situation becomes more complicated. For $f_\perp$, $g_\perp$, $f_{\perp\prime}$, and $g_{\perp\prime}$, the intermediate-$q^2$ region, roughly $q^2\in[0.1,0.6]~{\rm GeV}^2$, is important for all coefficients. For $f_+$ and $g_+$, the low-$q^2$ region gives the dominant contributions to the coefficients $L_{1(1)}$, $L_{1(2)}$, $L_{1(3)}$, $L_{4(1)}$, $L_{4(2)}$, and $L_{4(3)}$. For the remaining coefficients, the intermediate-$q^2$ region is important for determining $f_+$ and $g_+$.


\subsection{Implications of strong phases in form factors}

  From the discussion above, precise measurements of certain angular coefficients can provide access to the strong-phase differences between some form factors. A nonzero strong-phase difference, together with a nonzero weak-phase difference, is a necessary ingredient for generating direct CP violation~\cite{Wang:2024qff}. Although CP violation is not expected in these semileptonic decays, the strong phases extracted from them can still provide useful input for studies of CP-violating processes.
In this section, we analyze the strong phase differences between different form factors using these angular coefficients. When sufficient experimental data are available, these strong phase differences can be measured experimentally. 

The first set of angular coefficients that can reflect the strong phase differences consists of $L_{8(2)}$ and $L_{11(2)}$. Through an analytical calculation, the expressions of the angular coefficients $L_{8(2)}$ and $L_{11(2)}$ in terms of the form factors, after neglecting the lepton masses, are given as
\begin{align}
	L_{8(2)}=&
	-2\sqrt{s_+s_-q^2}\mathcal{A}_{\frac{3}{2}}^2|L_{\Lambda^*_{1520}}|^2
	\Big(
	(m_{\Lambda_c}-m_{\Lambda^*})|f_{\perp^\prime}^\frac{3}{2}||g_+^\frac{3}{2}|\cos(\delta^{f^\frac{3}{2}_{\perp^\prime}}-\delta^{g^\frac{3}{2}_+})-
	\nonumber\\
	&(m_{\Lambda_c}+m_{\Lambda^*})
	|f_+^\frac{3}{2}||g_{\perp^\prime}^\frac{3}{2}|\cos(\delta^{f_+^\frac{3}{2}}-\delta^{g_{\perp^\prime}^\frac{3}{2}})
	\Big),
	\nonumber \\
	L_{11(2)}=&-\sqrt{s_+s_-q^2}\mathcal{A}_{\frac{3}{2}}^2|L_{\Lambda^*_{1520}}|^2
	\Big(
	(m_{\Lambda_c}+m_{\Lambda^*})|f^\frac{3}{2}_+||g^\frac{3}{2}_{\perp^\prime}|\sin(\delta^{f^\frac{3}{2}_+}-\delta^{g^\frac{3}{2}_{\perp^\prime}}) \nonumber \\
	&+(m_{\Lambda_c}-m_{\Lambda^*})|f^\frac{3}{2}_{\perp^\prime}||g^\frac{3}{2}_+|\sin(\delta^{f^\frac{3}{2}_{\perp^\prime}}-\delta^{g^\frac{3}{2}_+})
	\Big).
\end{align}
It is straightforward to see from the above two expressions that, once the two angular coefficients and the magnitudes of the corresponding form factors are precisely measured, the strong phase differences $\delta^{f_{\perp^\prime}}-\delta^{g_+}$ and $\delta^{f_+}-\delta^{g_{\perp^\prime}}$ can be extracted. Similarly, through an analytical calculation of the angular coefficients $L_{9(2)}$ and $L_{10(2)}$, one obtains
\begin{align}
	L_{9(2)}=&
	2\sqrt{q^2}\mathcal{A}_{\frac{3}{2}}^2|L_{\Lambda^*_{1520}}|^2
	\Big(
	s_-(m_{\Lambda_c}-m_{\Lambda^*})|g^\frac{3}{2}_+||g^\frac{3}{2}_{\perp^\prime}|\sin(\delta^{g^\frac{3}{2}_+}-\delta^{g^\frac{3}{2}_{\perp^\prime}})-
	\nonumber\\
	&s_+(m_{\Lambda_c}+m_{\Lambda^*})
	|f^\frac{3}{2}_+||f^\frac{3}{2}_{\perp^\prime}|\sin(\delta^{f^\frac{3}{2}_+}-\delta^{f^\frac{3}{2}_{\perp^\prime}})
	\Big),
	\nonumber \\
	L_{10(2)}=&
	\sqrt{q^2}\mathcal{A}_{\frac{3}{2}}^2|L_{\Lambda^*_{1520}}|^2
	\Big(
	s_-(m_{\Lambda_c}-m_{\Lambda^*})|g^\frac{3}{2}_+||g^\frac{3}{2}_{\perp^\prime}|\cos(\delta^{g^\frac{3}{2}_+}-\delta^{g^\frac{3}{2}_{\perp^\prime}})-
	\nonumber\\
	&s_+(m_{\Lambda_c}+m_{\Lambda^*})
	|f^\frac{3}{2}_+||f^\frac{3}{2}_{\perp^\prime}|\cos(\delta^{f^\frac{3}{2}_+}-\delta^{f^\frac{3}{2}_{\perp^\prime}})
	\Big).
\end{align}
Using the above two sets of coefficients, one can determine the strong phase differences $\delta^{g_+}-\delta^{g_{\perp^\prime}}$ and $\delta^{f_+}-\delta^{f_{\perp^\prime}}$. In addition, one can see that the coefficients $L_{9(2)}$ and $L_{11(2)}$ are sensitive to the strong phase: they
  vanish when the strong phase is zero. Therefore, measurements of these coefficients can provide a direct test of the presence of a strong phase.

Meanwhile, we also find that the parts of the angular coefficients $L_{2(1)}$ and $L_{2(3)}$ that are proportional to the lepton mass terms can also provide information on the strong phase. These two parts can be expressed in terms of the form factors as follows,
\begin{align}
	L_{2(1)}^{m_\ell}=& -\frac{m_\ell^2(m_{\Lambda_c}^2-m_{\Lambda^*}^2)\sqrt{s_+s_-}}{q^2}
	\Big[\frac{5}{3}\mathcal{A}_{\frac{3}{2}}^2|L_{\Lambda^*_{1520}}|^2
	\Big(
	|f^\frac{3}{2}_0||f^\frac{3}{2}_+|\cos(\delta^{f^\frac{3}{2}_0}-\delta^{f^\frac{3}{2}_+})-
	\nonumber\\
	&
	|g^\frac{3}{2}_0||g^\frac{3}{2}_+|\cos(\delta^{g^\frac{3}{2}_0}-\delta^{g^\frac{3}{2}_+}) 
	\Big) 
	+16\mathcal{A}_{\frac{1}{2}}^2|L_{\Lambda^*_{1405}}|^2
	\Big(
	|f^\frac{1}{2}_0||f^\frac{1}{2}_+|\cos(\delta^{f^\frac{1}{2}_0}-\delta^{f^\frac{1}{2}_+})-
	\nonumber\\
	&
	|g^\frac{1}{2}_0||g^\frac{1}{2}_+|\cos(\delta^{g^\frac{1}{2}_0}-\delta^{g^\frac{1}{2}_+}) 
	\Big) 
	\Big],
	\nonumber\\
	L_{2(3)}^{m_\ell}=&-\frac{m_\ell^2(m_{\Lambda_c}^2-m_{\Lambda^*}^2)\sqrt{s_+s_-}}{q^2}\mathcal{A}_{\frac{3}{2}}^2|L_{\Lambda^*_{1520}}|^2
	\Big(
	|f^\frac{3}{2}_0||f^\frac{3}{2}_+|\cos(\delta^{f^\frac{3}{2}_0}-\delta^{f^\frac{3}{2}_+})-
	\nonumber\\
	&
	|g^\frac{3}{2}_0||g^\frac{3}{2}_+|\cos(\delta^{g^\frac{3}{2}_0}-\delta^{g^\frac{3}{2}_+}) 
	\Big).
\end{align}
In the above expression, we use the superscript $m_{\ell}$ to indicate the part of the angular coefficients that involves the lepton mass. These strong phase differences cannot be extracted directly from the expressions as in the above two cases, but they may be obtained through a global experimental fit. Experimentally, one can first measure the values of these angular coefficients using the electron decay channel. Since the electron is very light, the contribution proportional to the square of the lepton mass is almost negligible, effectively providing information on the squared-modulus terms of the helicity amplitudes. Then, by using the $\tau$ decay channel and comparing it with the $e$ channel, one can obtain the magnitudes of those interference terms and thereby fit the strong phase differences.
\section{Summary and outlook}
\label{sec:conclusion}

In this work, we have studied the semileptonic cascade decay
$\Lambda_c^+\to\Lambda^*_{1405/1520}(\to\Sigma\pi)\ell^+\nu_\ell$ within the
helicity-amplitude formalism. The four-body phase space was described by
$q^2$, $m_{\Sigma\pi}^2$, and the three helicity angles
$(\theta_\Lambda,\theta_\ell,\phi)$. For the intermediate resonances, we used a
Flatte-type lineshape for $\Lambda^*_{1405}$ and a Breit-Wigner form for
$\Lambda^*_{1520}$. The spin-$1/2$ and spin-$3/2$ resonance contributions, as
well as their interference terms, were included coherently. We derived the full
angular distribution and expanded it in terms of angular coefficients
$L_{i(j)}$. The complete expressions of these coefficients were collected in
the appendix, and simplified formulae were also given for the case of real
helicity amplitudes and negligible lepton mass.

We analyzed the invariant-mass spectrum and the one-dimensional angular
distributions in $\cos\theta_\Lambda$, $\cos\theta_\ell$, and $\phi$. The
$\theta_\Lambda$ distribution is useful for resolving the resonance structure:
$L_\Lambda$ receives contributions from both resonances, $L_{\Lambda 2c}$ is
generated by the $\Lambda^*_{1520}$ contribution, and $L_{\Lambda c}$ reflects
the interference between $\Lambda^*_{1405}$ and $\Lambda^*_{1520}$. The
$\theta_\ell$ distribution probes the interference between vector and
axial-vector helicity amplitudes. The $\phi$ distribution is especially
sensitive to interference effects, since the coefficients
$L_{\phi c}$, $L_{\phi 2c}$, $L_{\phi s}$, and $L_{\phi 2s}$ probe the real and
imaginary parts of the relevant interference terms.

To make these interference effects experimentally accessible, we constructed
several normalized angular asymmetries. For the forward-backward asymmetries,
we obtained
$\langle A^\Lambda_{FB}\rangle=-0.079$ and
$\langle A^\ell_{FB}\rangle=-0.122$. In the real-form-factor limit, the zero
points of $A^\Lambda_{FB}$ are tied to the central masses of the intermediate
resonances through the real part of the $\Lambda^*_{1405}$--$\Lambda^*_{1520}$
interference. A nonzero strong phase in the
$\Lambda_c^+\to\Lambda^*\ell^+\nu_\ell$ amplitudes can shift these zero points, making
the $m_{\Sigma\pi}$ dependence of $A^\Lambda_{FB}$ a useful probe of strong
phase effects. We also introduced the $A_{\pi/3}$ asymmetries, finding
$\langle A^\Lambda_{\pi/3}\rangle=0.063$ and
$\langle A^\ell_{\pi/3}\rangle=-0.252$, as well as several $\phi$-dependent
asymmetries that directly project out the coefficients associated with the
$\cos\phi$, $\sin\phi$, $\cos2\phi$, and $\sin2\phi$ terms.

We further investigated the sensitivity of the angular coefficients to the
transition form factors. Because of lepton-mass suppression, the scalar form
factors $f_0$ and $g_0$ give negligible contributions in the electron mode. For
the $\Lambda_c^+\to\Lambda^*_{1405}\ell^+\nu_\ell$ transition, the coefficients
$L_{8(1)}$ and $L_{11(1)}$ provide strong constraints on $g_+$ and $f_+$,
whereas $g_\perp$ is mainly constrained by $L_{2(1)}$, $L_{8(1)}$, and
$L_{11(1)}$, and $f_\perp$ by $L_{2(1)}$ and $L_{2(2)}$. For the
$\Lambda_c^+\to\Lambda^*_{1520}\ell^+\nu_\ell$ transition, $f_+$ and $g_+$ are mainly
constrained by $L_{8(1)}$, $f_\perp$ by $L_{2(3)}$ and $L_{3(1)}$,
$g_\perp$ by $L_{2(2)}$, $L_{2(3)}$, $L_{8(1)}$, and $L_{11(1)}$, while
$f_{\perp\prime}$ and $g_{\perp\prime}$ are primarily constrained by
$L_{2(3)}$. The $q^2$-differential sensitivity shows that, for
$\Lambda^*_{1405}$, the low-$q^2$ region is important for
$L_{1(1)}$, $L_{1(2)}$, $L_{4(1)}$, and $L_{4(2)}$, whereas the other
coefficients receive their dominant sensitivity from the intermediate-$q^2$
region. For $\Lambda^*_{1520}$, the intermediate-$q^2$ region is important for
$f_\perp$, $g_\perp$, $f_{\perp\prime}$, and $g_{\perp\prime}$, while the
low-$q^2$ region gives the dominant contribution to $f_+$ and $g_+$ for the
coefficients $L_{1(1)}$, $L_{1(2)}$, $L_{1(3)}$, $L_{4(1)}$, $L_{4(2)}$, and
$L_{4(3)}$.

Finally, we discussed how strong phases in the form factors can be probed
through angular coefficients. The coefficients $L_{8(2)}$ and $L_{11(2)}$
contain the phase differences
$\delta^{f_{\perp\prime}}-\delta^{g_+}$ and
$\delta^{f_+}-\delta^{g_{\perp\prime}}$, while $L_{9(2)}$ and $L_{10(2)}$
are sensitive to
$\delta^{g_+}-\delta^{g_{\perp\prime}}$ and
$\delta^{f_+}-\delta^{f_{\perp\prime}}$. In particular, $L_{9(2)}$ and
$L_{11(2)}$ vanish when the strong phase is zero, so measurements of these
coefficients would provide a direct test of the presence of strong phases. We
also found that the lepton-mass-dependent parts of $L_{2(1)}$ and $L_{2(3)}$
contain additional strong-phase information, which may be accessed through a
combined analysis of light-lepton and $\tau$ modes.

The results of this work provide a systematic angular-analysis framework for
future studies of $\Lambda_c^+\to\Sigma\pi\ell^+\nu_\ell$. A full
multi-dimensional fit to the variables
$(q^2,m_{\Sigma\pi}^2,\theta_\Lambda,\theta_\ell,\phi)$ would allow the
angular coefficients to be measured directly and would give complementary
constraints on the $\Lambda_c^+\to\Lambda^*_{1405}$ and
$\Lambda_c^+\to\Lambda^*_{1520}$ form factors. With improved experimental
statistics, especially from BESIII, Belle~II, and LHCb, these observables can
be used to test the resonance composition of the $\Sigma\pi$ spectrum, refine
the form-factor inputs, and search for strong-phase effects in charmed-baryon
semileptonic decays.

\section*{Acknowledgments}
 We thank Professor Xiao-Rui Lyu for useful discussion. The work of Ruilin Zhu is supported by NSFC under grant No. 12322503 and No. 12075124, and by Natural Science Foundation of Jiangsu under Grant No. BK20211267. The work of Zhi-Peng Xing is supported by NSFC under grant No.12375088 ,No. 12335003 and No. 12405113. This work is supported in part by National Key R$\&$D Program of China under Contracts Nos. 2025YFA1613900.

\appendix

\section{Helicity amplitudes}
\label{sec:app_helicity_amplitudes}

The hadronic helicity amplitudes are evaluated from
$H_{\lambda}=H_V(\lambda)-H_A(\lambda)$, where $H_V$ and $H_A$ denote the
vector and axial-vector contributions. In the following formula, the three quantum numbers inside the parentheses of the hadronic helicity amplitude denote the helicity quantum numbers of $\Lambda_c^+$, $\Lambda^*$ and $W$, respectively. For the
$\Lambda_c^+(\frac{1}{2}^+) \to \Lambda^*(\frac{1}{2}^-)$ transition, using the
form factors of Eq.~(\ref{eq:ff_12}), one obtains
\begin{eqnarray}
	H_V(\frac{1}{2},\frac{1}{2},0) &=& -H_V(-\frac{1}{2},-\frac{1}{2},0)
	= \sqrt{\frac{s_+}{q^2}}\frac{g_1m_{\Lambda_c}(m_{\Lambda_c}-m_{\Lambda^*})+g_2q^2}{m_{\Lambda_c}},
	\notag \\
	H_V(\frac{1}{2},-\frac{1}{2},-1) &=& -H_V(-\frac{1}{2},\frac{1}{2},1)
	= -\sqrt{2s_+}\frac{g_1m_{\Lambda_c}+g_2(m_{\Lambda_c}-m_{\Lambda^*})}{m_{\Lambda_c}},
	\notag \\
	H_V(\frac{1}{2},\frac{1}{2},t) &=& -H_V(-\frac{1}{2},-\frac{1}{2},t)
	=\sqrt{\frac{s_-}{q^2}}\frac{g_1m_{\Lambda_c}(m_{\Lambda_c}+m_{\Lambda^*})-g_3q^2}{m_{\Lambda_c}},
	\notag \\
	H_A(\frac{1}{2},\frac{1}{2},0) &=& H_A(-\frac{1}{2},-\frac{1}{2},0)
	= \sqrt{\frac{s_-}{q^2}}\frac{f_1m_{\Lambda_c}(m_{\Lambda_c}+m_{\Lambda^*})-f_2q^2}{m_{\Lambda_c}},
	\notag \\
	H_A(\frac{1}{2},-\frac{1}{2},-1) &=& H_A(-\frac{1}{2},\frac{1}{2},1)
	= \sqrt{2s_-}\frac{f_1m_{\Lambda_c}-f_2(m_{\Lambda_c}+m_{\Lambda^*})}{m_{\Lambda_c}},
	\notag \\
	H_A(\frac{1}{2},\frac{1}{2},t) &=& H_A(-\frac{1}{2},-\frac{1}{2},t)
	= \sqrt{\frac{s_+}{q^2}}\frac{f_1m_{\Lambda_c}(m_{\Lambda_c}-m_{\Lambda^*})+f_3q^2}{m_{\Lambda_c}}.
	\label{eq:HV_HA_12}
\end{eqnarray}
Equivalently, using the helicity-basis form factors of Eq.~(\ref{eq:helicity_ff_12}), these amplitudes take the compact form
\begin{eqnarray}
	H_V(\frac{1}{2},\frac{1}{2},0) &=&
	-H_V(-\frac{1}{2},-\frac{1}{2},0)
	= \sqrt{\frac{s_+}{q^2}}g_+(m_{\Lambda^*}-m_{\Lambda_c}),
	\notag \\
	H_V(\frac{1}{2},-\frac{1}{2},-1) &=&
	-H_V(-\frac{1}{2},\frac{1}{2},1)
	= \sqrt{2s_+}g_{\perp},
	\notag \\
	H_V(\frac{1}{2},\frac{1}{2},t) &=&
	-H_V(-\frac{1}{2},-\frac{1}{2},t)
	= -\sqrt{\frac{s_-}{q^2}}g_0(m_{\Lambda^*}+m_{\Lambda_c}),
	\notag \\
	H_A(\frac{1}{2},\frac{1}{2},0) &=&
	H_A(-\frac{1}{2},-\frac{1}{2},0)
	= -\sqrt{\frac{s_-}{q^2}}f_+(m_{\Lambda^*}+m_{\Lambda_c}),
	\notag \\
	H_A(\frac{1}{2},-\frac{1}{2},-1) &=&
	H_A(-\frac{1}{2},\frac{1}{2},1)
	= -\sqrt{2s_-}f_{\perp},
	\label{eq:HV_HA_12_helicity} \\
	H_A(\frac{1}{2},\frac{1}{2},t) &=&
	H_A(-\frac{1}{2},-\frac{1}{2},t)
	= \sqrt{\frac{s_+}{q^2}}f_0(m_{\Lambda^*}-m_{\Lambda_c}).
	\notag
\end{eqnarray}
For the $\Lambda_c^+(\frac{1}{2}^+) \to \Lambda^*(\frac{3}{2}^-)$ transition, the helicity amplitudes are
\begin{eqnarray}
	H_V(\frac{1}{2},\frac{3}{2},1) &=&
	H_V(-\frac{1}{2},-\frac{3}{2},-1)
	= -f_{\perp'}\sqrt{s_+},
	\notag \\
	H_V(\frac{1}{2},\frac{1}{2},0) &=&
	H_V(-\frac{1}{2},-\frac{1}{2},0)
	= \sqrt{\frac{s_+}{6q^2}}f_+(m_{\Lambda_c}+m_{\Lambda^*}),
	\notag \\
	H_V(\frac{1}{2},\frac{1}{2},t) &=&
	H_V(-\frac{1}{2},-\frac{1}{2},t)
	= \sqrt{\frac{s_-}{6q^2}}f_0(m_{\Lambda_c}-m_{\Lambda^*}),
	\notag \\
	H_V(\frac{1}{2},-\frac{1}{2},-1) &=&
	H_V(-\frac{1}{2},\frac{1}{2},1)
	= \frac{f_{\perp}\sqrt{s_+}}{\sqrt{3}},
	\notag \\
	H_A(\frac{1}{2},\frac{3}{2},1) &=&
	-H_A(-\frac{1}{2},-\frac{3}{2},-1)
	= g_{\perp'}\sqrt{s_-},
	\notag \\
	H_A(\frac{1}{2},\frac{1}{2},0) &=&
	-H_A(-\frac{1}{2},-\frac{1}{2},0)
	= \sqrt{\frac{s_-}{6q^2}}g_+(m_{\Lambda_c}-m_{\Lambda^*}),
	\notag \\
	H_A(\frac{1}{2},\frac{1}{2},t) &=&
	-H_A(-\frac{1}{2},-\frac{1}{2},t)
	= \sqrt{\frac{s_+}{6q^2}}g_0(m_{\Lambda_c}+m_{\Lambda^*}),
	\notag \\
	H_A(\frac{1}{2},-\frac{1}{2},-1) &=&
	-H_A(-\frac{1}{2},\frac{1}{2},1)
	= -\frac{g_{\perp}\sqrt{s_-}}{\sqrt{3}}.
	\label{eq:HV_HA_32}
\end{eqnarray}
The lepton helicity amplitudes in the dilepton rest frame are
\begin{eqnarray}
	L(\lambda=-1,s_\ell=\frac{1}{2}) &=& i\sqrt{2q^2}\,\beta\,(\cos\theta_\ell-1)\,e^{i\phi},
	L(\lambda=0,s_\ell=\frac{1}{2}) = i\,2\sqrt{q^2}\,\beta\,\sin\theta_\ell, \notag \\
	L(\lambda=1,s_\ell=\frac{1}{2}) &=& -i\sqrt{2q^2}\,\beta\,(\cos\theta_\ell+1)\,e^{-i\phi}, 
	L(\lambda=-1,s_\ell=-\frac{1}{2}) = i\sqrt{2}\,m_\ell\,\beta\,\sin\theta_\ell\,e^{i\phi}, \notag \\
	L(\lambda=0,s_\ell=-\frac{1}{2}) &=& -i\,2\,m_\ell\,\beta\,\cos\theta_\ell, 
	L(\lambda=1,s_\ell=-\frac{1}{2}) = -i\sqrt{2}\,m_\ell\,\beta\,\sin\theta_\ell\,e^{-i\phi}, \notag \\
	L(\lambda=t,s_\ell=-\frac{1}{2}) &=& -i\,2\,m_\ell\,\beta.
	\label{eq:lepton_amplitude}
\end{eqnarray}

\section{\texorpdfstring{Complete $L_{i(j)}$ angular coefficients}{Complete Lij angular coefficients}}
\label{sec:app_Lij}

In this appendix we list the full expressions for the angular coefficients $L_{i(j)}$ appearing in the $\cos\theta_\Lambda$ expansion, Eq.~(\ref{eq:angular_expansion}). These are expressed in terms of the helicity amplitudes $\mathcal{H}^{J}_{s_{\Lambda_c},\lambda}$ defined in Eq.~(\ref{eq:A_def}).

\begin{eqnarray}
	L_{1(1)} &=&
	2\big(q^2+m_\ell^2\big)
	\big(
	|\mathcal{H}^{\frac{1}{2}}_{-\frac{1}{2},0}|^2
	+|\mathcal{H}^{\frac{1}{2}}_{\frac{1}{2},0}|^2
	+\frac{5}{8}|\mathcal{H}^{\frac{3}{2}}_{-\frac{1}{2},0}|^2
	+\frac{5}{8}|\mathcal{H}^{\frac{3}{2}}_{\frac{1}{2},0}|^2
	\big)\notag \\
	&&
	+4m_\ell^2
	\big(
	|\mathcal{H}^{\frac{1}{2}}_{-\frac{1}{2},t}|^2
	+|\mathcal{H}^{\frac{1}{2}}_{\frac{1}{2},t}|^2
	+\frac{5}{8}|\mathcal{H}^{\frac{3}{2}}_{-\frac{1}{2},t}|^2
	+\frac{5}{8}|\mathcal{H}^{\frac{3}{2}}_{\frac{1}{2},t}|^2
	\big), \notag \\
	&&
	+\big(3q^2+m_\ell^2\big)
	\big(
	|\mathcal{H}^{\frac{1}{2}}_{\frac{1}{2},-1}|^2
	+|\mathcal{H}^{\frac{1}{2}}_{-\frac{1}{2},1}|^2
	+\frac{3}{8}|\mathcal{H}^{\frac{3}{2}}_{\frac{1}{2},1}|^2
	+\frac{3}{8}|\mathcal{H}^{\frac{3}{2}}_{-\frac{1}{2},-1}|^2
	\notag \\
	&&+\frac{5}{8}|\mathcal{H}^{\frac{3}{2}}_{-\frac{1}{2},1}|^2
	+\frac{5}{8}|\mathcal{H}^{\frac{3}{2}}_{\frac{1}{2},-1}|^2
	\big)
	\notag \\
	L_{1(2)} &=&
	4\big(q^2+m_\ell^2\big)
	\mathrm{Re}\big(
	\mathcal{H}^{\frac{1}{2}}_{-\frac{1}{2},0}
	\mathcal{H}^{\frac{3}{2}*}_{-\frac{1}{2},0}
	+\mathcal{H}^{\frac{1}{2}}_{\frac{1}{2},0}
	\mathcal{H}^{\frac{3}{2}*}_{\frac{1}{2},0}
	\big)+8m_\ell^2
	\mathrm{Re}\big(
	\mathcal{H}^{\frac{1}{2}}_{-\frac{1}{2},t}
	\mathcal{H}^{\frac{3}{2}*}_{-\frac{1}{2},t}
	\notag \\
	&& +\mathcal{H}^{\frac{1}{2}}_{\frac{1}{2},t}
	\mathcal{H}^{\frac{3}{2}*}_{\frac{1}{2},t}
	\big), 
	+2\big(3q^2+m_\ell^2\big)
	\mathrm{Re}\big(
	\mathcal{H}^{\frac{1}{2}}_{-\frac{1}{2},1}
	\mathcal{H}^{\frac{3}{2}*}_{-\frac{1}{2},1}
	+\mathcal{H}^{\frac{1}{2}}_{\frac{1}{2},-1}
	\mathcal{H}^{\frac{3}{2}*}_{\frac{1}{2},-1}
	\big)
	\notag \\
	L_{1(3)} &=&
	\frac{3}{4}\big(q^2+m_\ell^2\big)
	\big(
	|\mathcal{H}^{\frac{3}{2}}_{-\frac{1}{2},0}|^2
	+|\mathcal{H}^{\frac{3}{2}}_{\frac{1}{2},0}|^2
	\big) +\frac{3}{2}m_\ell^2
	\big(
	|\mathcal{H}^{\frac{3}{2}}_{-\frac{1}{2},t}|^2
	+|\mathcal{H}^{\frac{3}{2}}_{\frac{1}{2},t}|^2
	\big) \notag\\
	&&
	+\frac{3}{8}\big(3q^2+m_\ell^2\big)
	\big(
	|\mathcal{H}^{\frac{3}{2}}_{-\frac{1}{2},1}|^2
	-|\mathcal{H}^{\frac{3}{2}}_{\frac{1}{2},1}|^2
	+|\mathcal{H}^{\frac{3}{2}}_{\frac{1}{2},-1}|^2
	-|\mathcal{H}^{\frac{3}{2}}_{-\frac{1}{2},-1}|^2
	\big)\notag\\
	L_{2(1)} &=&
	4q^2
	\big(
	|\mathcal{H}^{\frac{1}{2}}_{-\frac{1}{2},1}|^2
	-|\mathcal{H}^{\frac{1}{2}}_{\frac{1}{2},-1}|^2
	+\frac{5}{8}|\mathcal{H}^{\frac{3}{2}}_{-\frac{1}{2},1}|^2
	-\frac{5}{8}|\mathcal{H}^{\frac{3}{2}}_{\frac{1}{2},-1}|^2
	+\frac{3}{8}|\mathcal{H}^{\frac{3}{2}}_{\frac{1}{2},1}|^2
	-\frac{3}{8}|\mathcal{H}^{\frac{3}{2}}_{-\frac{1}{2},-1}|^2
	\big)
	\notag \\
	&&
	-8m_\ell^2\,\mathrm{Re}
	\big(
	\mathcal{H}^{\frac{1}{2}}_{-\frac{1}{2},t}
	\mathcal{H}^{\frac{1}{2}*}_{-\frac{1}{2},0}
	+\mathcal{H}^{\frac{1}{2}}_{\frac{1}{2},t}
	\mathcal{H}^{\frac{1}{2}*}_{\frac{1}{2},0}
	+\frac{5}{8}\mathcal{H}^{\frac{3}{2}}_{-\frac{1}{2},t}
	\mathcal{H}^{\frac{3}{2}*}_{-\frac{1}{2},0}
	+\frac{5}{8}\mathcal{H}^{\frac{3}{2}}_{\frac{1}{2},t}
	\mathcal{H}^{\frac{3}{2}*}_{\frac{1}{2},0}
	\big), \notag \\
	L_{2(2)} &=&
	-8m_\ell^2\,\mathrm{Re}
	\big(
	\mathcal{H}^{\frac{1}{2}}_{\frac{1}{2},t}
	\mathcal{H}^{\frac{3}{2}*}_{\frac{1}{2},0}
	+\mathcal{H}^{\frac{1}{2}}_{-\frac{1}{2},t}
	\mathcal{H}^{\frac{3}{2}*}_{-\frac{1}{2},0}
	+\mathcal{H}^{\frac{3}{2}}_{\frac{1}{2},t}
	\mathcal{H}^{\frac{1}{2}*}_{\frac{1}{2},0}
	+\mathcal{H}^{\frac{3}{2}}_{-\frac{1}{2},t}
	\mathcal{H}^{\frac{1}{2}*}_{-\frac{1}{2},0}
	\big), \notag \\
	&&+ 8q^2\,\mathrm{Re}
	\big(
	\mathcal{H}^{\frac{1}{2}}_{-\frac{1}{2},1}
	\mathcal{H}^{\frac{3}{2}*}_{-\frac{1}{2},1}
	-\mathcal{H}^{\frac{1}{2}}_{\frac{1}{2},-1}
	\mathcal{H}^{\frac{3}{2}*}_{\frac{1}{2},-1}
	\big)
	\notag \\
	L_{2(3)} &=&
	\frac{3}{2}q^2
	\big(
	|\mathcal{H}^{\frac{3}{2}}_{-\frac{1}{2},1}|^2
	-|\mathcal{H}^{\frac{3}{2}}_{\frac{1}{2},1}|^2
	+|\mathcal{H}^{\frac{3}{2}}_{-\frac{1}{2},-1}|^2
	-|\mathcal{H}^{\frac{3}{2}}_{\frac{1}{2},-1}|^2
	\big)
	-3m_\ell^2\,\mathrm{Re}
	\big(
	\mathcal{H}^{\frac{3}{2}}_{-\frac{1}{2},t}
	\mathcal{H}^{\frac{3}{2}*}_{-\frac{1}{2},0}
	\notag \\
	&&+\mathcal{H}^{\frac{3}{2}}_{\frac{1}{2},t}
	\mathcal{H}^{\frac{3}{2}*}_{\frac{1}{2},0}
	\big) \notag\\
	L_{3(1)} &=&
	-\frac{\sqrt{3}}{4}\big(q^2-m_\ell^2\big)\,\mathrm{Re}
	\big(
	\mathcal{H}^{\frac{3}{2}}_{-\frac{1}{2},-1}
	\mathcal{H}^{\frac{3}{2}*}_{-\frac{1}{2},1}
	+\mathcal{H}^{\frac{3}{2}}_{\frac{1}{2},-1}
	\mathcal{H}^{\frac{3}{2}*}_{\frac{1}{2},1}
	\big), \notag \\
	L_{3(2)} &=& L_{5(1)} = -L_{3(1)} = -L_{5(2)}\notag\\
	L_{4(1)} &=&
	\big(q^2-m_\ell^2\big)
	\big(
	|\mathcal{H}^{\frac{1}{2}}_{\frac{1}{2},-1}|^2
	+|\mathcal{H}^{\frac{1}{2}}_{-\frac{1}{2},1}|^2
	+\frac{5}{8}|\mathcal{H}^{\frac{3}{2}}_{-\frac{1}{2},1}|^2
	+\frac{5}{8}|\mathcal{H}^{\frac{3}{2}}_{\frac{1}{2},-1}|^2
	+\frac{3}{8}|\mathcal{H}^{\frac{3}{2}}_{-\frac{1}{2},-1}|^2
	+\frac{3}{8}|\mathcal{H}^{\frac{3}{2}}_{\frac{1}{2},1}|^2
	\big)
	\notag \\
	&&\quad
	-2\big(q^2-m_\ell^2\big)\big(
	|\mathcal{H}^{\frac{1}{2}}_{-\frac{1}{2},0}|^2
	+|\mathcal{H}^{\frac{1}{2}}_{\frac{1}{2},0}|^2
	+\frac{5}{8}|\mathcal{H}^{\frac{3}{2}}_{-\frac{1}{2},0}|^2
	+\frac{5}{8}|\mathcal{H}^{\frac{3}{2}}_{\frac{1}{2},0}|^2
	\big), \notag \\[6pt]
	L_{4(2)} &=&
	2\big(q^2-m_\ell^2\big)\,\mathrm{Re}
	\big(
	\mathcal{H}^{\frac{1}{2}}_{-\frac{1}{2},1}
	\mathcal{H}^{\frac{3}{2}*}_{-\frac{1}{2},1}
	+\mathcal{H}^{\frac{1}{2}}_{\frac{1}{2},-1}
	\mathcal{H}^{\frac{3}{2}*}_{\frac{1}{2},-1}
	\big)\notag\\
	&&
	-4\big(q^2-m_\ell^2\big)\,\mathrm{Re}
	\big(
	\mathcal{H}^{\frac{1}{2}}_{-\frac{1}{2},0}
	\mathcal{H}^{\frac{3}{2}*}_{-\frac{1}{2},0}
	+\mathcal{H}^{\frac{1}{2}}_{\frac{1}{2},0}
	\mathcal{H}^{\frac{3}{2}*}_{\frac{1}{2},0}
	\big), \notag \\
	L_{4(3)} &=&
	\frac{3}{8}\big(q^2-m_\ell^2\big)
	\big(
	|\mathcal{H}^{\frac{3}{2}}_{-\frac{1}{2},1}|^2
	-|\mathcal{H}^{\frac{3}{2}}_{\frac{1}{2},1}|^2
	+|\mathcal{H}^{\frac{3}{2}}_{\frac{1}{2},-1}|^2
	-|\mathcal{H}^{\frac{3}{2}}_{-\frac{1}{2},-1}|^2
	\big)
	\notag \\
	&&
	-\frac{3}{4}\big(q^2-m_\ell^2\big)
	\big(
	|\mathcal{H}^{\frac{3}{2}}_{-\frac{1}{2},0}|^2
	+|\mathcal{H}^{\frac{3}{2}}_{\frac{1}{2},0}|^2
	\big) \notag\\
	L_{6(1)} &=& L_{7(2)} = -L_{6(2)}= -L_{7(1)} = -L_{5(1)}(\mathrm{Re}\rightarrow\mathrm{Im}), \notag \\[6pt]
	L_{8(1)} &=&-2\sqrt{2}q^2\,\mathrm{Re}
	\big(
	\mathcal{H}^{\frac{3}{2}}_{-\frac{1}{2},0}
	\mathcal{H}^{\frac{1}{2}*}_{-\frac{1}{2},1}
	+\mathcal{H}^{\frac{3}{2}}_{\frac{1}{2},-1}
	\mathcal{H}^{\frac{1}{2}*}_{\frac{1}{2},0}
	+\sqrt{3}\mathcal{H}^{\frac{3}{2}}_{-\frac{1}{2},-1}
	\mathcal{H}^{\frac{1}{2}*}_{-\frac{1}{2},0}
	\notag \\
	&&
	-\mathcal{H}^{\frac{1}{2}}_{-\frac{1}{2},0}
	\mathcal{H}^{\frac{3}{2}*}_{-\frac{1}{2},1}
	-\mathcal{H}^{\frac{1}{2}}_{\frac{1}{2},-1}
	\mathcal{H}^{\frac{3}{2}*}_{\frac{1}{2},0}
	-\sqrt{3}\mathcal{H}^{\frac{1}{2}}_{\frac{1}{2},0}
	\mathcal{H}^{\frac{3}{2}*}_{\frac{1}{2},1}
	\big)
	\notag \\
	&& +2\sqrt{2}m_\ell^2\,\mathrm{Re}
	\big(
	\mathcal{H}^{\frac{1}{2}}_{-\frac{1}{2},t}
	\mathcal{H}^{\frac{3}{2}*}_{-\frac{1}{2},1}
	-\mathcal{H}^{\frac{1}{2}}_{\frac{1}{2},-1}
	\mathcal{H}^{\frac{3}{2}*}_{\frac{1}{2},t}
	+\sqrt{3}\mathcal{H}^{\frac{1}{2}}_{\frac{1}{2},t}
	\mathcal{H}^{\frac{3}{2}*}_{\frac{1}{2},1}
	+\mathcal{H}^{\frac{3}{2}}_{\frac{1}{2},-1}
	\mathcal{H}^{\frac{1}{2}*}_{\frac{1}{2},t}
	\notag \\
	&&
	-\mathcal{H}^{\frac{3}{2}}_{-\frac{1}{2},t}
	\mathcal{H}^{\frac{1}{2}*}_{-\frac{1}{2},1}
	+\sqrt{3}\mathcal{H}^{\frac{3}{2}}_{-\frac{1}{2},-1}
	\mathcal{H}^{\frac{1}{2}*}_{-\frac{1}{2},t}
	\big) \notag\\
	L_{8(2)} &=&
	\sqrt{6}q^2\,\mathrm{Re}
	\big(
	\mathcal{H}^{\frac{3}{2}}_{\frac{1}{2},0}
	\mathcal{H}^{\frac{3}{2}*}_{\frac{1}{2},1}
	-\mathcal{H}^{\frac{3}{2}}_{-\frac{1}{2},-1}
	\mathcal{H}^{\frac{3}{2}*}_{-\frac{1}{2},0}
	\big)
	+\sqrt{6}m_\ell^2\,\mathrm{Re}
	\big(
	\mathcal{H}^{\frac{3}{2}}_{-\frac{1}{2},-1}
	\mathcal{H}^{\frac{3}{2}*}_{-\frac{1}{2},t}
	+\mathcal{H}^{\frac{3}{2}}_{\frac{1}{2},1}
	\mathcal{H}^{\frac{3}{2}*}_{\frac{1}{2},t}
	\big) \notag\\
	L_{9(1)} &=& -L_{8(1)}(\mathrm{Re}\rightarrow\mathrm{Im}), \;\;L_{9(2)} = -L_{8(2)}(\mathrm{Re}\rightarrow\mathrm{Im}) \notag\\
	L_{10(1)} &=&
	\sqrt{2}(q^2-m_\ell^2)\,\mathrm{Re}
	\big(
	\mathcal{H}^{\frac{1}{2}}_{-\frac{1}{2},0}
	\mathcal{H}^{\frac{3}{2}*}_{-\frac{1}{2},1}
	-\mathcal{H}^{\frac{1}{2}}_{\frac{1}{2},-1}
	\mathcal{H}^{\frac{3}{2}*}_{\frac{1}{2},0}
	+\sqrt{3}\mathcal{H}^{\frac{1}{2}}_{\frac{1}{2},0}
	\mathcal{H}^{\frac{3}{2}*}_{\frac{1}{2},1}
	\notag \\
	&&
	+\mathcal{H}^{\frac{3}{2}}_{\frac{1}{2},-1}
	\mathcal{H}^{\frac{1}{2}*}_{\frac{1}{2},0}
	-\mathcal{H}^{\frac{3}{2}}_{-\frac{1}{2},0}
	\mathcal{H}^{\frac{1}{2}*}_{-\frac{1}{2},1}
	+\sqrt{3}\mathcal{H}^{\frac{3}{2}}_{-\frac{1}{2},-1}
	\mathcal{H}^{\frac{1}{2}*}_{-\frac{1}{2},0}
	\big), \notag \\[6pt]
	L_{10(2)} &=&
	\frac{\sqrt{6}}{2}q^2\,\mathrm{Re}
	\big(
	\mathcal{H}^{\frac{3}{2}}_{-\frac{1}{2},-1}
	\mathcal{H}^{\frac{3}{2}*}_{-\frac{1}{2},0}
	+\mathcal{H}^{\frac{3}{2}}_{\frac{1}{2},0}
	\mathcal{H}^{\frac{3}{2}*}_{\frac{1}{2},1}
	\big)
	-\frac{\sqrt{6}}{2}m_\ell^2\,\mathrm{Re}
	\big(
	\mathcal{H}^{\frac{3}{2}}_{-\frac{1}{2},-1}
	\mathcal{H}^{\frac{3}{2}*}_{-\frac{1}{2},0}
	+\mathcal{H}^{\frac{3}{2}}_{\frac{1}{2},0}
	\mathcal{H}^{\frac{3}{2}*}_{\frac{1}{2},1}
	\big), \notag \\[6pt]
	L_{11(1)} &=& -L_{10(1)}(\mathrm{Re}\rightarrow\mathrm{Im}), \;\;L_{11(2)} = -L_{10(2)}(\mathrm{Re}\rightarrow\mathrm{Im}) \notag
\end{eqnarray}

	\section{Simplified angular coefficients with real amplitudes}
\label{sec:app_simple}

If the terms of order $m_\ell$ are neglected ($m_\ell\to 0$, $\beta\to 1$), the angular coefficients simplify to:

\begin{eqnarray}
	L_{1(1)} &=&
	2q^2
	\big(
	|\mathcal{H}^{\frac{1}{2}}_{-\frac{1}{2},0}|^2
	+|\mathcal{H}^{\frac{1}{2}}_{\frac{1}{2},0}|^2
	+\frac{5}{8}|\mathcal{H}^{\frac{3}{2}}_{-\frac{1}{2},0}|^2
	+\frac{5}{8}|\mathcal{H}^{\frac{3}{2}}_{\frac{1}{2},0}|^2
	\big)
	+3q^2
	\big(
	|\mathcal{H}^{\frac{1}{2}}_{\frac{1}{2},-1}|^2
	+|\mathcal{H}^{\frac{1}{2}}_{-\frac{1}{2},1}|^2
	\notag \\
	&&+\frac{3}{8}|\mathcal{H}^{\frac{3}{2}}_{\frac{1}{2},1}|^2
	+\frac{3}{8}|\mathcal{H}^{\frac{3}{2}}_{-\frac{1}{2},-1}|^2
	+\frac{5}{8}|\mathcal{H}^{\frac{3}{2}}_{-\frac{1}{2},1}|^2
	+\frac{5}{8}|\mathcal{H}^{\frac{3}{2}}_{\frac{1}{2},-1}|^2
	\big), \notag \\[4pt]
	L_{1(2)} &=&
	4q^2
	\mathrm{Re}\big(
	\mathcal{H}^{\frac{1}{2}}_{-\frac{1}{2},0}
	\mathcal{H}^{\frac{3}{2}*}_{-\frac{1}{2},0}
	+\mathcal{H}^{\frac{1}{2}}_{\frac{1}{2},0}
	\mathcal{H}^{\frac{3}{2}*}_{\frac{1}{2},0}
	\big)
	+6q^2
	\mathrm{Re}\big(
	\mathcal{H}^{\frac{1}{2}}_{-\frac{1}{2},1}
	\mathcal{H}^{\frac{3}{2}*}_{-\frac{1}{2},1}
	+\mathcal{H}^{\frac{1}{2}}_{\frac{1}{2},-1}
	\mathcal{H}^{\frac{3}{2}*}_{\frac{1}{2},-1}
	\big), \notag \\[4pt]
	L_{1(3)} &=&
	\frac{3}{4}q^2
	\big(
	|\mathcal{H}^{\frac{3}{2}}_{-\frac{1}{2},0}|^2
	+|\mathcal{H}^{\frac{3}{2}}_{\frac{1}{2},0}|^2
	\big)
	+\frac{9}{8}q^2
	\big(
	|\mathcal{H}^{\frac{3}{2}}_{-\frac{1}{2},1}|^2
	-|\mathcal{H}^{\frac{3}{2}}_{\frac{1}{2},1}|^2
	+|\mathcal{H}^{\frac{3}{2}}_{\frac{1}{2},-1}|^2
	-|\mathcal{H}^{\frac{3}{2}}_{-\frac{1}{2},-1}|^2
	\big) \notag\\
	L_{2(1)} &=&
	4q^2
	\big(
	|\mathcal{H}^{\frac{1}{2}}_{-\frac{1}{2},1}|^2
	-|\mathcal{H}^{\frac{1}{2}}_{\frac{1}{2},-1}|^2
	+\frac{5}{8}|\mathcal{H}^{\frac{3}{2}}_{-\frac{1}{2},1}|^2
	-\frac{5}{8}|\mathcal{H}^{\frac{3}{2}}_{\frac{1}{2},-1}|^2
	+\frac{3}{8}|\mathcal{H}^{\frac{3}{2}}_{\frac{1}{2},1}|^2
	-\frac{3}{8}|\mathcal{H}^{\frac{3}{2}}_{-\frac{1}{2},-1}|^2
	\big), \notag \\
	L_{2(2)} &=&
	8q^2\,\mathrm{Re}
	\big(
	\mathcal{H}^{\frac{1}{2}}_{-\frac{1}{2},1}
	\mathcal{H}^{\frac{3}{2}*}_{-\frac{1}{2},1}
	-\mathcal{H}^{\frac{1}{2}}_{\frac{1}{2},-1}
	\mathcal{H}^{\frac{3}{2}*}_{\frac{1}{2},-1}
	\big), \notag \\
	L_{2(3)} &=&
	\frac{3}{2}q^2
	\big(
	|\mathcal{H}^{\frac{3}{2}}_{-\frac{1}{2},1}|^2
	-|\mathcal{H}^{\frac{3}{2}}_{\frac{1}{2},1}|^2
	+|\mathcal{H}^{\frac{3}{2}}_{-\frac{1}{2},-1}|^2
	-|\mathcal{H}^{\frac{3}{2}}_{\frac{1}{2},-1}|^2
	\big), \notag \\
	L_{3(1)} &=&
	-\frac{\sqrt{3}}{4}q^2\,\mathrm{Re}
	\big(
	\mathcal{H}^{\frac{3}{2}}_{-\frac{1}{2},-1}
	\mathcal{H}^{\frac{3}{2}*}_{-\frac{1}{2},1}
	+\mathcal{H}^{\frac{3}{2}}_{\frac{1}{2},-1}
	\mathcal{H}^{\frac{3}{2}*}_{\frac{1}{2},1}
	\big)\notag\\
	L_{4(1)} &=&
	q^2
	\big(
	|\mathcal{H}^{\frac{1}{2}}_{\frac{1}{2},-1}|^2
	+|\mathcal{H}^{\frac{1}{2}}_{-\frac{1}{2},1}|^2
	+\frac{5}{8}|\mathcal{H}^{\frac{3}{2}}_{-\frac{1}{2},1}|^2
	+\frac{5}{8}|\mathcal{H}^{\frac{3}{2}}_{\frac{1}{2},-1}|^2
	+\frac{3}{8}|\mathcal{H}^{\frac{3}{2}}_{-\frac{1}{2},-1}|^2
	+\frac{3}{8}|\mathcal{H}^{\frac{3}{2}}_{\frac{1}{2},1}|^2
	\big)
	\notag \\
	&&
	-2q^2\big(
	|\mathcal{H}^{\frac{1}{2}}_{-\frac{1}{2},0}|^2
	+|\mathcal{H}^{\frac{1}{2}}_{\frac{1}{2},0}|^2
	+\frac{5}{8}|\mathcal{H}^{\frac{3}{2}}_{-\frac{1}{2},0}|^2
	+\frac{5}{8}|\mathcal{H}^{\frac{3}{2}}_{\frac{1}{2},0}|^2
	\big), \notag \\[4pt]
	L_{4(2)} &=&
	2q^2\,\mathrm{Re}
	\big(
	\mathcal{H}^{\frac{1}{2}}_{-\frac{1}{2},1}
	\mathcal{H}^{\frac{3}{2}*}_{-\frac{1}{2},1}
	+\mathcal{H}^{\frac{1}{2}}_{\frac{1}{2},-1}
	\mathcal{H}^{\frac{3}{2}*}_{\frac{1}{2},-1}
	\big)
	-4q^2\,\mathrm{Re}
	\big(
	\mathcal{H}^{\frac{1}{2}}_{-\frac{1}{2},0}
	\mathcal{H}^{\frac{3}{2}*}_{-\frac{1}{2},0}
	+\mathcal{H}^{\frac{1}{2}}_{\frac{1}{2},0}
	\mathcal{H}^{\frac{3}{2}*}_{\frac{1}{2},0}
	\big), \notag \\
	L_{4(3)} &=&
	\frac{3}{8}q^2
	\big(
	|\mathcal{H}^{\frac{3}{2}}_{-\frac{1}{2},1}|^2
	-|\mathcal{H}^{\frac{3}{2}}_{\frac{1}{2},1}|^2
	+|\mathcal{H}^{\frac{3}{2}}_{\frac{1}{2},-1}|^2
	-|\mathcal{H}^{\frac{3}{2}}_{-\frac{1}{2},-1}|^2
	\big)
	-\frac{3}{4}q^2
	\big(
	|\mathcal{H}^{\frac{3}{2}}_{-\frac{1}{2},0}|^2
	+|\mathcal{H}^{\frac{3}{2}}_{\frac{1}{2},0}|^2
	\big) \notag\\
	L_{8(1)} &=&
	-2\sqrt{2}q^2\,\mathrm{Re}
	\big(
	\mathcal{H}^{\frac{3}{2}}_{-\frac{1}{2},0}
	\mathcal{H}^{\frac{1}{2}*}_{-\frac{1}{2},1}
	+\mathcal{H}^{\frac{3}{2}}_{\frac{1}{2},-1}
	\mathcal{H}^{\frac{1}{2}*}_{\frac{1}{2},0}
	+\sqrt{3}\mathcal{H}^{\frac{3}{2}}_{-\frac{1}{2},-1}
	\mathcal{H}^{\frac{1}{2}*}_{-\frac{1}{2},0}
	\notag \\
	&&\quad
	-\mathcal{H}^{\frac{1}{2}}_{-\frac{1}{2},0}
	\mathcal{H}^{\frac{3}{2}*}_{-\frac{1}{2},1}
	-\mathcal{H}^{\frac{1}{2}}_{\frac{1}{2},-1}
	\mathcal{H}^{\frac{3}{2}*}_{\frac{1}{2},0}
	-\sqrt{3}\mathcal{H}^{\frac{1}{2}}_{\frac{1}{2},0}
	\mathcal{H}^{\frac{3}{2}*}_{\frac{1}{2},1}
	\big), \notag \\[4pt]
	L_{8(2)} &=&
	\sqrt{6}q^2\,\mathrm{Re}
	\big(
	\mathcal{H}^{\frac{3}{2}}_{\frac{1}{2},0}
	\mathcal{H}^{\frac{3}{2}*}_{\frac{1}{2},1}
	-\mathcal{H}^{\frac{3}{2}}_{-\frac{1}{2},-1}
	\mathcal{H}^{\frac{3}{2}*}_{-\frac{1}{2},0}
	\big), \notag \\[4pt]
	L_{10(1)} &=&
	\sqrt{2}q^2\,\mathrm{Re}
	\big(
	\mathcal{H}^{\frac{1}{2}}_{-\frac{1}{2},0}
	\mathcal{H}^{\frac{3}{2}*}_{-\frac{1}{2},1}
	-\mathcal{H}^{\frac{1}{2}}_{\frac{1}{2},-1}
	\mathcal{H}^{\frac{3}{2}*}_{\frac{1}{2},0}
	+\sqrt{3}\mathcal{H}^{\frac{1}{2}}_{\frac{1}{2},0}
	\mathcal{H}^{\frac{3}{2}*}_{\frac{1}{2},1}
	\notag \\
	&&
	+\mathcal{H}^{\frac{3}{2}}_{\frac{1}{2},-1}
	\mathcal{H}^{\frac{1}{2}*}_{\frac{1}{2},0}
	-\mathcal{H}^{\frac{3}{2}}_{-\frac{1}{2},0}
	\mathcal{H}^{\frac{1}{2}*}_{-\frac{1}{2},1}
	+\sqrt{3}\mathcal{H}^{\frac{3}{2}}_{-\frac{1}{2},-1}
	\mathcal{H}^{\frac{1}{2}*}_{-\frac{1}{2},0}
	\big), \notag \\[4pt]
	L_{10(2)} &=&
	\frac{\sqrt{6}}{2}q^2\,\mathrm{Re}
	\big(
	\mathcal{H}^{\frac{3}{2}}_{-\frac{1}{2},-1}
	\mathcal{H}^{\frac{3}{2}*}_{-\frac{1}{2},0}
	+\mathcal{H}^{\frac{3}{2}}_{\frac{1}{2},0}
	\mathcal{H}^{\frac{3}{2}*}_{\frac{1}{2},1}
	\big) \notag\\
	L_{9(1)}&=&-L_{8(1)}(Re \rightarrow Im),    \;\; L_{11(1)}=-L_{10(1)}(Re \rightarrow Im) 
\end{eqnarray}
Since all form factors are real, the imaginary part of the amplitude originates entirely from the lineshape, so the amplitude can be further simplified to
\begin{eqnarray}
	L_{1(1)} &=&
	2q^2
	\Big(
	\mathcal{A}_{\frac{1}{2}}^2 |L_{\Lambda^*_{1405}}|^2
	\big(
	|H^{\frac{1}{2}}_{-\frac{1}{2}0}|^2
	+|H^{\frac{1}{2}}_{\frac{1}{2},0}|^2
	\big)
	+\mathcal{A}_{\frac{3}{2}}^2 |L_{\Lambda^*_{1520}}|^2
	\big(
	\frac{5}{8}|H^{\frac{3}{2}}_{-\frac{1}{2},0}|^2
	+\frac{5}{8}|H^{\frac{3}{2}}_{\frac{1}{2},0}|^2
	\big)
	\Big)
	\notag \\
	&&\quad
	+3q^2
	\Big(
	\mathcal{A}_{\frac{1}{2}}^2 |L_{\Lambda^*_{1405}}|^2
	\big(
	|H^{\frac{1}{2}}_{\frac{1}{2},-1}|^2
	+|H^{\frac{1}{2}}_{-\frac{1}{2},1}|^2
	\big)
	+\mathcal{A}_{\frac{3}{2}}^2 |L_{\Lambda^*_{1520}}|^2
	\big(
	\frac{3}{8}|H^{\frac{3}{2}}_{\frac{1}{2},1}|^2
	+\frac{3}{8}|H^{\frac{3}{2}}_{-\frac{1}{2},-1}|^2
	\notag \\
	&&\quad
	+\frac{5}{8}|H^{\frac{3}{2}}_{-\frac{1}{2},1}|^2
	+\frac{5}{8}|H^{\frac{3}{2}}_{\frac{1}{2},-1}|^2
	\big)
	\Big), \notag \\[4pt]
	L_{1(2)} &=&
	4q^2\mathcal{A}_{\frac{3}{2}}\mathcal{A}_{\frac{1}{2}}
	\mathrm{Re}
	\big(
	L^*_{\Lambda^*_{1520}}L_{\Lambda^*_{1405}}
	\big)
	\big(
	H^{\frac{1}{2}}_{-\frac{1}{2},0}
	H^{\frac{3}{2}}_{-\frac{1}{2},0}
	+H^{\frac{1}{2}}_{\frac{1}{2},0}
	H^{\frac{3}{2}}_{\frac{1}{2},0}
	+\frac{3}{2}H^{\frac{1}{2}}_{-\frac{1}{2},1}
	H^{\frac{3}{2}}_{-\frac{1}{2},1}
	\notag \\
	&&+\frac{3}{2}H^{\frac{1}{2}}_{\frac{1}{2},-1}
	H^{\frac{3}{2}}_{\frac{1}{2},-1}
	\big), \notag \\
	L_{1(3)} &=&
	\frac{3}{4}q^2\mathcal{A}_{\frac{3}{2}}^2|L_{\Lambda^*_{1520}}|^2
	\big(
	|H^{\frac{3}{2}}_{-\frac{1}{2},0}|^2
	+|H^{\frac{3}{2}}_{\frac{1}{2},0}|^2
	\big)
	+\frac{9}{8}q^2\mathcal{A}_{\frac{3}{2}}^2|L_{\Lambda^*_{1520}}|^2
	\big(
	|H^{\frac{3}{2}}_{-\frac{1}{2},1}|^2
	-|H^{\frac{3}{2}}_{\frac{1}{2},1}|^2
	\notag \\
	&&\quad
	+|H^{\frac{3}{2}}_{\frac{1}{2},-1}|^2
	-|H^{\frac{3}{2}}_{-\frac{1}{2},-1}|^2
	\big),  \notag \\[4pt]
	L_{2(1)} &=&
	4q^2
	\Big(
	\mathcal{A}_{\frac{1}{2}}^2|L_{\Lambda^*_{1405}}|^2
	\big(
	|H^{\frac{1}{2}}_{-\frac{1}{2},1}|^2
	-|H^{\frac{1}{2}}_{\frac{1}{2},-1}|^2
	\big)
	+\mathcal{A}_{\frac{3}{2}}^2|L_{\Lambda^*_{1520}}|^2
	\big(
	\frac{5}{8}|H^{\frac{3}{2}}_{-\frac{1}{2},1}|^2
	-\frac{5}{8}|H^{\frac{3}{2}}_{\frac{1}{2},-1}|^2
	\notag \\
	&&\quad
	+\frac{3}{8}|H^{\frac{3}{2}}_{\frac{1}{2},1}|^2
	-\frac{3}{8}|H^{\frac{3}{2}}_{-\frac{1}{2},-1}|^2
	\big)
	\Big), \notag \\[4pt]
	L_{2(2)} &=&
	8q^2\mathcal{A}_{\frac{3}{2}}\mathcal{A}_{\frac{1}{2}}
	\mathrm{Re}
	\big(
	L^*_{\Lambda^*_{1520}}L_{\Lambda^*_{1405}}
	\big)
	\big(
	H^{\frac{1}{2}}_{-\frac{1}{2},1}
	H^{\frac{3}{2}}_{-\frac{1}{2},1}
	-H^{\frac{1}{2}}_{\frac{1}{2},-1}
	H^{\frac{3}{2}}_{\frac{1}{2},-1}
	\big), \notag \\[4pt]
	L_{2(3)} &=&
	\frac{3}{2}q^2\mathcal{A}_{\frac{3}{2}}^2|L_{\Lambda^*_{1520}}|^2
	\big(
	|H^{\frac{3}{2}}_{-\frac{1}{2},1}|^2
	-|H^{\frac{3}{2}}_{\frac{1}{2},1}|^2
	+|H^{\frac{3}{2}}_{-\frac{1}{2},-1}|^2
	-|H^{\frac{3}{2}}_{\frac{1}{2},-1}|^2
	\big), \notag \\[4pt]
	L_{3(1)} &=&
	-\frac{\sqrt{3}}{4}q^2\mathcal{A}_{\frac{3}{2}}^2|L_{\Lambda^*_{1520}}|^2
	\big(
	H^{\frac{3}{2}}_{-\frac{1}{2},-1}
	H^{\frac{3}{2}}_{-\frac{1}{2},1}
	+H^{\frac{3}{2}}_{\frac{1}{2},-1}
	H^{\frac{3}{2}}_{\frac{1}{2},1}
	\big) \notag \\[4pt]
	L_{4(1)} &=&
	q^2
	\Big(
	\mathcal{A}_{\frac{1}{2}}^2|L_{\Lambda^*_{1405}}|^2
	\big(
	|H^{\frac{1}{2}}_{\frac{1}{2},-1}|^2
	+|H^{\frac{1}{2}}_{-\frac{1}{2},1}|^2
	\big)
	+
	\mathcal{A}_{\frac{3}{2}}^2|L_{\Lambda^*_{1520}}|^2
	\big(
	\frac{5}{8}|H^{\frac{3}{2}}_{-\frac{1}{2},1}|^2
	+\frac{5}{8}|H^{\frac{3}{2}}_{\frac{1}{2},-1}|^2
	\notag \\
	&&
	+\frac{3}{8}|H^{\frac{3}{2}}_{-\frac{1}{2},-1}|^2
	+\frac{3}{8}|H^{\frac{3}{2}}_{\frac{1}{2},1}|^2
	\big)
	\Big)
	-2q^2\Big(
	\mathcal{A}_{\frac{1}{2}}^2|L_{\Lambda^*_{1405}}|^2
	\big(
	|H^{\frac{1}{2}}_{-\frac{1}{2},0}|^2
	+|H^{\frac{1}{2}}_{\frac{1}{2},0}|^2
	\big)
	\notag \\
	&&+
	\mathcal{A}_{\frac{3}{2}}^2|L_{\Lambda^*_{1520}}|^2
	\big(
	\frac{5}{8}|H^{\frac{3}{2}}_{-\frac{1}{2},0}|^2
	+\frac{5}{8}|H^{\frac{3}{2}}_{\frac{1}{2},0}|^2
	\big)
	\Big), \notag \\[4pt]
	L_{4(2)} &=&
	2q^2\mathcal{A}_{\frac{3}{2}}\mathcal{A}_{\frac{1}{2}}
	\mathrm{Re}
	\big(
	L^*_{\Lambda^*_{1520}}L_{\Lambda^*_{1405}}
	\big)
	\big(
	H^{\frac{1}{2}}_{-\frac{1}{2},1}
	H^{\frac{3}{2}}_{-\frac{1}{2},1}
	+H^{\frac{1}{2}}_{\frac{1}{2},-1}
	H^{\frac{3}{2}}_{\frac{1}{2},-1}
	-2H^{\frac{1}{2}}_{-\frac{1}{2},0}
	H^{\frac{3}{2}}_{-\frac{1}{2},0}
	\notag \\
	&&-2H^{\frac{1}{2}}_{\frac{1}{2},0}
	H^{\frac{3}{2}}_{\frac{1}{2},0}
	\big), \notag \\[4pt]
	L_{4(3)} &=&
	\frac{3}{8}q^2\mathcal{A}_{\frac{3}{2}}^2|L_{\Lambda^*_{1520}}|^2
	\big(
	|H^{\frac{3}{2}}_{-\frac{1}{2},1}|^2
	-|H^{\frac{3}{2}}_{\frac{1}{2},1}|^2
	+|H^{\frac{3}{2}}_{\frac{1}{2},-1}|^2
	-|H^{\frac{3}{2}}_{-\frac{1}{2},-1}|^2
	-2|H^{\frac{3}{2}}_{-\frac{1}{2},0}|^2
	\notag \\
	&&-2|H^{\frac{3}{2}}_{\frac{1}{2},0}|^2
	\big) \notag \\
	L_{8(1)} &=&
	-2\sqrt{2}q^2\mathcal{A}_{\frac{3}{2}}\mathcal{A}_{\frac{1}{2}}\mathrm{Re}
	\big(
	L^*_{\Lambda^*_{1520}}L_{\Lambda^*_{1405}}
	\big)
	\big(
	H^{\frac{3}{2}}_{-\frac{1}{2},0}
	H^{\frac{1}{2}}_{-\frac{1}{2},1}
	+H^{\frac{3}{2}}_{\frac{1}{2},-1}
	H^{\frac{1}{2}}_{\frac{1}{2},0}
	\notag \\
	&&
	+\sqrt{3}H^{\frac{3}{2}}_{-\frac{1}{2},-1}
	H^{\frac{1}{2}}_{-\frac{1}{2},0}
	-H^{\frac{1}{2}}_{-\frac{1}{2},0}
	H^{\frac{3}{2}}_{-\frac{1}{2},1}
	-H^{\frac{1}{2}}_{\frac{1}{2},-1}
	H^{\frac{3}{2}}_{\frac{1}{2},0}
	-\sqrt{3}H^{\frac{1}{2}}_{\frac{1}{2},0}
	H^{\frac{3}{2}}_{\frac{1}{2},1}
	\big), \notag \\[4pt]
	L_{8(2)} &=&
	\sqrt{6}q^2\mathcal{A}_{\frac{3}{2}}^2|L_{\Lambda^*_{1520}}|^2
	\big(
	H^{\frac{3}{2}}_{\frac{1}{2},0}
	H^{\frac{3}{2}}_{\frac{1}{2},1}
	-H^{\frac{3}{2}}_{-\frac{1}{2},-1}
	H^{\frac{3}{2}}_{-\frac{1}{2},0}
	\big), \notag \\[4pt] 
	L_{9(1)} &=&
	-2\sqrt{2}q^2\mathcal{A}_{\frac{3}{2}}\mathcal{A}_{\frac{1}{2}}\mathrm{Im}
	\big(
	L^*_{\Lambda^*_{1520}}L_{\Lambda^*_{1405}}
	\big)
	\big(
	H^{\frac{3}{2}}_{-\frac{1}{2},0}
	H^{\frac{1}{2}}_{-\frac{1}{2},1}
	+H^{\frac{3}{2}}_{\frac{1}{2},-1}
	H^{\frac{1}{2}}_{\frac{1}{2},0}
	\notag \\
	&&\quad
	+\sqrt{3}H^{\frac{3}{2}}_{-\frac{1}{2},-1}
	H^{\frac{1}{2}}_{-\frac{1}{2},0}
	+H^{\frac{1}{2}}_{-\frac{1}{2},0}
	H^{\frac{3}{2}}_{-\frac{1}{2},1}
	+H^{\frac{1}{2}}_{\frac{1}{2},-1}
	H^{\frac{3}{2}}_{\frac{1}{2},0}
	+\sqrt{3}H^{\frac{1}{2}}_{\frac{1}{2},0}
	H^{\frac{3}{2}}_{\frac{1}{2},1}
	\big), \notag \\[4pt]
	L_{10(1)} &=&
	\sqrt{2}q^2\mathcal{A}_{\frac{3}{2}}\mathcal{A}_{\frac{1}{2}}\mathrm{Re}
	\big(
	L^*_{\Lambda^*_{1520}}L_{\Lambda^*_{1405}}
	\big)
	\big(
	H^{\frac{1}{2}}_{-\frac{1}{2},0}
	H^{\frac{3}{2}}_{-\frac{1}{2},1}
	-H^{\frac{1}{2}}_{\frac{1}{2},-1}
	H^{\frac{3}{2}}_{\frac{1}{2},0}
	\notag \\
	&&\quad
	+\sqrt{3}H^{\frac{1}{2}}_{\frac{1}{2},0}
	H^{\frac{3}{2}}_{\frac{1}{2},1}
	+H^{\frac{3}{2}}_{\frac{1}{2},-1}
	H^{\frac{1}{2}}_{\frac{1}{2},0}
	-H^{\frac{3}{2}}_{-\frac{1}{2},0}
	H^{\frac{1}{2}}_{-\frac{1}{2},1}
	+\sqrt{3}H^{\frac{3}{2}}_{-\frac{1}{2},-1}
	H^{\frac{1}{2}}_{-\frac{1}{2},0}
	\big), \notag \\[4pt]
	L_{10(2)} &=&
	\frac{\sqrt{6}}{2}q^2\mathcal{A}_{\frac{3}{2}}^2|L_{\Lambda^*_{1520}}|^2
	\big(
	H^{\frac{3}{2}}_{-\frac{1}{2},-1}
	H^{\frac{3}{2}}_{-\frac{1}{2},0}
	+H^{\frac{3}{2}}_{\frac{1}{2},0}
	H^{\frac{3}{2}}_{\frac{1}{2},1}
	\big) \notag \\[4pt]
	L_{11(1)} &=&
	-\sqrt{2}q^2\mathcal{A}_{\frac{3}{2}}\mathcal{A}_{\frac{1}{2}}\mathrm{Im}
	\big(
	L^*_{\Lambda^*_{1520}}L_{\Lambda^*_{1405}}
	\big)
	\big(
	H^{\frac{1}{2}}_{-\frac{1}{2},0}
	H^{\frac{3}{2}}_{-\frac{1}{2},1}
	-H^{\frac{1}{2}}_{\frac{1}{2},-1}
	H^{\frac{3}{2}}_{\frac{1}{2},0}
	\notag \\
	&&\quad
	+\sqrt{3}H^{\frac{1}{2}}_{\frac{1}{2},0}
	H^{\frac{3}{2}}_{\frac{1}{2},1}
	-H^{\frac{3}{2}}_{\frac{1}{2},-1}
	H^{\frac{1}{2}}_{\frac{1}{2},0}
	+H^{\frac{3}{2}}_{-\frac{1}{2},0}
	H^{\frac{1}{2}}_{-\frac{1}{2},1}
	-\sqrt{3}H^{\frac{3}{2}}_{-\frac{1}{2},-1}
	H^{\frac{1}{2}}_{-\frac{1}{2},0}
	\big),
\end{eqnarray}
In the above expression, to label the corresponding helicity amplitudes, we use the notation $H^{J}_{s_{\Lambda_c},\lambda}$.

\end{document}